\UseRawInputEncoding 
\documentclass[journal]{IEEEtran}
\usepackage{bm}

\usepackage{cite}
\usepackage{amssymb}

\ifCLASSINFOpdf
\usepackage[pdftex]{graphicx}
\else
\fi

\usepackage{caption}
\usepackage{amsmath,amsthm,bm,amssymb}

\newtheorem{prop}{Proposition}

\newtheorem{thm}{Theorem}

\usepackage{amsfonts} 
\usepackage{algorithm}  
\usepackage{algorithmic}
\usepackage{color}
\usepackage[dvipsnames]{xcolor}

\usepackage{array}
\usepackage{subfigure}
\usepackage{url}
\begin{document}
	\title{Virtual Energy Storage Sharing and Capacity Allocation}
	\author{Dongwei~Zhao,~\IEEEmembership{Student~Member,~IEEE,}
		Hao~Wang,~\IEEEmembership{Member,~IEEE,}\\
		Jianwei~Huang,~\IEEEmembership{Fellow,~IEEE,}
		and~Xiaojun~Lin,~\IEEEmembership{Fellow,~IEEE}
		\vspace{-2mm}
		\thanks{This work is supported by the Presidential Fund from the Chinese University of Hong Kong, Shenzhen, China, and in part by the NSF awards: ECCS-1509536. Part of the results have appeared in IEEE ICC 2017 \cite{Dongwei}}
		\thanks{Dongwei Zhao is with the Department of Information Engineering, The Chinese University of Hong Kong, Hong Kong, China (e-mail: zd015@ie.cuhk.edu.hk). Hao Wang is with the Department of Civil and Environmental Engineering and the Stanford Sustainable Systems Lab, Stanford University, CA 94305 USA (e-mail: hwang16@uw.edu). Jianwei Huang is with the  School of Science and Engineering, The Chinese University of Hong Kong, Shenzhen, China and the Department of Information Engineering, The Chinese University of Hong Kong, Hong Kong,  China (e-mail: jianweihuang@cuhk.edu.cn). Xiaojun Lin is with the School of Electrical and Computer Engineering, Purdue University, West Lafayette, IN 47907, USA (e-mail: linx@ecn.purdue.edu).}
		
	}
	\maketitle

	\begin{abstract} 
		Energy storage can play an important role in energy management of end users. To promote an  efficient utilization of energy storage, we develop a novel business model to enable virtual storage sharing among a group of users. Specifically, a storage aggregator invests and operates the central physical storage unit, by virtualizing it into separable virtual capacities and  selling to users. Each user purchases the virtual  capacity, and utilize it to reduce the energy cost. We formulate the interaction between the aggregator and users as a two-stage optimization problem. In Stage 1, over the investment horizon, the aggregator determines the investment and pricing decisions. 
		In Stage 2, in each operational horizon, each user decides the virtual capacity to purchase together with the operation of the virtual storage. {We characterize a stepwise form of the optimal solution of  Stage-2 Problem and a piecewise linear structure of the optimal profit of Stage-1 Problem,  both with respect to the  virtual capacity price. Based on the solution structure, we design an algorithm to attain the optimal solution of the two-stage problem.} In our simulation results, the proposed storage virtualization model can reduce the physical energy storage investment of the aggregator by 54.3\% and reduce the users' total costs by 34.7\%, compared to the case where users acquire their own physical storage.                                                   
	\end{abstract}
	\vspace{-1mm}

	\IEEEpeerreviewmaketitle

	\vspace{-1mm}	
	\section{Introduction}
	\vspace{-1mm}
	\subsection{Background and motivation}
	\vspace{-1mm}
	Energy storage is becoming a crucial element to ensure the stable and efficient operation of the new-generation of power systems. The benefits of the energy storage at the grid side have been well-recognized (e.g., for generation backup, transmission support, voltage control, and frequency regulation) \cite{overview1}. Recently, there has also been an increasing interest in leveraging energy storage for end users (e.g., by harvesting distributed generations, and cutting electrical bill) \cite{overview1}. However, deploying energy storage at the end-user side also faces challenges. On one hand, the current commercial storage products for end users often have high price tags.\footnote{A Tesla Powerwall storage with a capacity of 13.5 kWh costs \$6200\cite{Teslap}.}  Further, since a storage product lasts for years, it is challenging for a user to decide the storage size due to the uncertainty of future energy demand. In fact, the Tesla Powerwall only provides one or two choices of storage size for users.  Both of these factors can discourage  users from purchasing such storage products and enjoying the benefits. On the other hand, if many users invest in energy storage, it is possible for them to cooperate  and share the benefits of storage due to complementary charge and discharge needs. The above considerations  motivate us to study the following problem in the paper: \textit{what would be a good business model that promotes users' more efficient use of energy storage?}
	
	In our work, we develop a novel business model to virtualize and allocate central energy storage resources to end users through a pricing mechanism. This is analogous to the practice of cloud service providers, who set prices for virtualized computing resources shared by end users \cite{cloud2}. In the power system, we can also envision that a storage aggregator invests in a central physical storage unit and then virtualizes it into separable virtual storage capacities that are sold to end users at a suitable price. Users purchase the virtual storage to reduce the energy cost. 
	
	One key advantage of our storage virtualization framework is the ability to leverage users' complementary charge and discharge profiles. Note that the aggregator only cares about the net power flowing in and out the storage. As some users may choose to charge while others choose to discharge in the same time slot, some requests will cancel out at the aggregated level. This suggests that even if all the users are fully utilizing their virtual storage capacity, it is possible to support users' needs by using a smaller central storage comparing with the total virtual storage capacities sold to users. Such complementary charge and discharge profiles can arise in practice due to the diverse load and renewable generation profiles of end users. Specifically, as the most promising sources of clean and sustainable energy, solar and wind energy have both been increasingly adopted by households, commercial buildings, and residential communities \cite{2018solar}\cite{windreport}. Studies in  \cite{haojoint,windz,solarforecast} showed that solar and wind energy exhibit diverse and locational-dependent generation profiles. Similarly, end users' load profiles can also be significantly diverse even in a localized region \cite{pecan}.\footnote{We show the diversity of users' load profiles in the online Appendix O of the technical report \cite{reportjournal} based on the data from\cite{pecan}.}
	
	Another key advantage of storage virtualization is that a user can flexibly change the amount of virtual capacity to purchase  over time based on his varying demand. Such flexibility is difficult to realize if the user owns physical storage by himself, and encourages the users to take advantage of the energy storage. The above key advantages can further increase users' demand for the storage and reduce the aggregator's investment cost, which can increase the aggregator's profit.
	
	To rigorously study such benefits of storage virtualization, in this paper, we consider two possible types of aggregators. The first possibility is a profit-seeking aggregator.  In a deregulated energy market, the profit-seeking storage aggregator can decide whether or not and how much storage capacity to invest in, so as to maximize her profits. Such deregulated markets can be found in the U.S. and  many European countries, and third parties are encouraged to participate in the market to provide different services for the grid and end users \cite{usdereg}\cite{thirdparty}.  The second possibility is that the aggregator is regulated by the system operators or regulatory agents, which may have the goal of maximizing the benefit of end users subject to a nonnegative profit.

	\vspace{-1mm}
	\subsection{Main results and contributions}
	\vspace{-1mm}
	{To the best of our knowledge, our paper is the first work that develops a pricing mechanism for the storage virtualization and sharing.} In such a framework, a storage aggregator invests in a central physical storage unit and then virtualizes it into separable virtual storage capacities that are sold to end users at a suitable price. Users purchase the virtual storage to reduce the energy cost. 
	
	A new question for this storage virtualization  model is how the  aggregator's  investment and pricing decisions are coupled with the users' purchase and storage operation decisions. To answer this question, {we formulate a two-stage optimization problem for the interactions between the aggregator and users at two different horizons: the investment horizon divided into many operational horizons. Over the investment horizon (e.g., 15 years), the aggregator determines the size of the physical  storage for virtualization and the price of the virtual storage.} At the beginning of each operational horizon (e.g., one day), each user determines the virtual capacity to purchase as well as the charge and discharge decision. The aggregator chooses a price of the virtual storage to balance her profit and users' benefits. For a profit-seeking aggregator,  we aim to find the optimal-profit price to maximize her profit. For an aggregator that is regulated by the system operator or regulatory agents,  we aim to find the lowest-nonnegative-profit price, which can give the most benefits to users while maintaining a nonnegative profit for the aggregator. We demonstrate that such a virtualization leads to more efficient use of the physical energy storage, compared with the case where each user acquires his own physical storage.
	
	The main contributions of this paper are as follows:
	\begin{itemize}
		\item \textit{Storage virtualization framework}: In Section \uppercase\expandafter{\romannumeral2}, we develop a storage virtualization and sharing framework. To the best of our knowledge, this is the first work that develops a pricing mechanism for storage virtualization and sharing.
		\item \textit{Pricing-based virtual capacity allocation}: In Section \uppercase\expandafter{\romannumeral3}, we formulate a  two-stage optimization problem between the aggregator and users. In Stage 1,  the aggregator determines the pricing and investment of the storage.   In Stage 2,  each user decides his purchase decision and the  storage schedule. We consider two pricing strategies for the aggregator: one  maximizes the aggregator's profit while the other  gives the highest benefits to users. 	
		\item \textit{{Threshold-based search algorithm}}: In Section \uppercase\expandafter{\romannumeral4}, we resolve a multi-optima issue of Stage 2 by introducing a penalty on users' charge and discharge power. As the penalty approaches zero, we characterize a stepwise structure of users' optimal solutions, and show a piecewise linear structure of the aggregator's optimal profit with respect to the  virtual storage price. {This structure then allows us to iteratively search for near-optimal investment and pricing strategies within an arbitrary precision.}
		\item \textit{Realistic-data simulations}: In Section \uppercase\expandafter{\romannumeral5}, we conduct the simulation using realistic load data from PG\&E Corporation and meteorology data from Hong Kong Observatory. We show that our model enables the aggregator to save the physical storage investment cost  by 54.3\% and the users to reduce energy costs by 34.7\%, compared with the case where users acquire their own physical storage.
	\end{itemize}

	\vspace{-1mm}
	\subsection{Related works}
	\vspace{-1mm}
	There have been several studies on the deployment of energy storage at the end-user side\cite{operationstorage,stodemand,stodemand5,DRShareSto,storagesharinggame,central3,central2,stosharingauc,bistoragesharing,cloudstorage}. In   \cite{operationstorage,stodemand,stodemand5,DRShareSto}, each user only utilizes  his own energy storage units for demand management without mutual sharing, which may lead to inefficient use of the storage. In contrast, the works in \cite{storagesharinggame, central3,central2} considered user sharing of a central storage without considering the investment issue of the central storage and the potential impact on the users. In  \cite{stosharingauc} and \cite{bistoragesharing},  end users share the energy storage with a third party. However, both works allocate the physical storage capacities (instead of virtual capacities) to the users, which does not take advantage of the complementariness of users' profiles.
	
	The work that is most closely related to ours is \cite{cloudstorage}, in which the authors proposed a business model to enable users to share the central storage. Our work differs from \cite{cloudstorage} in several crucial ways. First, in \cite{cloudstorage} the storage sizing decisions are not coordinated between the storage aggregator and users.  More specifically, the aggregator needs to invest in a sufficiently large capacity to satisfy users' needs, which is not cost-efficient. In contrast, in our work, the aggregator can adjust the price of virtual storage to influence  the demand of storage, and effectively coordinate the benefit sharing of virtual storage between the aggregator and users. Second, the model in \cite{cloudstorage} assumes that a user's purchased virtual storage cannot change on a daily basis. In contrast, our model allows  a user to flexibly choose the amount of virtual storage to purchase every day, depending on his daily renewable generation and load demand. This additional level of flexibility further explores the potential of storage virtualization and reduces users' cost.
	
	Our work models the virtual storage sharing framework as a two-stage optimization problem. Such multi-stage problems have been studied in smart grid systems (e.g.,\cite{bistoragesharing,retailer,haojoint,haobargain}). The work  \cite{bistoragesharing}  built a two-stage optimization problem for the  sharing of a  central storage unit between a distribution company and customers. The work  \cite{retailer} proposed a two-stage model for the energy pricing and dispatch problem of the electricity retailers. Both works  \cite{bistoragesharing} and \cite{retailer}  solved the two-stage problem by constructing a single optimization problem, which requires the operator to know all the users' private information. Compared with \cite{bistoragesharing} and \cite{retailer} that require complete network information, our work designs a distributed algorithm based on the information exchange between users and  the  aggregator.
	
	The works  \cite{haojoint} and \cite{haobargain}  designed distributed algorithms based on the information exchange with an aggregator to coordinate the decisions among different users or microgrids. Such distributed algorithms can be used to realize the concept of  transactive energy in the smart grid, which can achieve an equilibrium by exchanging value-based information \cite{localenergymarkette}. In such a transactive energy  framework, the  agents of   mid- or small-sized energy resources can automatically negotiate  with each other as well as exchange information with the main grid through advanced energy management and control system. In our work, users can actively and automatically  respond to the price signal from the aggregator, which is supported by the transactive energy framework.  Compared with  \cite{haojoint} and \cite{haobargain}, our work focuses on the pricing mechanism of the aggregator who sells the virtual storage capacities to users and seeks the profits, while the works \cite{haojoint} and \cite{haobargain} focused on the  coordination between users (or microgrids) under the help of the aggregator to reach the social optima or consensus.

	\vspace{-1mm}		
	\section{System Overview}
\begin{figure}[t]
		\centering
		\includegraphics[width=2.4in]{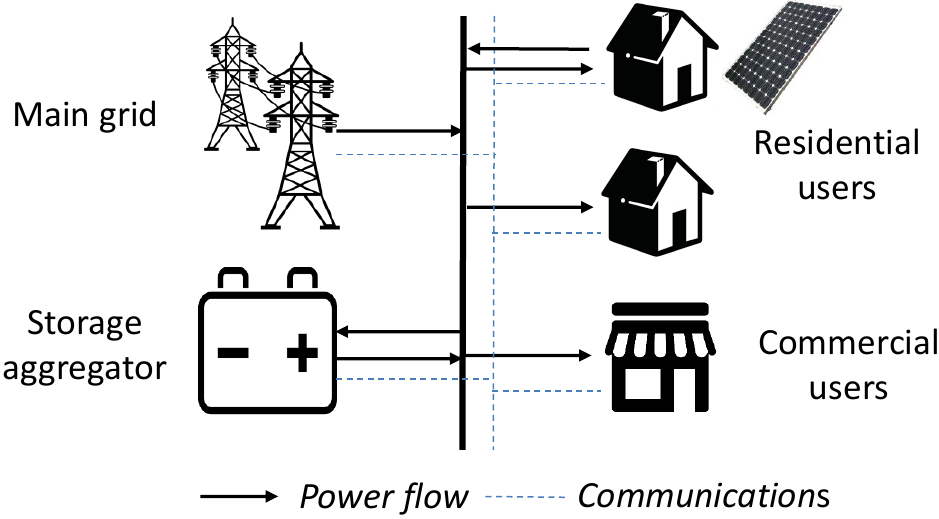}
		\vspace{-1mm}
		\caption{\small System structure.}
		\label{fig:structure}
		\vspace{-3mm}
	\end{figure}

	\vspace{-1mm}	
	Figure \ref{fig:structure}  illustrates the system model, where a community of users are connected with a central storage unit and the main grid (and with each other) through power and communication infrastructures.\footnote{We assume that the grid constraints are not stringent, so that we can focus on how the aggregator sets the price of the virtual storage and how storage virtualization reduces the requirement of physical storage and the users' costs. We will consider the grid constraints in the future work.} Each user has his load demand and may also own some local renewables. An aggregator invests and operates the central storage unit. Next, we introduce the models of the users and the aggregator in more details. 
	\vspace{-1mm}
	\subsection{Users}
	\vspace{-1mm}
	We consider a set of users $\mathcal{I}=\{1,\dots,I\}$ whose energy load profiles can be different. Users may own renewables of solar and wind energy. To satisfy the demand, a user can use the locally generated renewable energy,  purchase energy from the main grid, or use the energy from the energy storage. Next, we first introduce the user's electricity bill, and then discuss how storage can be used to reduce the electricity bill.
	
	We adopt a peak-based demand charge tariff for the electricity bill. {Peak-based demand charge has been widely adopted for commercial and industrial consumers in order to reduce the system peak and recover grid costs. Thanks to the increasingly more advanced metering infrastructure, such a demand charge scheme has also been  offered to residential customers by some utilities in the United States \cite{wood2016recovery} \cite{dech}.} For a billing cycle $\mathcal{T}=\{1,2,...,T\}$ of $T$ time slots, if user $i$ consumes electricity $p_i^g[t]$ from the grid in time slot $t$, his electricity bill\cite{peak} in $\mathcal{T}$ is calculated by:
	\vspace{-1mm}
	\begin{align}
	\pi_b \sum_{t\in \mathcal{T}}p_i^g[t]+\pi_p \max_{t\in \mathcal{T}}\ p_i^g[t],\label{eq:tariff}
	\end{align}\par	\vspace{-1mm}
	\noindent where $\pi_b$ is the unit energy price and $\pi_p$ is the unit price for peak consumption in the billing cycle. To reduce users' peak demand, the utility  usually sets $\pi_p$ much higher than $\pi_b$ \cite{dech}. 
	
	Demand charge tariff in \eqref{eq:tariff} provides a strong incentive for users to utilize energy storage to shave their peak loads. Specifically, users can proactively charge their storage using energy from the grid, and discharge to meet the peak load so as to reduce the electricity bill. Furthermore, if a user owns the renewables, he can store excessive renewable energy in the storage for later use. We assume that users can sell back renewable energy to the grid and the unit feed-in price $\pi_s$ satisfies $\pi_s < \pi_b$, such that users prefer to first use the locally generated renewable energy to serve their loads rather than to directly sell to the grid.\footnote{It is common that the renewable feed-in tariff is lower than the consumption tariff, for example, in Germany and  some states of the U.S.\cite{nature}.}
	
	\vspace{-1mm}
	\subsection{Storage aggregator}
	The aggregator invests and operates the central physical storage. She virtualizes the physical storage into separable virtual capacities and sells them to users. Since users can't control the central storage directly, they report their charge and discharge decisions to the aggregator, and the aggregator dispatches the central storage on behalf of users accordingly. Further, the aggregator can coordinate users' charge and discharge decisions by setting the price of virtual storage, which will ensure that users' charge and discharge decisions are well accommodated by the physical storage.
	
	We assume that there is a billing arrangement among the utility, the storage aggregator and users such that when the utility calculates  users' electricity bill, it will count both the physical load and the virtual storage charge/discharge.  Thus, even though users don't own and operate their physical storage, users can use the virtual storage to achieve a peak load reduction and reduce the electricity bill.
	
	As we have discussed  in Section \uppercase\expandafter{\romannumeral1}, our storage virtualization model can lead to a  more efficient use of physical storage due to two reasons: (i) the complementarity of different users' charge and discharge decisions, and (ii) the flexibility in purchasing different amounts of virtual capacities on different days. The aggregator's investment in the physical storage will take advantage of these aspects while satisfying users' demand.     However, occasionally there can be very high aggregate demand from users. Satisfying such demand with a fixed physical storage investment will lead to low efficiency due to either over-investment or over-pricing. Thus, we further generalize our model  by  allowing the aggregator to use additional energy resources other than the physical storage  to meet users' demand. For example, the aggregator can contract with other generators (or consumers) to purchase  additional energy (or sell surplus energy)  to serve users' demand.\footnote{Such a generalization can be supported by the works \cite{cheng2016virtual} and \cite{cheng2017virtual}, which propose a control framework for the general energy storage system  by aggregating other energy resources, e.g., demand responses in addition to the physical energy storage.}
	
	\vspace{-1mm}				
	\section{Two-stage Formulation}
\begin{figure}[t]
		\centering
		\includegraphics[width=2.6in]{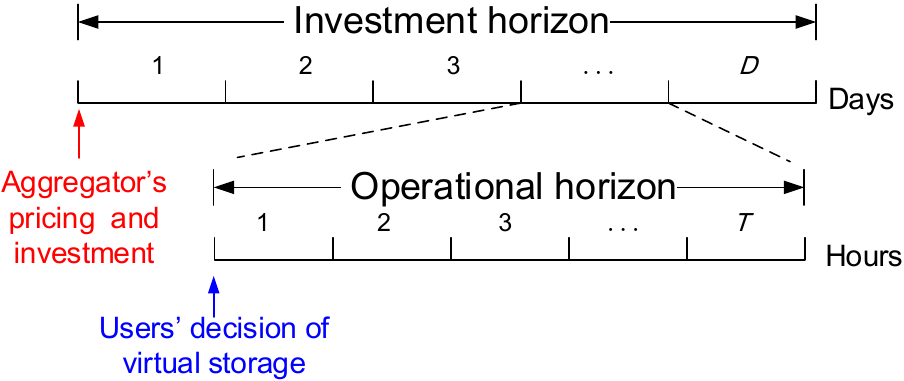}
		\vspace{-1mm}
		\caption{\small Decision making over two timescales.}
		\label{fig:time0}
		\vspace{-1mm}
	\end{figure}

\begin{figure}[t]
		\centering
		\includegraphics[width=3in]{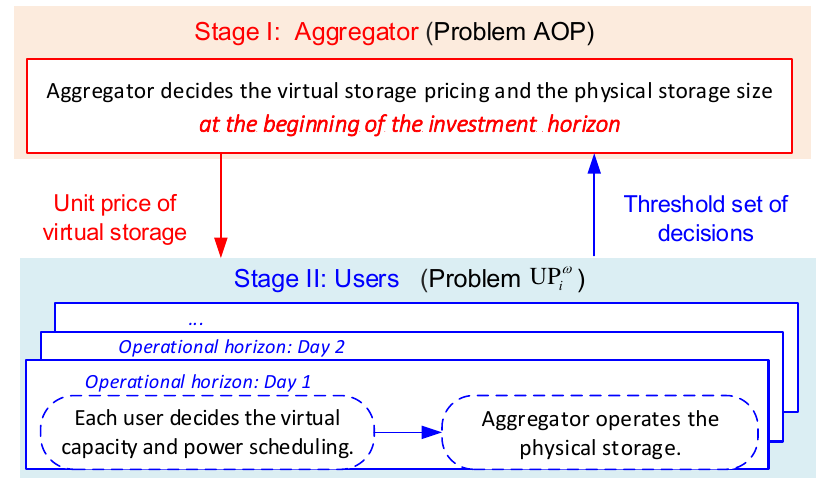}
		\caption{\small Two-stage optimization.}
		\label{fig:sos0}
		\vspace{-1mm}
	\end{figure}

	\vspace{-1mm}	
	Figure \ref{fig:time0} illustrates two timescales of decision making in our model. Figure \ref{fig:sos0} illustrates a two-stage problem for the interactions between the aggregator and users.  In Stage 1, at the beginning of an investment horizon $\mathcal{D}\hspace{-1mm}=\hspace{-1mm}\{1,2,...,D\}$ of $D$ days (e.g., $D$ corresponding to many years), the aggregator determines  the size of the physical storage  and the unit price of the virtual storage. The investment horizon is divided into many operational horizons, i.e.,  each $d\in \mathcal{D}$ corresponds to one operational horizon, which is further divided into many time slots $\mathcal{T}\hspace{-1mm}=\hspace{-1mm}\{1,2,...,T\}$ (e.g., 24 time slots corresponding to 24 hours).  In Stage 2, at the beginning of each operational horizon,  given the unit price of virtual storage, each user decides the optimal capacity to purchase  and the corresponding charge and discharge profiles over the operational horizon, based on the prediction of their loads and renewable generations.\footnote{Since the focus of our work is on the design of the virtual storage sharing framework, we have initially chosen to assume that users can perfectly predict their renewable generations and loads. We include the discussions about the impact of uncertainties on the users' decisions in the online Appendix N \cite{reportjournal}.} Then, the aggregator operates the physical storage by aggregating all the users' charge and discharge decisions. Note that we consider a daily operation of virtual storage sharing as well as the daily demand charge tariff for users' electricity bills, because users' electricity loads reflect their activities which are often periodic on a daily basis (see, e.g., extensive studies in\cite{loadcurve} and\cite{dailyload2}). Furthermore, users' load profiles can differ from one day to another (e.g., the differences between weekdays and weekends). The operation and billing cycle on a daily basis can leverage the diversity in users' loads and provide flexibility to users, such that users can purchase a different amount of virtual storage on different days to minimize their costs.
	
	In order to solve the aggregator's investment and pricing problem over the investment horizon, the aggregator needs to incorporate users' responses across different operational horizons in the entire investment horizon. Since users' responses depend on different operational conditions (e.g., local renewable generations and loads), we use historical data to build a set of \emph{scenarios} $\Omega$ that  empirically models the joint distribution of all users' load and renewable generation profiles. For each operational horizon $d$, scenario $\omega \in \Omega$ occurs with a probability $\rho^\omega$.  In scenario $\omega$, we denote user $i$'s load profile as $\bm{P}_i^{\omega,l}\hspace{-1mm}=\{P_i^{\omega,l}[t],~\forall t \in \mathcal{T}\}$ and his renewable profile as $\bm{P}_i^{\omega,r}\hspace{-1mm}=\{P_i^{\omega,r}[t],~\forall t \in \mathcal{T}\}$. Each user will report a set of the threshold decisions (explained in detail later in Algorithm 1) in each operational scenario to the aggregator. Based on users' reported information, the aggregator makes the investment and pricing decisions over the investment horizon  by considering the expected profit over the scenarios. 
	
	Furthermore, note that the aggregator's decision and users' decisions in two stages are coupled. On the one hand, the aggregator's virtual storage pricing will affect the users' decisions of virtual storage, and the aggregator's invested physical storage size will constraint the aggregated charge and discharge decisions of users.  On the other hand, the aggregated charge and discharge decisions of users will determine the aggregator's  operation of the physical storage.  Such a coupled two-stage problem needs to be solved through backward induction. Thus, in the next two subsections, we will first explain users' model in Stage 2 and then explain the aggregator's  model in Stage 1. 
	
	
	\vspace{-1mm}			
	\subsection{Stage 2: User's model }
	\vspace{-1mm}			
	\subsubsection{User's power scheduling in each operational horizon}
\begin{figure}[t]
		\centering
		\includegraphics[width=2.7in]{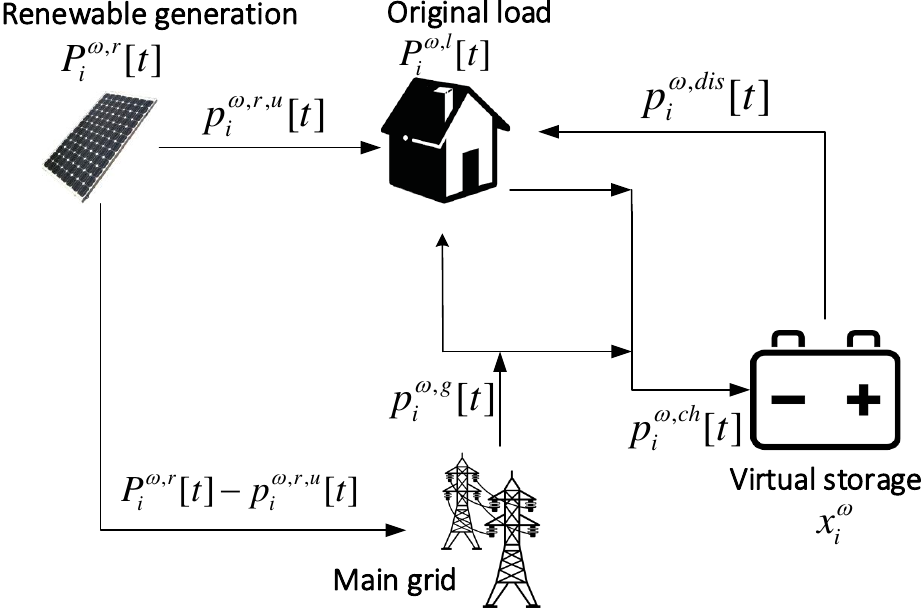}
		\vspace{-1mm}
		\caption{\small Users' power scheduling.}
		\label{fig:power}
		\vspace{-4mm}
	\end{figure}

	Given the price $q$ of the virtual capacity, user $i$ decides  the virtual  capacity $x_i^\omega$  and the corresponding power scheduling (as illustrated in Figure \ref{fig:power}). We then explain the power scheduling in time slot $t$.  Assume that user $i$ has locally generated renewable energy $P_i^{w,r}[t]$. He decides the amount of self-used renewable energy $p_i^{\omega,r,u}[t]$ that will serve his load or charge into his  virtual storage. He sells back the remaining renewable energy $P_i^{\omega,r}[t]-p_i^{\omega,r,u}[t]$ to the grid.\footnote{We assume that users only feed in the unused locally-generated renewable energy to the grid. We do not  consider the feed-in power from the storage to the grid, which may complicate users' decisions.} User $i$ can purchase the amount of energy $p_i^{\omega,g}[t]$ from the grid. Part of  energy $p_i^{\omega,g}[t]$ may serve his own load, and the remaining part will be charged into his virtual storage.\footnote{We assume that users can have an extra meter and  wires to connect the renewable generators to the grid. Hence, it is technically feasible for a user to sell back the renewable energy to the grid while consuming energy from the grid.} Finally, user $i$ can discharge the amount of energy $p_i^{\omega,dis}[t]$ from his virtual storage to serve his load. To balance the power, we can express user $i$'s energy purchase from the grid in time slot $t$ as follows:
	\vspace{-1mm}
	\begin{align}
	p_i^{\omega,g}[t]= P_i^{\omega,l}[t]-p_i^{\omega,r,u}[t]-p_i^{\omega,dis}[t]+p_i^{\omega,ch}[t]\label{eq:2}.
	\end{align} \par \vspace{-1mm}
	\noindent If a user has no renewables, the  charged energy is only from the purchase from the grid, i.e.,   $p_i^{\omega,r,u}[t]=0$ in \eqref{eq:2}. We denote  $\bm{p}_i^{\omega,g}=\{{p}_i^{\omega,g}[t],\forall t \in \mathcal{T}\}$,  $\bm{p}_i^{\omega,r,u}=\{{p}_i^{\omega,r,u}[t],\forall t \in \mathcal{T}\}$, $\bm{p}_i^{\omega,ch}=\{{p}_i^{\omega,ch}[t],\forall t \in \mathcal{T}\}$, and $\bm{p}_i^{\omega,dis}=\{{p}_i^{\omega,dis}[t],\forall t \in \mathcal{T}\}$.
	
	User $i$'s charge and discharge decision should satisfy the constraint of the virtual capacity:
	\vspace{-1mm}
	\begin{align}
	&e_i^\omega[t]=e_i^\omega[t-1]+\eta^c p_i^{\omega,ch}[t]-p_i^{\omega,dis}[t]/\eta^d,\ \forall t \in \mathcal{T}, \label{eq:6}\\
	& 0 \leq e_i^\omega[t]\leq x_i^\omega,~\forall t \in \mathcal{T'},\label{eq:7}　\\
	&e_i^\omega[0]=e_i^\omega[T].\label{eq:8}
	\end{align}\par \vspace{-1mm}
	\noindent We let  $\bm{e}_i^\omega=\{e_i^\omega[t],~ \forall t \in \mathcal{T}'\}$ denote the energy level in the storage over the operational horizon, where $\mathcal{T}'=\{0\}\bigcup\mathcal{T}$ and $e_i^\omega[0]$ denotes the initial energy level.  Since the user's storage is virtual, we allow user $i$  to optimize $e_i^\omega[0]$ in each operational horizon.  We let $\eta^c$ and $\eta^d$ denote the virtual charge and discharge efficiency rate respectively. We assume that the aggregator enforces the  same virtual efficiency rate as the physical one. {We model the charge and discharge efficiencies for the physical storage (e.g., Li-ion batteries) as constant values. Such an assumption has been widely used in the literature (e.g., \cite{overview1}\cite{operationstorage}) and can capture the key characteristics of energy loss during the charging and discharging process.} Constraint \eqref{eq:8} ensures  the independent operation of the virtual storage across operational horizons \cite{hao2}. Other power-related variables are constrained as follows:
	\vspace{-1mm}
	\begin{align}
	&p_i^{\omega,g}[t] \geq 0,~p_i^{\omega,ch}[t]\geq 0,~p_i^{\omega,dis}[t]\geq 0,~\forall t \in\mathcal{T}, \label{eq:9}\\  &0\leq p_i^{\omega,r,u}[t]\leq P_i^{\omega,r}[t],~\forall t \in\mathcal{T}.\label{eq:10}
	\end{align}
	
	\vspace{-1mm}
	\subsubsection{User's net cost in each operational horizon}
	Each user minimizes his net cost in each operational horizon. The cost includes the payment  for the  virtual  capacity and the  electricity bill. The revenue is from selling back the renewable energy. 
	
	Specifically, given the  virtual storage price $q$, over the operational horizon of scenario $\omega$, user $i$'s \textit{payment  to the aggregator} for purchasing the capacity $x_i^\omega$  is $C_i^s(x_i^\omega)=qx_i^\omega.$
	The \textit{ electricity bill} for the consumption from the grid is
	\vspace{-1mm}
	\begin{align}
	C_i^e(\bm{p}_i^{\omega,g})=\pi_b\sum_{t\in \mathcal{T}}	p_i^{\omega,g}[t]+\pi_p \max_{t\in \mathcal{T}} 	 p_i^{\omega,g}[t],\label{eq:3}
	\end{align}\par	\vspace{-1mm} 
	\noindent and we can substitute the variable $p_i^{\omega,g}[t]$ from \eqref{eq:2} and denote the electricity bill as $C_i^e(\bm{p}_i^{\omega,r,u},\bm{p}_i^{\omega,ch},\bm{p}_i^{\omega,dis})$. User $i$'s  \textit{revenue of  selling back renewable energy}  is 
	\vspace{-1mm}
	\begin{align}
	R_i^r(\bm{p}_i^{\omega,r,u})=\pi_s \sum_{t\in \mathcal{T}}{ ({P}_i^{\omega,r}[t]-p_i^{\omega,r,u}[t])}.\label{eq:4}
	\end{align}\par	\vspace{-1mm} 
	\noindent Thus, the \textit{net cost}  in the operational horizon of scenario $\omega$ is
	\vspace{-1mm} 
	\begin{align}
	C_i^s(x_i^\omega)+C_i^e(\bm{p}_i^{\omega,r,u},\bm{p}_i^{\omega,ch},\bm{p}_i^{\omega,dis})-R_i^r(\bm{p}_i^{\omega,r,u}).
	\end{align}\par	\vspace{-1mm} 
	
	We then formulate user $i$'s Problem $\textbf{UP}_i^\omega$ that minimizes the net cost in the operational horizon of scenario $\omega$  as follows.
	
	\textbf{Stage 2: User $i$ optimization problem $\textbf{UP}_i^\omega$}
	\vspace{-1mm}
	\begin{align*}
	\min  \ \ 	&C_i^s(x_i^\omega)+C_i^e(\bm{p}_i^{\omega,r,u},\bm{p}_i^{\omega,ch},\bm{p}_i^{\omega,dis})-R_i^r(\bm{p}_i^{\omega,r,u})\\
	\text{s.t.} \ \ &\eqref{eq:2}, \eqref{eq:6}-\eqref{eq:10},\\
	\text{var:}\ \  &x_i^\omega,\bm{p}_i^{\omega,r,u},\bm{p}_i^{\omega,ch},\bm{p}_i^{\omega,dis}, \bm{e}_i^\omega,
	\end{align*}
	where the price $q$ is  determined by the aggregator in Stage 1.  We denote the optimal solution to Problem $\textbf{UP}_i^\omega $  by $\left(x_i^{\omega\ast}(q), \bm{p}_i^{\omega,r,u*}(q), \bm{p}_i^{\omega,ch*}(q), \bm{p}_i^{\omega,dis*}(q),  \bm{e}_i^{\omega*}(q)\right)$. Note that the aggregator sets the same price for all users without any price discrimination, and each user $i$ makes his own purchase decision to minimize his cost by solving Problem $\textbf{UP}_i^\omega$. Therefore,  if the virtual storage can bring higher revenues to some users (e.g., those who have more renewable energy), then these users have higher demands and purchase more virtual capacities (than the users who benefit less using the storage). In this sense,  our business model is fair for all users.

	\vspace{-1mm}		
	\subsection{Stage 1: Aggregator's model}
	\vspace{-1mm}
	
	In Stage 1,  at the investment phase, the aggregator decides the unit price $q$ of the virtual capacity, as well as the capacity $X$ and power rating $P$ of the physical storage.
	
	\subsubsection{Aggregator's power scheduling over each operational horizon}
	Each user decides the charge and discharge profiles (as in Stage 2) and then reports them to the aggregator. The aggregator aggregates these charge and discharge decisions and obtains the net charge $\bm{p}_a^{\omega,ch}(q)=\{{p}_a^{\omega,ch}[t],~\forall t \in \mathcal{T}\}$ and discharge $\bm{p}_a^{\omega,dis}(q)=\{{p}_a^{\omega,dis}[t],~\forall t \in \mathcal{T}\}$ as follows:
	\vspace{-1mm}
	\begin{align}
	&p_a^{\omega,ch}[t](q)=\left[\sum_{i\in \mathcal{I}}p_i^{\omega, ch*}[t](q)-\sum_{i\in \mathcal{I}} p_i^{\omega,dis*}[t](q)\right]^+,\label{eq:12}\\
	&p_a^{\omega,dis}[t](q)=\left[\sum_{i\in \mathcal{I}} p_i^{\omega,dis*}[t](q)-\sum_{i\in \mathcal{I}}p_i^{\omega, ch*}[t](q)\right]^+, \label{eq:13}
	\end{align}\par\vspace{-1mm}
	\noindent $\forall t \in \mathcal{T},~\forall \omega \in {\Omega}$, where we define $ [f]^+=\max\{f,0\}$. {Note that \eqref{eq:12} and \eqref{eq:13} ensure that $p_a^{\omega,ch}[t](q)$ and $p_a^{\omega,dis}[t](q)$ cannot be positive at the same time, i.e., the physical storage cannot be charged and discharged simultaneously.}
	
	The aggregator can use the physical storage and additional resources to satisfy users' requirement. We denote the charge and discharge requirement served by the physical storage as $\bm{p}_a^{\omega, ch,s}\hspace{-1mm}=\hspace{-1mm}\{{p}_a^{\omega, ch,s}[t],\forall t\in \mathcal{T}\}$, $\bm{p}_a^{\omega,dis,s}\hspace{-1mm}=\hspace{-1mm}\{{p}_a^{\omega, dis,s}[t],\forall t\in \mathcal{T}\}$. Similarly, we denote the charge discharge and requirement supported by the additional resources as $\bm{p}_a^{\omega, ch,a}\hspace{-1mm}=\hspace{-1mm}\{{p}_a^{\omega, ch,a}[t],~\forall t\in \mathcal{T}\}$, $\bm{p}_a^{\omega,dis,a}\hspace{-1mm}=\hspace{-1mm}\{{p}_a^{\omega, dis,a}[t],~\forall t\in \mathcal{T}\}$. They satisfy the following constraints for users' demand:
	\vspace{-1mm}
	\begin{align}
	&{p}_a^{\omega, ch,s}[t]+{p}_a^{\omega, ch,a}[t]=p_a^{\omega,ch}[t](q),\forall t \in \mathcal{T},\forall \omega \in {\Omega},\label{eq:14}\\
	&{p}_a^{\omega, dis,s}[t]+{p}_a^{\omega, dis,a}[t]=p_a^{\omega,dis}[t](q),\forall t \in \mathcal{T},\forall \omega \in {\Omega}.\label{eq:15}
	\end{align}\par \vspace{-1mm}
	
	The aggregator's charge and discharge scheduling is constrained by the physical storage size as follows: 
	\vspace{-1mm}
	\begin{align}
	&e_a^\omega[t]=e_a^\omega[t-1]+\eta_a^c p_a^{\omega,ch,s}[t]-p_a^{\omega,dis,s}[t]/\eta_a^d,\notag \\&\ \ \ \ \ \ \ \ \ \ \ \ \ \ \ \ \ \  \ \ \ \ \ \ \ \ \   \ \ \ \ \ \ \ \ \   \ \ \ \ \ \ \ \forall t \in \mathcal{T}, \forall \omega \in {\Omega},\label{eq:27}\\
	& \gamma^{\text{min}} X \leq e_a^\omega [t]\leq \gamma^{\text{max}} X,\forall t \in \mathcal{T'},\forall \omega \in {\Omega},\label{eq:28}\\
	&{p}_{a}^{\omega,ch,s}[t]\leq P,{p}_{a}^{\omega,dis,s}[t]\leq P, \forall t \in \mathcal{T},\forall \omega \in {\Omega}, \label{eq:30}\\
	&e_a^\omega[0]=e_a^\omega[T], \forall \omega \in {\Omega}.\label{eq:ief}
	\end{align}\par \vspace{-1mm}
	\noindent We let  $\bm{e}_a^\omega=\{e_a^\omega[t], \forall t \in \mathcal{T}'\}$ denote the energy level in the physical storage.  We let $\eta_a^c$ and $\eta_a^d$ denote the charge and discharge efficiency rate respectively. The fraction coefficients $\gamma^{\text{min}}$ and $\gamma^{\text{max}}$ correspond to the minimum and maximum energy levels that can be stored in the storage, respectively. Constraint \eqref{eq:ief} ensures the independent operation of storage in each operational horizon.\footnote{We assume that the aggregator can adjust the  initial energy level of the  storage by purchasing or selling the energy with the same price through an external market. Since the storage operational constraint in \eqref{eq:ief} restricts the terminal level to be equal to the initial level, the total adjustment and its cost is negligible in the long run. In addition, the case where the initial energy level is fixed can be viewed as a special case of our problem.}	
	
	\subsubsection{Aggregator's profit over the investment phase (scaled in one operational horizon)} Over the investment phase, the aggregator bears the storage capital cost which includes the capacity cost and power rating cost \cite{overview1}. In each operational horizon, the aggregator obtains the revenue from selling the virtual capacity  but also bears the operational cost of the physical storage and additional resources.
	
	Specifically, the \textit{capital cost $C_a^{cap}(X,P)$}  over the investment phase (scaled into one operational horizon) is: 
	\vspace{-1mm}
	\begin{align}
	C_a^{cap}(X,P)=\kappa c^{X} X+\kappa c^{P} P,
	\end{align}\par\vspace{-0.5mm}
	\noindent	where the coefficient $c^X$ and $c^P$ are the unit cost of the capacity and power rating over the investment phase, respectively.\footnote{Here we consider a single type of storage technology for the central storage. We can easily generalize this model to incorporate  multiple types.}  {The coefficient $\kappa$ is the scaling factor that is illustrated in the online Appendix L \cite{reportjournal}.
		
		The expectation of the  \textit{revenue  $R_a^{v}(q)$ of selling the virtual storage capacity} over all scenarios is
		\vspace{-0.5mm}
		\begin{align}
		R_a^{v}(q)=\mathbb{E}_{\omega \in \Omega}[q\sum_{i\in \mathcal{I}} x_i^{\omega*}(q)]=\sum_{\omega \in \Omega}\rho^\omega q\sum_{i\in \mathcal{I}} x_i^{\omega*}(q).\label{eq:rev}
		\end{align}	\par\vspace{-1mm}
		
		The expectation of the \textit{storage operational cost $C_a^{op}(\bm{p}_a^{\omega, ch,s},\bm{p}_a^{\omega,dis,s})$}  over all scenarios is
		\vspace{-0.5mm}
		\begin{align}
		\sum_{\omega \in \Omega }\rho^\omega c^s \sum_{t\in \mathcal{T}} ( {p}_a^{\omega, ch,s}[t]+{p}_a^{\omega, dis,s}[t]),
		\end{align}\par{\vspace{-1mm}}
		\noindent where $c^s$ is the unit cost of the charge and discharge, which models the cost of degradation of the storage. We adopt the linear cost model that is widely used in the literature\cite{hao2}\cite{datasto}. 		
		
		The  expected \textit{operational cost  of  additional resources} $C_a^{ ad}(\bm{p}_a^{\omega, ch,a},\bm{p}_a^{\omega,dis,a})$ over all scenarios is
		\vspace{-1mm}
		\begin{align}
		\sum_{\omega \in \Omega }\rho^\omega \sum_{t\in \mathcal{T}}(c_a^{c} {p}_a^{\omega, ch,a}[t]\hspace{-1mm}+\hspace{-1mm}c_a^{d}{p}_a^{\omega, dis,a}[t]),
		\end{align}\par\vspace{-1mm}
		\noindent where $c_a^{c}$ is the unit cost of absorbing  users' charge demand and $c_a^{d}$ is the unit cost of acquiring the energy to support users' discharge demand. 		
		
		Thus, the aggregator's expected \textit{profit} over the investment phase, scaled into one operational horizon, is 
		\vspace{-1mm}
		\begin{align}
		&R_a^{v}(q)-C_a^{cap}(X,P)-C_a^{op}(\bm{p}_a^{\omega, ch,s},\bm{p}_a^{\omega,dis,s}) \notag \\&\ \ \ \ \ \ \ \ \ \ \ \ \ \ \ \ \ \ \ \ \ \ \ \ \ \ \ \ \ \ \  -C_a^{ad}(\bm{p}_a^{\omega, ch,a},\bm{p}_a^{\omega,dis,a}).\label{eq:profit}
		\end{align} \par		\vspace{-1mm}
		
		The aggregator's power scheduling is determined  by price $q$ and investment decisions $X$ and $P$. Thus we denote the operational cost of storage and additional resources  as  functions of $(q, X, P)$ respectively, i.e.,  $C_a^{op}(q,X,P)$ and  $C_a^{ad}(q,X,P)$.
		
		\subsubsection{Aggregator's pricing}
		We consider two pricing strategies for the aggregator: the \textit{optimal-profit price} (OP price) and the  \textit{lowest-nonnegative-profit price} (LNP price). The \textit{OP price} maximizes the aggregator's profit, and the \textit{LNP price} is the minimal  price that keeps the aggregator's profit nonnegative.  The latter is reasonable when the aggregator  is regulated and required to provide the most benefit for users. We formulate Problem $\textbf{AOP}$ to obtain the OP price $q^\star$ as follows. Due to the page limit, we enclose all the discussions about the  problem of  the LNP price $q^l$ in the online Appendix J\cite{reportjournal}. 
		
		\noindent\textbf{Stage 1: Aggregator's Optimal-profit Price Problem ($\textbf{AOP}$)}
		\vspace{-4mm}
		\begin{align*}
		\max\  &{R_a^{pf}}  \hspace{-1mm}:=\hspace{-1mm} R_a^{v}(q)\hspace{-1mm}-\hspace{-1mm}C_a^{cap}(X,P)   \hspace{-1mm}-\hspace{-1mm}C_a^{op}(q,X,P)\hspace{-1mm}-\hspace{-1mm}C_a^{ad}(q,X,P)\\
		\text{s.t.} \  & \eqref{eq:12}-\eqref{eq:15},\eqref{eq:27}-\eqref{eq:ief},\\
		\text{var:}\   &q, X, P.
		\end{align*}
		
		\vspace{-1mm}	
		\section{Solving Two-stage Problem}
		\vspace{-1mm}
		{The two-stage problem is challenging to solve due to its non-convex nature. To solve the problem, we first characterize the properties of each user's optimal solution (in Stage 2) under a fixed price $q$ in Proposition \ref{prop:capacity} and Theorem 1, and then incorporate users' decisions into Stage 1 to characterize the properties of the aggregator's profit in Proposition 3. Based on the properties of Stage 2 (in Proposition \ref{prop:capacity} and Theorem 1) and Stage 1 (in Proposition 3), we propose Algorithm 1 to solve the two-stage problem, which determines the aggregator's  optimal pricing and investment decisions.} 
		\vspace{-1mm}
		\subsection{Solution of Stage 2}
		For Stage 2, we first prove that the optimal capacity $x_i^{\omega*}(q)$ that user $i$ purchases is stepwise in price $q$.  Then we add a small penalty on the user's cost to solve the issue of multi-optima of the charge and discharge decision. We show that the user's optimal decision has a simple stepwise structure over price $q$ as the penalty approaches zero.  
		
		\subsubsection{{Stepwise structure of  $x_i^{\omega\ast}(q)$}}
		The optimal capacity $x_i^{\omega\ast}(q)$ is stepwise over price $q$ as shown in Proposition 1:
		
		\begin{prop}[Stepwise property of virtual capacity] \label{prop:capacity}
			The optimal capacity $x_i^{\omega\ast}(q)$ of Problem $\textbf{UP}_i^\omega$ is a non-increasing and stepwise correspondence of the price $q$. Specifically, there exists the set of  $K_i^\omega+1$ threshold prices  $\mathcal{Q}_i^\omega=\{q_i^{\omega_0},q_i^{\omega_1},q_i^{\omega_2},...,q_i^{\omega_{K_i^\omega}}\}$ and $0=q_i^{\omega_0}<q_i^{\omega_1}<\cdots <q_i^{\omega_{K_i^\omega}}$, such that $x_i^{\omega\ast}(q)$  is given by
			\vspace{-1mm}
			\begin{equation}
			x_i^{\omega\ast}(q)=\left \{
			\begin{aligned}
			&x_i^{\omega_0},q \in (q_i^{\omega_0},q_i^{\omega_1}),\\
			\vspace{-1mm}
			&x_i^{\omega_1},q \in (q_i^{\omega_1},q_i^{\omega_2}),\\
			\vspace{-2mm}
			&...\\
			\vspace{-2mm}
			&x_i^{\omega_{K_i^\omega}},q \in (q_i^{\omega_{K_i^\omega}},\infty),\\
			\end{aligned}
			\right.
			\end{equation}\par \vspace{-1mm}		
			\noindent	where  $x_i^{\omega_0}>x_i^{\omega_1}>\cdots >x_i^{\omega_{K_i^\omega}}=0$. For any threshold price $q_i^{\omega_k}>0$, $x_i^{\omega\ast}(q)$ can be any value in $ [x_i^{\omega_{k-1}},x_i^{\omega_k}]$. For $q_i^{\omega_0}=0$, $x_i^{\omega\ast}(q)$ can achieve any value in $ [x_i^{\omega_0},\infty)$.  We denote user $i$'s optimal capacity set as $\mathcal{X}_i^\omega=\{x_i^{\omega_0},x_i^{\omega_1},...,x_i^{\omega_{K_i^\omega}}\}$.
		\end{prop}

		We illustrate Proposition \ref{prop:capacity} in Figure \ref{fig:propo}(a). As price $q$ increases, the optimal capacity that a user purchases decreases.  If the price is higher than the threshold $q_i^{\omega_{K_i^\omega}}$, the user will purchase none.  Between two adjacent threshold prices, the optimal capacity remains the same. In the online Appendix C \cite{reportjournal}, we prove Proposition \ref{prop:capacity}. In the  Appendix D \cite{reportjournal}, we present Algorithm 3 for computing the sets $\mathcal{X}_i^\omega$ and $\mathcal{Q}_i^\omega$ .	
		
		\subsubsection{{Solving multi-optima problem}}
		Although we have characterized a user $i$'s optimal capacity decision in Proposition \ref{prop:capacity}, we still face the difficulty  of multi-optima. More specifically, even at  a fixed optimal capacity $x_i^{\omega*}(q)$, there may  still be multiple virtual charge and discharge solutions $\big(\bm{p}_i^{\omega,ch},\bm{p}_i^{\omega,dis}\big)$ {(the set of which is denoted as  $\big(\bm{P}_i^{\omega,ch*}(q), \bm{P}_i^{\omega,dis*}(q)\big)$)} that lead to the same  net cost to the user. As a result, the aggregator's cost is not well-defined, because the cost is due to users' charge and discharge demand. To address this difficulty, we  introduce a small positive penalty coefficient $\varepsilon$  on the user’s net cost in Problem $\textbf{UP}_i^\omega$  as follows:
		\vspace{-1mm}
		\begin{align}
		C_i^q(\bm{p}_i^{\omega,ch},\bm{p}_i^{\omega,dis})=\varepsilon\sum_{t\in \mathcal{T}} ((p_i^{\omega,ch}[t])^2+(p_i^{\omega,dis}[t])^2).\label{penalty}
		\end{align}\par  \vspace{-1mm}
		
		After including penalty \eqref{penalty}  to the objective function of Problem $\textbf{UP}_i^\omega$, we obtain a  new modified  problem denoted as $\textbf{UPP}_i^\omega$, {which is a quadratic programming problem}. We denote the optimal solution to Problem $\textbf{UPP}_i^\omega$ as $\left(x_i^{\omega\star}(q,\varepsilon), \bm{p}_i^{\omega,r,u\star}(q,\varepsilon), \bm{p}_i^{\omega,ch\star}(q,\varepsilon), \bm{p}_i^{\omega,dis\star}(q,\varepsilon), \bm{e}_i^{\omega\star}(q,\varepsilon)\right)$ for any price $q>0$ and penalty $\varepsilon >0$. Proposition \ref{prop:unique} below shows such a solution of Problem $\textbf{UPP}_i^\omega$ is unique, which is proved in  Appendix E\cite{reportjournal}. To ensure users' unique decisions, we let the aggregator choose  $q>0$ and $\varepsilon>0$.\footnote{Note that when $q=0$, even when we set $\varepsilon>0$, users can purchase an arbitrarily large capacity beyond their minimum needs because they bear no capacity costs.}
		
		\begin{prop}[Uniqueness] \label{prop:unique}
			For any $\varepsilon>0$ and any $q> 0$, the optimal solution to Problem $\textbf{UPP}_i^\omega$ is unique.	
		\end{prop}

		\begin{figure}[t]
			\centering
			\subfigure[]{
				\label{fig:subfig:a} 
				\raisebox{-7mm}{\includegraphics[width=1.65in]{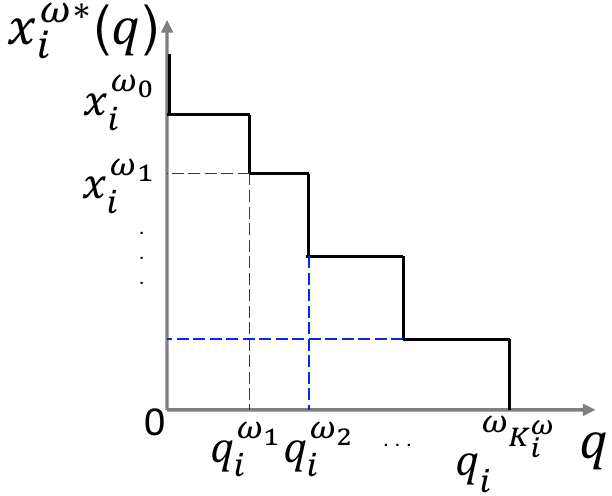}}}
			\hspace{-2ex}
			\subfigure[]{
				\label{fig:subfig:b} 
				\raisebox{-7mm}{\includegraphics[width=1.65in]{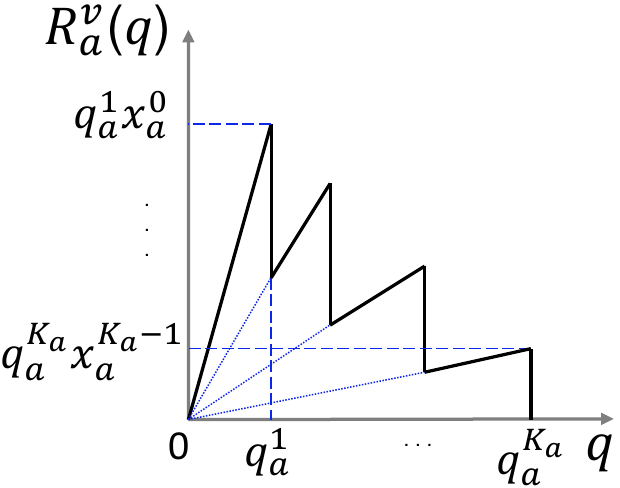}}}
			\vspace{-3mm}
			\caption{\small (a) User $i$'s optimal capacity $x_i^{\omega *}(q)$; (b) Revenue $R_a^v(q)$. }
			\label{fig:propo}
			\vspace{-4mm}
		\end{figure} 
		
		\subsubsection{Asymptotic solution}
		Later in Section B-3, we will choose a small $\varepsilon$ to obtain a near-optimal solution for Stage 1. Intuitively, when $\varepsilon$ approaches zero, one would expect that the solution to Problem $\textbf{UPP}_i^\omega$ approaches the solution to Problem $\textbf{UP}_i^\omega$. Since the optimal charge and discharge decision of Problem $\textbf{UP}_i^\omega$ has multiple optima, it would seem that the same difficulty will persist as $\varepsilon$ approaches zero. Surprisingly, we show below  that the limit of the solution to Problem $\textbf{UPP}_i^\omega$  exists and can be uniquely determined as $\varepsilon$ approaches zero.  
		
		\begin{thm}[Asymptotic solution]\label{thm:solution}
			For any price  $ q \notin \mathcal{Q}_i^\omega$ (as defined in Proposition \ref{prop:capacity}),  as $\varepsilon$ approaches zero, \\
			\noindent {(a) there is a unique limit of the optimal charge and discharge decision  $\big(\bm{p}_i^{\omega,ch\star}(q,\varepsilon), \bm{p}_i^{\omega,dis\star}(q,\varepsilon)\big)$. This limit  belongs to the  set $\big(\bm{P}_i^{\omega,ch*}(q), \bm{P}_i^{\omega,dis*}(q)\big)$, i.e.,}
			\vspace{-2mm}
			\begin{align*}
			\hspace{-1mm} &\lim_{\varepsilon \rightarrow 0^+}\big(\bm{p}_i^{\omega,ch\star}(q,\varepsilon),\bm{p}_i^{\omega,dis\star}(q,\varepsilon)\big)\hspace{-1mm}\in\hspace{-1mm} \big(\bm{P}_i^{\omega,ch*}(q), \bm{P}_i^{\omega,dis*}(q)\big),
			\end{align*} \par  \vspace{-2mm}
			\noindent {and it remains constant for all prices within each threshold price interval $(q_i^{\omega_k},q_i^{\omega_{k+1}})$, where $q_i^{\omega_k},q_i^{\omega_{k+1}}\in \mathcal{Q}_i^\omega$;}\\
			\noindent {(b) the  optimal  capacity $x_i^{\omega\star}(q,\varepsilon)$ approaches the  optimal solution $x_i^{\omega*}(q)$ (given in Proposition \ref{prop:capacity}) with $\varepsilon=0$, i.e.,}
			\vspace{-2mm}
			\begin{align*}
			\hspace{-1mm} &\lim_{\varepsilon \rightarrow 0^+}x_i^{\omega\star}(q,\varepsilon)=x_i^{\omega*}(q).
			\end{align*} \par  \vspace{0mm}
		\end{thm}	
		\vspace{-3mm}	
		
		Theorem \ref{thm:solution}(a) is highly non-trivial due to multiple optimal solutions $\big(\bm{p}_i^{\omega,ch*}(q),\bm{p}_i^{\omega,dis*}(q)\big)$ of Problem $\textbf{UP}_i^\omega$.  We prove Theorem \ref{thm:solution}(a) based on the Maximum Theorem \cite{maximum} by showing the uniqueness of the optimal objective value of Problem $\textbf{UP}_i^\omega$,  and prove Theorem \ref{thm:solution}(b) by showing the uniqueness of $x_i^{\omega*}(q)$. We present the detailed proof in Appendix G\cite{reportjournal}.
		
		Based on Theorem \ref{thm:solution}(b), we can compute the limit $\lim_{\varepsilon \rightarrow 0^+}x_i^{\omega\star}(q,\varepsilon)$ by the optimal solution $x_i^{\omega*}(q)$ of Problem $\textbf{UP}_i^\omega$, which takes discrete values from the set $\mathcal{X}_i^\omega$ defined in Proposition \ref{prop:capacity}.  Then, given each  element in $\mathcal{X}_i^\omega$, we can solve  optimization problems (by Algorithm 4 of Appendix H\cite{reportjournal}) that minimize the penalty term to obtain the limiting optimal charge and discharge decision as $\varepsilon$ approaches zero. In other words, the limiting optimal charge and discharge solution is a function of each element in $\mathcal{X}_i^\omega$. The above results show that as $\varepsilon$ approaches zero, the user's solution in Stage 2 has a stepwise structure, which can be efficiently computed.
		
		\vspace{-4mm}
		\subsection{Solution of Stage 1}
		\vspace{-1mm}
		
		Based on the asymptotic structure of the users' decisions in stage 2 as shown in Theorem \ref{thm:solution},  we further analyze the structure of the aggregator's optimal profit as a function of  price $q$ as $\varepsilon$ approaches zero. We propose an algorithm to derive the near-optimal profit price $\hat{q}^{\star}$ by considering $\varepsilon$ that is small enough. We use a similar method to compute the LNP price and present it in Appendix J\cite{reportjournal}.
		
		\subsubsection{The optimal profit given $(q,\varepsilon)$}Given $(q,\varepsilon)$, the aggregator can vary the investment $X$ and $P$ to optimize her profit as follows. First, the aggregator can compute the revenue $R_a^{v}(q,\epsilon)$ as in \eqref{eq:rev} with $ x_i^{\omega*}(q)$ replaced by $ x_i^{\omega\star}(q,\varepsilon)$. Second, given users' optimal charge and discharge decision $\big(\bm{p}_i^{\omega,ch\star}(q,\varepsilon), \bm{p}_i^{\omega,dis\star}(q,\varepsilon)\big),\forall i,\omega$, the aggregator can solve Problem $\textbf{CO}$ to compute her optimal cost $C_a(q,\varepsilon)$ as follows.
		\vspace{-6mm}
		\begin{align*}
		\textbf{CO:\hspace{0.5mm}}C_a(q,\varepsilon)\hspace{-1mm}:=\hspace{-1mm}\min~\hspace{-1mm} &C_a^{cap}(X,P)\hspace{-1mm} +\hspace{-1mm}C_a^{op}(q,X,P)\hspace{-1mm}+\hspace{-1mm}C_a^{ad}(q,X,P)\\
		\text{s.t.} \  & \eqref{eq:12}-\eqref{eq:15},\eqref{eq:27}-\eqref{eq:ief},\\
		\vspace{-1mm}
		\text{var:}\   &X, P.
		\end{align*}\par \vspace{-3mm}
		\noindent {This is a linear programming problem, which can be solved efficiently using the simplex  method\cite{dantzig}. Finally, we compute the  optimal profit for any given $(q,\varepsilon)$ by
			\vspace{-2mm}
			\begin{align}
			R_a^{pf}(q,\varepsilon)=R_a^{v}(q,\epsilon)- C_a(q,\varepsilon)\label{eq:pr}.
			\end{align}}}
	
	\vspace{-2mm}
	\subsubsection{The limit of the optimal profit as $\varepsilon$ approaches zero}{{In Proposition \ref{prop:profit} below, as $\varepsilon$ approaches zero,  we first show in part (a) that the revenue $R_a^{v}(q,\varepsilon)$ approaches a limit $R_a^{v}(q)$ as in \eqref{eq:rev} without $\varepsilon$. We further show in (b) that the optimal cost ${C}_a(q,\varepsilon)$ approaches a limit $C_a(q)$, which is stepwise over the threshold price set $\mathcal{Q}_a=\bigcup_{i,\omega}{Q}_i^\omega$. Finally, we show in (c) that the optimal profit $R_a^{pf}(q,\varepsilon)$ approaches $R_a^{v}(q)-C_a(q)$ denoted as ${R}_a^{pf}(q)$ as $\varepsilon$ approaches zero.} We present the proof of Proposition \ref{prop:profit} in Appendix I\cite{reportjournal}.
		
		\begin{prop}[Asymptotic profit]\label{prop:profit}
			For any $q \notin \mathcal{Q}_a$, we have
			\vspace{-6mm}
			\begin{align*}
			&(a) \lim_{\varepsilon \rightarrow 0^+}R_a^{v}(q,\varepsilon)=R_a^{v}(q),~(b) \lim_{\varepsilon \rightarrow 0^+}{C}_a(q,\varepsilon)=  C_a(q),
			\end{align*}\par \vspace{-2mm}
			\noindent{where  $C_a(q)$ denotes the optimal value of Problem $\textbf{CO}$ given the limit of each user $i$'s optimal charge and discharge decision  $\left(\lim_{\varepsilon \rightarrow 0^+}(\bm{p}_i^{\omega,ch\star}(q,\varepsilon), \lim_{\varepsilon \rightarrow 0^+}\bm{p}_i^{\omega,dis\star}(q,\varepsilon)\right),\forall i,\omega$,}
			\vspace{-2mm}
			\begin{align*}
			\hspace{-11mm}~(c) \lim_{\varepsilon \rightarrow 0^+}{R}_a^{pf}(q,\varepsilon)=R_a^{v}(q)-C_a(q)\triangleq R_a^{pf}(q).
			\end{align*}
		\end{prop}

		Based on Proposition \ref{prop:capacity}, the revenue $R_a^{v}(q)$ is piecewise linear  over the threshold price set $\mathcal{Q}_a$ as depicted in Figure \ref{fig:propo}(b):  The revenue increases linearly from each threshold price and then decreases vertically at the next adjacent threshold price.  The slopes of $R_a^{v}(q)$ over different price intervals are  determined by $\sum_{\omega}\rho^\omega\sum_{i} x_i^{\omega*}(q)$. The limiting cost  $C_a(q)$ is a stepwise function over the threshold price set $\mathcal{Q}_a$ as shown in Figure \ref{fig:cost}. Hence the limiting profit ${R}_a^{pf}(q)$ is  a piecewise linear function of price $q$ as shown in Figure \ref{fig:profit} with multiple local optima.\footnote{Although in Proposition 3 these limiting values are only defined for $q \notin \mathcal{Q}_a$,  we can use the left-handed limits of  ${R}_a^{pf}(q)$ as the function values  for $q \in \mathcal{Q}_a$. In this way,  the function ${R}_a^{pf}(q)$  is well-defined for all $q$.} Finally, based on each user $i$'s limiting decision $\lim_{\varepsilon \rightarrow 0^+}x_i^{\omega\star}(q,\varepsilon)$ and $\lim_{\varepsilon \rightarrow 0^+}\big(\bm{p}_i^{\omega,ch\star}(q,\varepsilon), \bm{p}_i^{\omega,dis\star}(q,\varepsilon)\big)$ in Theorem \ref{thm:solution}, we can  compute  the  limiting revenue $R_a^{v}(q)$,  the  limiting cost $C_a(q)$, and thus the limiting profit ${R}_a^{pf}(q)$ as in Lines 2-8 in Algorithm 1 below. 
		\begin{figure}[t]
			\centering
			\subfigure[]{
				\label{fig:cost} 
				\raisebox{-4mm}{\includegraphics[width=1.45in]{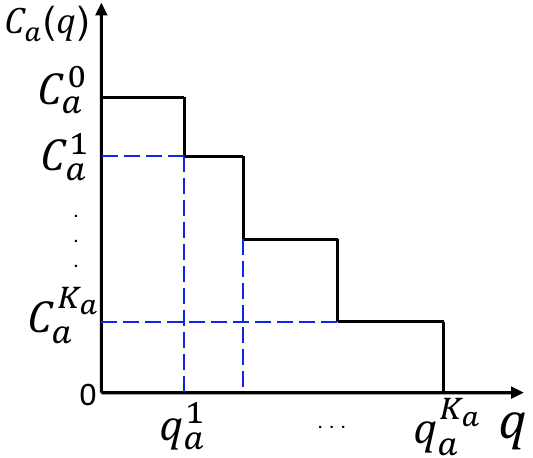}}}
			\hspace{-1ex}
			\subfigure[]{
				\label{fig:profit} 
				\raisebox{-4mm}{\includegraphics[width=1.39in]{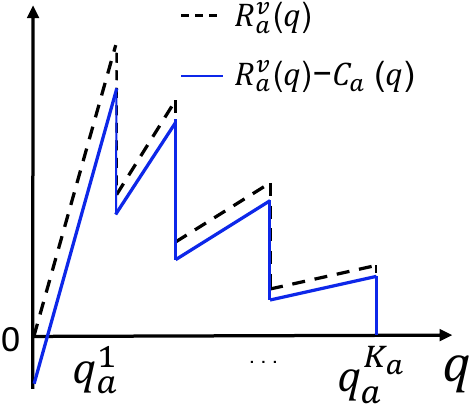}}}
			\vspace{-3mm}
			\caption{\small (a) Cost $C_a(q)$ ; (b) Revenue ${R}_a^{v}(q)$ and profit ${R}_a^{v}(q)-{C}_a(q)$. }
			\label{fig:prop}
			\vspace{-2mm}
		\end{figure}

		\subsubsection{The algorithm to compute the near-OP  price $\hat{q}^\star$}{
			The above results demonstrate important  solution structures  of Stage 2 and Stage 1 as $\varepsilon$ goes to zero. Unfortunately, the aggregator cannot choose $\varepsilon=0$  (which has the multi-optima problem as mentioned in Section A). Thus, below we propose an iterative procedure in Algorithm 1 (i.e., Lines 9-16) for computing a sufficiently small $\varepsilon$ so that (i) it is close enough to the above limit, and (ii) the near-OP price $\hat{q}^{\star}$ can be efficiently  computed.} Specifically, first, the aggregator computes the left-handed limit of profit ${R}_a^{pf}(q)$ at each threshold price $q_a\in \mathcal{Q}_a$, and selects the price $q_a^m$ that achieves the global optimum (Line 9).  Second, since profit ${R}_a^{pf}(q)$ is linearly increasing  over each threshold price interval, the aggregator chooses a near-OP price  $\hat{q}^\star$ slightly lower than $q_a^{m}$ such that the profit ${R}_a^{pf}(\hat{q}^\star)$ approximates  the maximum profit ${R}_a^{pf}(q_a^{m})$ within the given accuracy $err_1$ (Lines 10-11). Finally, the aggregator chooses a small $\varepsilon$ in an iterative fashion so that the profit ${R}_a^{pf}(\hat{q}^\star,\varepsilon)$ approximates ${R}_a^{pf}(\hat{q}^\star)$ within the given accuracy $err_2$  (Lines 12-16), where each user solves the quadratic programming problem $\textbf{UPP}_i^\omega$ in each iteration.\footnote{The quadratic programming problem can be efficiently solved by the interior point method\cite{quad}.}  After computing the  near-OP price $\hat{q}^\star$ and $\varepsilon$, we can obtain the near-optimal investment decision by Subroutine 1 accordingly.
		
		{Algorithm 1 is scalable in terms of the number of users and scenarios. On the users' side, they compute their decisions in parallel, which will not increase the execution time as the number of users increases. On the aggregator's  side, she searches the threshold price set for the maximum profit, where the number of threshold prices is simply linear in the product of the  number of users and the number of scenarios. Therefore, the execution time of Algorithm 1 is proportional to the number of users and scenarios, and thus scales well. }
		\vspace{-1mm}
		\begin{algorithm}  
			\caption{Search of  the near-optimal-price $\hat{q}^\star$}  
			\label{alg:B}  
			\begin{algorithmic}[1] 
				\STATE \textbf{initialization}: set iteration index $k=0$, $\varepsilon^0>0$, relative error $err_1$ and  $err_2$;
				\FOR {user $i\in \mathcal{I}$ \textbf{in parallel}} 
				\STATE Compute the  set   $\mathcal{Q}_i^\omega$ and $\mathcal{X}_i^\omega$ by Algorithm 3 of Appendix D\cite{reportjournal}, and compute the limiting charge/discharge decision by Algorithm 4 of Appendix H\cite{reportjournal}, $\forall \omega\in {\Omega}$;
				\STATE  Report  the computation results to the aggregator;	
				\ENDFOR
				\FOR {each $q_a \in \mathcal{Q}_a=\bigcup_{i,\omega}\mathcal{Q}_i^\omega$} 
				\STATE The aggregator computes the left-handed limit of profit $R_a^{pf}(q_a)=R_a^{v}(q_a)-C_a(q_a)$, with  $R_a^{v}(q_a)$  computed by  \eqref{eq:rev} and $C_a(q_a)$ computed as in Proposition \ref{prop:profit}(b);
				\ENDFOR
				\STATE  The aggregator searches the threshold price set $\mathcal{Q}_a$ and chooses the price $q_a^m \in \mathcal{Q}_a$   that  maximizes $R_a^{pf}(q)$;
				\STATE The aggregator calculates the slope  $slp(q_a^m)$  of $R_a^{pf}(q)$ over  the threshold price interval $(q_a^{m-1},q_a^m)$:
				$$slp(q_a^m)=\sum_{\omega}\rho^\omega\sum_{i} x_i^{\omega*}\left(\frac{q_a^{m-1}+q_a^m}{2}\right);$$\par \vspace{-1mm}
				\STATE The aggregator  computes a  price $\hat{q}^\star$ (lower than $q_a^{m}$) such that  $R_a^{pf}(\hat{q}^\star)$  approximates $R_a^{pf}(q_a^m)$ within the relative error $err_1$: 
				\vspace{-2mm}
				\begin{align*}
				\hat{q}^\star= q_a^{m}-\frac{R_a^{pf}(q_a^m) err_1}{slp(q_a^m)}; 
				\end{align*}\par \vspace{-1mm}
				\vspace{-1mm}
				\REPEAT   
				\STATE $k:=k+1$;
				\STATE $\varepsilon^k=\varepsilon^{k-1}/10$;
				\STATE 
				$\left(\hspace{-0.3mm}R_a^{pf}(\hat{q}^\star,\varepsilon^k),X(\hat{q}^\star,\varepsilon^k),P(\hat{q}^\star,\varepsilon^k)\hspace{-0.3mm}\right )\hspace{-1mm}=\hspace{-1mm}CU(\hat{q}^\star,\varepsilon^k)$;	
				\UNTIL{	$$\frac{R_a^{pf}(\hat{q}^\star,\varepsilon^k)-R_a^{pf}({\hat{q}^\star})}{R_a^{pf}(\hat{q}^\star)} \leq err_2;$$ }
				\STATE {\textbf{output}}: \big($\hat{q}^\star, \varepsilon^k, P(\hat{q}^\star,\varepsilon^k),  X(\hat{q}^\star,\varepsilon^k)$\big);
			\end{algorithmic}  
		\end{algorithm} 
		
		\vspace{-2mm}
		
		\begin{algorithm}  
			\setcounter{algorithm}{0}
			\floatname{algorithm}{Subroutine}
			\caption{Communication unit ${CU}(q,\varepsilon)$}  
			\label{alg:A}  
			\begin{algorithmic}[1]
				\STATE \textbf{input}:  $(q,\varepsilon)$  announced by the aggregator;
				\FOR  {each user $i\in \mathcal{I}$ \textbf{in parallel}}
				\STATE  Solve Problem $\textbf{UUP}_i^\omega$ given $(q,\varepsilon)$,  and  reports to the  aggregator the optimal decisions of  $x_i^\omega(q,\varepsilon)$, $\bm{p}_i^{\omega,ch}(q,\varepsilon)$, $\bm{p}_i^{\omega,dis}(q,\varepsilon)$,  for all $\omega\in {\Omega}$; 
				\ENDFOR
				\STATE The aggregator determines the investment decision  $X(q,\varepsilon),P(q,\varepsilon)$ by solving Problem $\mathbf{CO}$,  and computes the profit $R_a^{pf}(q,\varepsilon)$ by \eqref{eq:pr};
				\STATE \textbf{output}:  $(R_a^{pf}(q,\varepsilon),X(q,\varepsilon),P(q,\varepsilon))$;
			\end{algorithmic}  
		\end{algorithm}

		\section{Numerical Study}
		\begin{figure*}[t]
			\centering
			\subfigure[]{
				\label{fig:subfig:load1} 
				\raisebox{-1mm}{\includegraphics[width=1.7in]{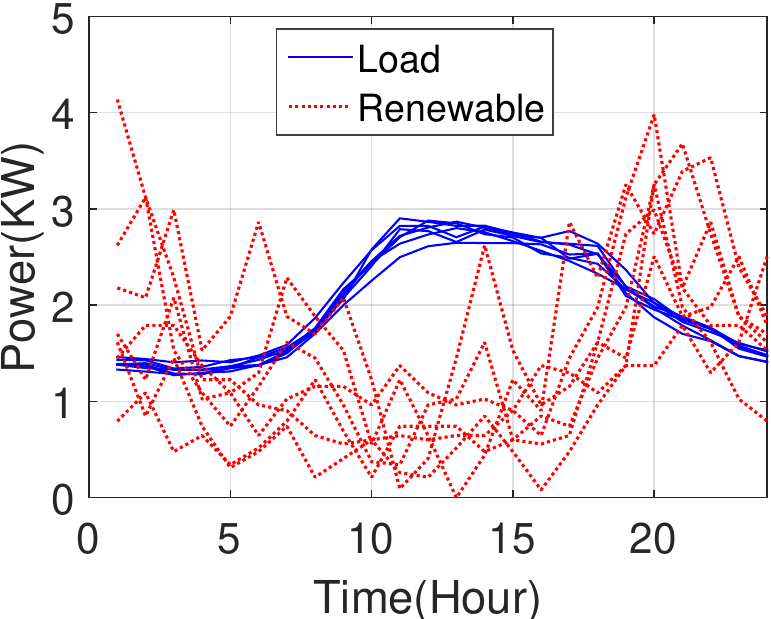}}}
			\hspace{-1mm}
			\subfigure[]{
				\label{fig:subfig:load2} 
				\raisebox{-1mm}{\includegraphics[width=1.7in]{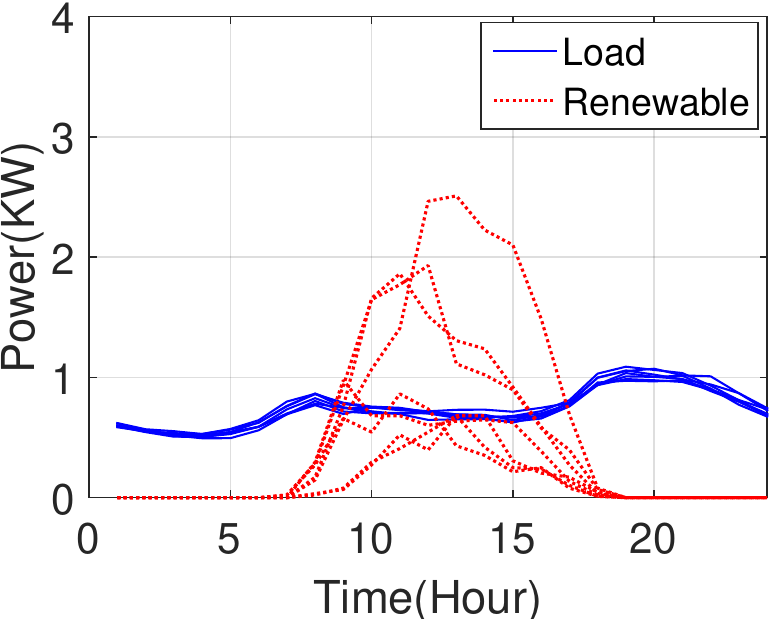}}}
			\hspace{-1mm}
			\subfigure[]{
				\label{fig:subfig:load3} 
				\raisebox{-1mm}{\includegraphics[width=1.7in]{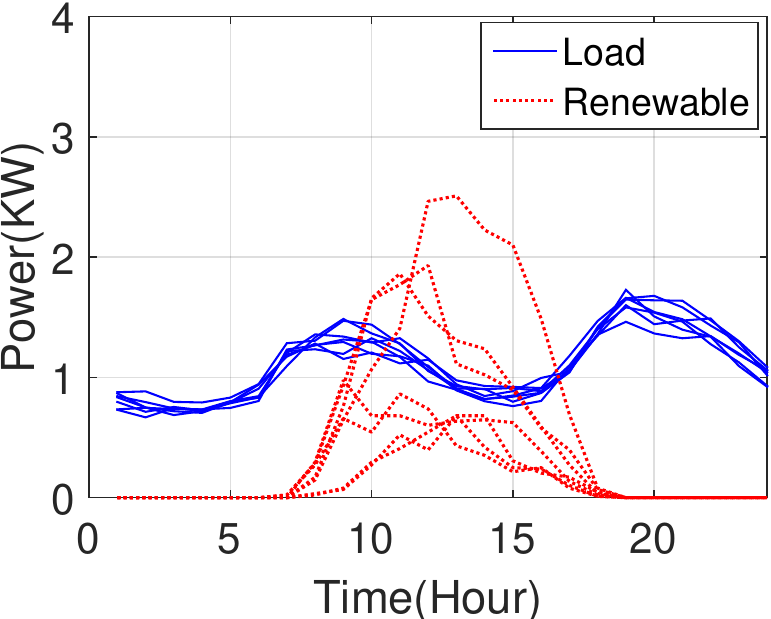}}}
			\vspace{-3mm}
			\caption{\small 7 typical load and renewable generation scenarios for (a) Type-1 user; (b) Type-2 user; (c) Type-3 user.  }
			\label{fig:load}
			\vspace{-3mm}
		\end{figure*} 
		
		We simulate a system based on realistic data where users have diverse load and renewable profiles. We demonstrate that, {as long as there are groups of users with diverse load and renewable profiles}, our virtualization model can promote more efficient use of the physical storage, offer flexibility to users in the choices of virtual capacities, and reduce users' costs compared with the case where users acquire their own physical storage.
		\vspace{-1mm}
		\subsection{{Simulation setup}}
		\vspace{-1mm}
		\subsubsection{\textbf{Parameters}}
		We consider the lithium-ion battery as the energy storage technology. We use realistic load data  from PG\&E Corporation in 2012 \cite{loaddata} to characterize users' load profiles, and we use wind speed and solar radiation data from Hong Kong Observatory \cite{hkob} to characterize users' renewable generations. To illustrate how our storage virtualization model works, we simulate a system as in Figure \ref{fig:structure} with three users of different types. Type-1 user owns {the} local wind generation, while Type-2 and Type-3 users own solar panels (with the same capacity).\footnote{Though the installation of wind turbines faces geographical restrictions, technological innovations (such as vertical axis wind turbines) have promoted the adoption of wind energy for households, commercial building, and residential communities \cite{windreport}\cite{risewind}\cite{windcomm}. Therefore, to capture a more complete picture of renewable energy deployment in practice, we consider both solar energy and wind energy in our simulation. We also conduct the simulations where Type-1 user has no renewable energy while Type-2 and Type-3 users have solar energy (with details in Appendix M\cite{reportjournal}), which shows that our framework still works well for  significantly reducing the invested physical storage capacity and the cost of users with solar energy.}

		We obtain the one-year historical data, namely 366 scenarios for users' load  and renewable  profile jointly. As a large number of scenarios lead to a high computational complexity, we choose a smaller subset of 7 typical scenarios that can approximate the original scenario set as shown in Figure \ref{fig:load}: Type-1 user has the peak load around noon (typical for some commercial users) and produces more wind power at night; Type-2 and Type-3 users have the peak load in the morning and evening (typical for residential users), and their solar power reaches peak supply around noon. For more details regarding the data of renewable generation and load, we include them in the online dataset \cite{datastorage}. For other parameters of electricity bill and energy storage, we present them in Appendix K\cite{reportjournal}.  Note that we do not make any assumptions on the correlation between users' load profiles and renewable generations; we just select an example with three types of users for illustration. Furthermore, since our framework for virtual energy storage sharing is general, interested readers can also use their own data as inputs to examine the performance of the framework.

		\subsubsection{\textbf{Benchmark}}
		To show the performance of our virtualization model, we consider a benchmark where each user invests in a physical storage product (e.g., Tesla Powerwall) by himself. He will optimize the (fixed) capacity of the physical storage, and can only use his own storage over the investment phase. Apart from the capacity cost, each user also bears the power rating cost and the operational cost by himself. Each user $i$' solves an optimization problem $\textbf{BM}_i$  to minimize his  cost and determines the storage size over the investment phase. We present the details of  Problem $\textbf{BM}_i$ in Appendix K \cite{reportjournal}.
		
		\subsubsection{\textbf{Storage price}}
		For the benchmark problem, we consider 2 different storage prices for users as follows:
		\vspace{-1mm}
		\begin{itemize}
			\item \textit{Production cost $\mathbf{c^p}(c_u^x,c_u^p)$}: Users pay the same storage production cost as the aggregator.
			\item \textit{Retailer price $\mathbf{c^r}(c_u^x, c_u^p)$}: Users pay a higher storage retailer price than the aggregator.
		\end{itemize}
		\vspace{-1mm}
		
		The \textit{production cost} indicates the minimal cost for users to acquire the energy storage, and the \textit{retailer price}  is a more realistic price for users to purchase the storage on the market. 
		
		The simulation is implemented using MATLAB 2015 on a computer with an Intel Core i7 of 3.6 GHz and 8 GB memory. The execution time is about 13s.
		
		\vspace{-1mm}	
		\subsection{Simulation results}
		\vspace{-1mm}
		We demonstrate several key benefits of  our virtualization business model as follows.
		
		\subsubsection{\textbf{More efficient use of storage}}	
		We show that our model can lead to more efficient use of energy storage, such that the aggregator can invest in a smaller physical storage capacity to support much larger virtual capacity allocation.
		
		Figure \ref{fig:vpc} shows the comparison between the aggregator's invested physical capacity and the sold expected virtual capacity when the price varies from the LNP price $q^l$ to the OP price $q^\star$. The sold virtual capacity is always much larger than the actual physical size. Specifically, compared with the  virtual capacity, the physical capacity is reduced by 42.5\% at the  price $q^l$ and 54.3\% at the price $q^\star$. To understand this, in Figure \ref{fig:chargedischarege}, we show the virtual charge and discharge profiles of three types of users (represented by different colors) in seven scenarios (shown as the seven curves for each type of user). For each curve, the positive value part corresponds to charging the virtual storage, and the negative value part corresponds to discharging the virtual storage. We can see that that Type-1 user discharges the storage around noon to serve his peak load and charges the storage around morning and night. On the other hand, Type-2 and Type-3 users charge the storage around noon to store excessive solar energy and discharge around morning and night to serve their peak loads. Therefore, these users' charge and discharge profiles can partially cancel out at the aggregated level, which reduces the need of the physical storage. We also see from Figure \ref{fig:vpc} that the sold virtual capacity and invested physical capacity decrease with the price since a higher storage price  generally reduces users' requirement for storage.
		
		\begin{figure}[t]
			\centering
			\hspace{-1ex}
			\subfigure[]{
				\label{fig:vpc} 
				\raisebox{-1.5mm}{\includegraphics[width=1.66in]{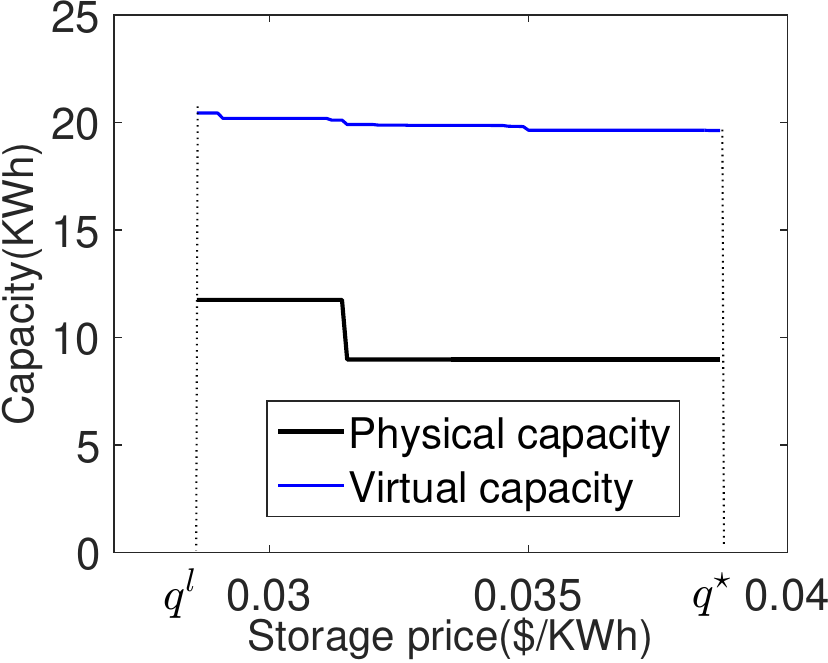}}}
			\hspace{-2.2ex}
			\subfigure[]{
				\label{fig:chargedischarege} 
				\raisebox{-1.5mm}{\includegraphics[width=1.7in]{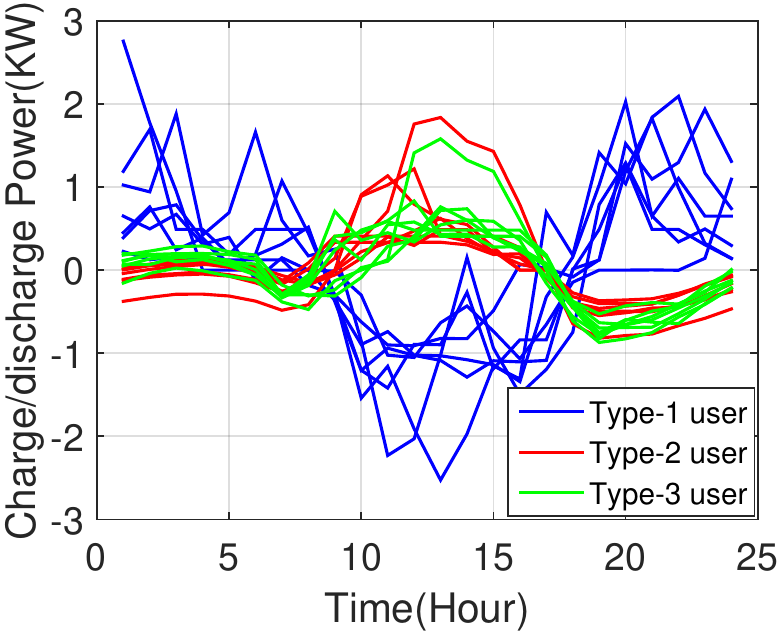}}}
			\vspace{-2.5mm}
			\caption{(a) \small Physical capacity and  virtual capacity from $q^l$ to $q^\star$; (b) Users' charge and discharge decisions.}
			\label{fig:pricerange}
			\vspace{-3mm}
		\end{figure}

		\subsubsection{\textbf{Purchasing flexible virtual capacity}}
		We show that our model enables users to purchase flexible capacities over different operational horizons. To demonstrate this benefit, we compare users' cost under two cases as follows:
		\vspace{-1mm}
		\begin{itemize}
			\item Case 1: Users purchase  flexible virtual capacities over different scenarios, as in the proposed framework
			\item Case 2: Users purchase  capacities that cannot change in  different scenarios.
		\end{itemize}
		\vspace{-1mm}
		
		Case 1 corresponds to the user's Problem $\textbf{UP}_i^\omega$ in our virtualization model.  Case 2 is similar to the user's benchmark problem $\textbf{BM}_i$ except that in Case 2 users will only pay for the capacity without paying for the power rating cost and operational cost. This may make the comparison between Case 1 and Case 2 more fair. To illustrate the benefits of flexibility in choosing virtual capacities, we focus on the  realistic load and renewable generation data of Type-1 user in one week (from 2012.10.1 to 2012.10.7), as shown in Figure \ref{fig:scenarioreduction}(a). We compare the user's cost during this week in both cases under the same storage price, and show the cost reduction (in percentage)  in Case 1 compared with Case 2 in Figure \ref{fig:scenarioreduction}(b). We can see that when the storage price is very low, the cost reduction is small since the user pays little for energy storage in both cases. When the storage price is very high, the user will not purchase storage and the cost reduction will be zero. The gain can be as high as 17\% when the price is medium.
		
		\begin{figure}[t]
			\centering
			\hspace{-1ex}
			\subfigure[]{
				\label{fig:loadx} 
				\raisebox{-2mm}{\includegraphics[width=1.75in]{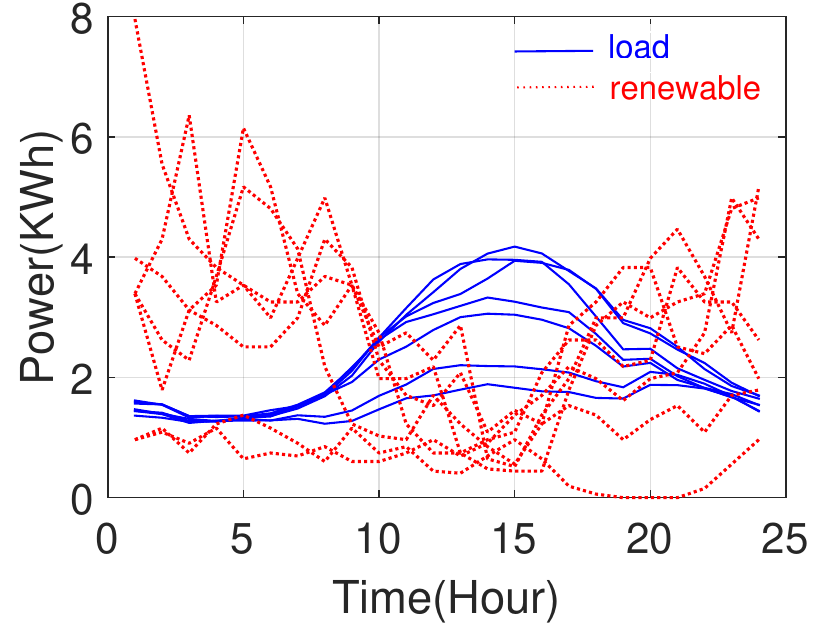}}}
			\hspace{-3ex}
			\subfigure[]{
				\label{fig:reduction} 
				\raisebox{-2mm}{\includegraphics[width=1.75in]{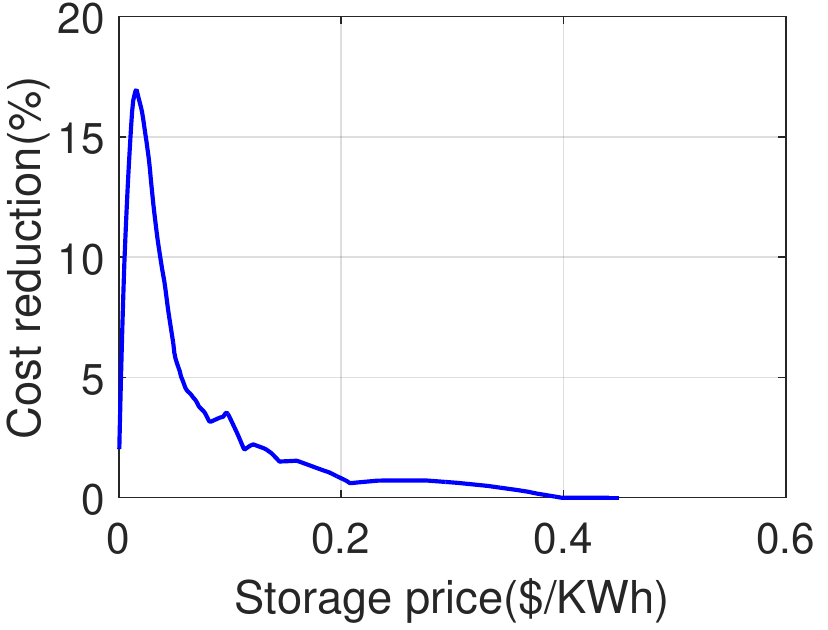}}}
			\vspace{-2mm}
			\caption{(a) \small One-week load and renewable  profiles of Type-1 user; (b) Cost reduction in Case 1 compared with Case 2.}
			\label{fig:scenarioreduction}
			\vspace{-2mm}
		\end{figure} 
		
		\begin{figure}[t]
			\centering
			\hspace{-1ex}
			\includegraphics[width=2.2in]{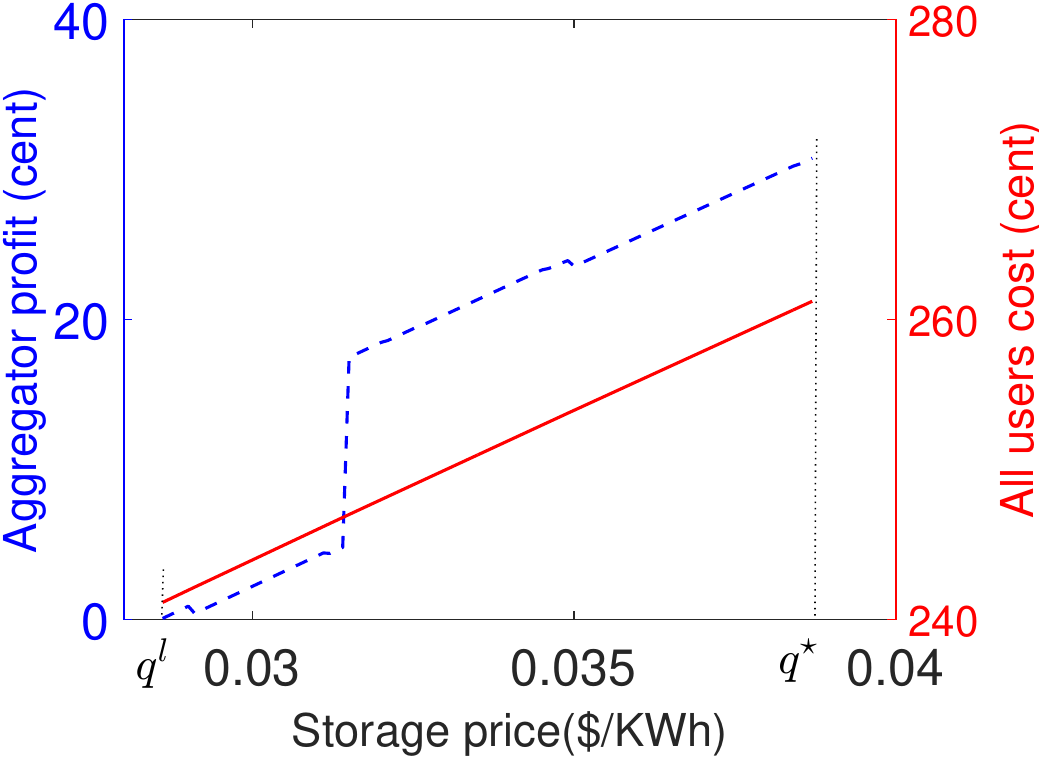}
			\vspace{-2mm}
			\caption{\small Aggregator's profit and users' cost from $q^l$ to $q^\star$.}
			\label{fig:split}
			\vspace{-3mm}
		\end{figure} 
		
		\begin{figure}[t]
			\centering
			\hspace{-1ex}
			\subfigure[]{
				\label{fig:simulation2} 
				\raisebox{-2mm}{\includegraphics[width=1.6in]{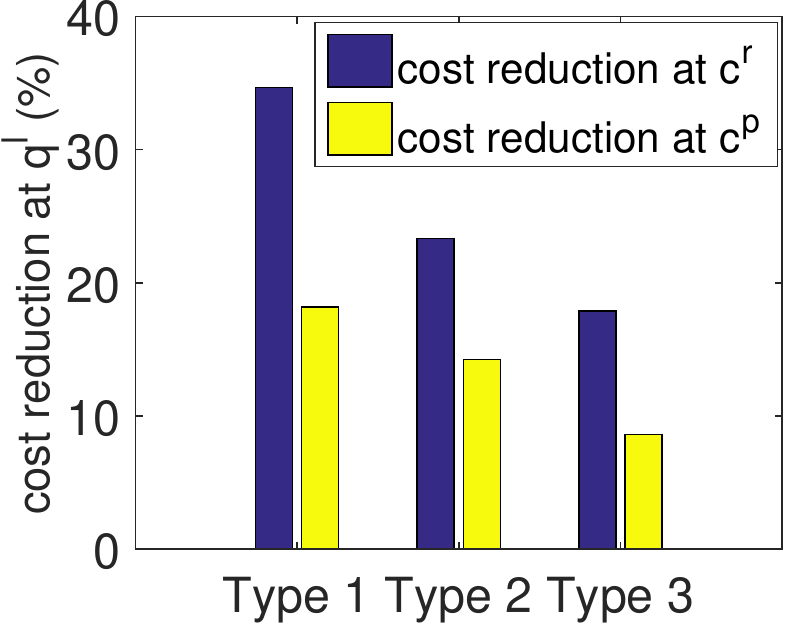}}}
			\hspace{-2ex}
			\subfigure[]{
				\label{fig:simulation3} 
				\raisebox{-2mm}{\includegraphics[width=1.6in]{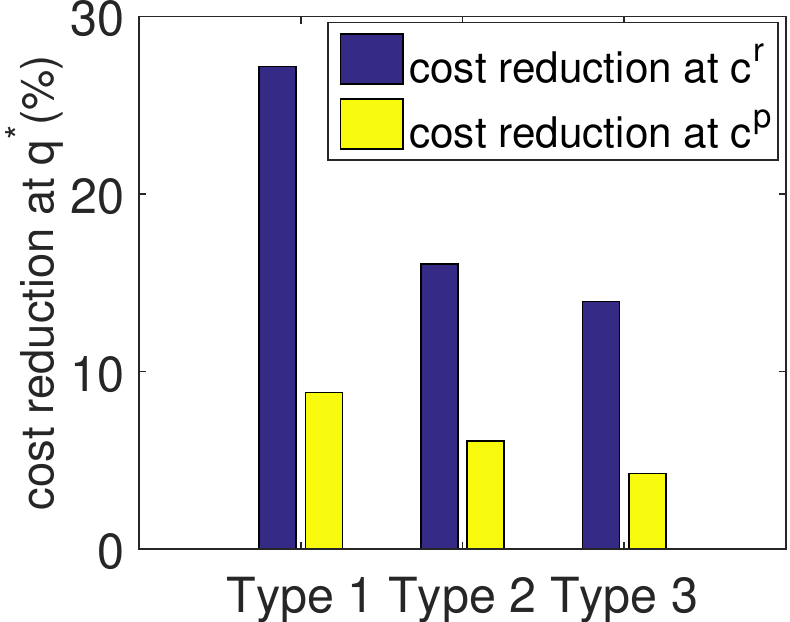}}}
			\vspace{-2mm}
			\caption{\small (a) Cost reduction at $q^l$; (b) Cost reduction at $q^\star$.}
			\label{fig:costreduction}
			\vspace{-2mm}
		\end{figure} 
		
		\subsubsection{\textbf{Benefits of reducing users' cost}}	
		In Figure \ref{fig:split}, we show how the users' cost and the aggregator's profit change from the LNP price $q^l$ to  the OP price $q^\star$. In Figure 8, we show users' cost reduction in our model compared with the benchmark under the price $q^l$ and $q^\star$.\footnote{We also numerically demonstrate in Appendix P\cite{reportjournal} that  our storage virtualization model not only helps users cut their electricity bill but also leads to a peak load reduction in the system. }
		
		In Figure \ref{fig:split}, we see that both the aggregator's profit and the users' total cost increase as the price increases from  $q^l$ to $q^\star$. The LNP price $q^l$ gives the aggregator a zero profit,  meanwhile leaves the most benefits to users. The OP price $q^\star$ gives the aggregator maximum profit at the expense of the maximum cost of users.
		
		We then  show each user's cost reduction in our virtualization model compared with the benchmark in Figure \ref{fig:costreduction}.	At the LNP price $q^l$ in Figure \ref{fig:costreduction}(a), a user's cost can be reduced by up to 34.7\% compared with the benchmark where the user affords retailer price $\mathbf{c^r}$, and the cost reduction can still be up to 18.2\% if the user pays the production cost $\mathbf{c^p}$ in the benchmark. At the OP price $q^\star$ as shown in Figure \ref{fig:costreduction}(b), the user's cost reduction is not significant (up to 8.8\%) if the user pays the production cost $\mathbf{c^p}$ in the benchmark. This result is natural, as the OP price maximizes the aggregator's profit at the expense of the users' benefit.  Nonetheless, compared with the user's cost in the benchmark when user affords the retailer price $\mathbf{c^r}$,  the user's cost can still be reduced by up to 27.2\%.  Furthermore, as shown in figures, Type-1 user obtains higher cost reduction than  Type-2 and Type-3 users. The intuition is that Type-1 user has higher renewable  penetration and high load, which increases the demand for the storage. 
		
		\section{Conclusion}
		This paper proposed a pricing-based virtual storage sharing scheme among a group of users. An aggregator invests and operates the physical energy storage and virtualizes the physical storage into separable virtual capacities, which can be sold to serve different users. We formulated a two-stage optimization problem for the interaction between the aggregator and users. Simulation results showed that energy storage virtualization can save investment in physical energy storage by 54.3\% and reduce users' costs by up to 34.7\%, compared with the case where users utilize their own physical storage.


\clearpage
 \newpage
				
\section*{Appendices}

The appendices below provide the proofs of our main results, as well as algorithms and simulation setup used in the main body of the paper. Specifically,  Appendix A provides equivalent reformulations of Problems $\textbf{UP}_i^\omega$ and $\textbf{UPP}_i^\omega$, which allows us to apply results in Appendix B from parametric linear programming. The results in Appendices A and B will be used to  in Appendix C for the proof of Proposition 1 as well as in Appendix D for Algorithm 3. Furthermore, Appendix F introduces the Maximum Theorem, which will be used in Appendix G for the proof of Theorem 1 as well as in Appendix I for the proof of Proposition 3. Based on the proof of Theorem 1 in Appendix G, we present Algorithm 4 in Appendix H. In addition, we prove Proposition 2 in Appendix E. We give the formulation and the solution method for finding the zero-profit price in Appendix J. We present the simulation setup in Appendix K and the daily capital recovery factor in Appendix L. {In Appendix M, N, and P, we show more simulation results based on the simulation settings in Section \uppercase\expandafter{\romannumeral 5} of the main text. In Appendix M, we conduct simulation for the case where one user has no renewable generation while the other two users have solar energy. In Appendix N, we conduct simulation to discuss the impact of the uncertainty of the load and renewable generation on users' decisions.} { In Appendix P, we numerically demonstrate that our storage virtualization model can lead to a peak load reduction in the system.} {Finally, in Appendix O, we refer to a residential load dataset to show the diversity of load profiles in a local community.}

\section*{Appendix A: Equivalent form of Stage-2 optimization problem}
{Note that the original Problem $\textbf{UP}_i^\omega$  of Stage 2 has the max-function $\max_{t} \{ p_i^{\omega,g}[t]\}$ in its objective function. We first transform it into an equivalent linear programming problem, so that we can use results for parametric linear programming in Appendix B. Towards this end, we introduce an auxiliary  variable $p_i^{\omega,m}:= \max_{t} \{p_i^{\omega,g}[t]\}$. Further, we eliminate  the variable $p_i^{\omega,g}$  by substituting Equation \text{(2)}. As a result, Problem $\textbf{UP}_i^\omega$ is equivalent to }
  \begin{align}
    \ \min \ 	&qx_i^\omega+\pi_b\sum_{t\in \mathcal{T}}(-p_i^{\omega,dis}[t]+p_i^{\omega,ch}[t])\notag\\&+\pi_p p_i^{\omega,m}+(\pi_s-\pi_b) \sum_{t\in \mathcal{T}}{ p_i^{\omega,r,u}[t]}\notag\\
    \text{s.t.} \ \ & \text{(3)},\text{(4)},\text{(5)},\text{(7)},\notag\\
    &0\leq P_i^{\omega,l}[t]-p_i^{\omega,r,u}[t]-p_i^{\omega,dis}[t]+p_i^{\omega,ch}[t],\notag\\&\ \ \ \ \ \ \ \ \ \ \ \ \ \ \ \ \ \ \ \ \ \ \ \ \ \ \ \ \ \ \ \ \ \ \ \ \ \ \ \ \forall t \in\mathcal{T}, \tag{27}\label{eq:37}\\
    &P_i^{\omega,l}[t]-p_i^{\omega,r,u}[t]-p_i^{\omega,dis}[t]+p_i^{\omega,ch}[t]\leq p_i^{\omega,m}\notag,\\&\ \ \ \ \ \ \ \ \ \ \ \ \ \ \ \ \ \ \ \ \ \ \ \ \  \ \ \ \ \ \ \ \ \ \ \ \ \ \ \ \forall t \in\mathcal{T}, \tag{28}\label{eq:38}\\
    &0\leq p_i^{\omega,ch}[t],~0\leq p_i^{\omega,dis}[t],\forall t \in\mathcal{T},\tag{29}\label{eq:39}\\
    \text{var:}\ \  &x_i^\omega ,p_i^{\omega,m},\bm{p}_i^{\omega,r,u},\bm{p}_i^{\omega,ch},\bm{p}_i^{\omega,dis},  \bm{e}_i^\omega.\notag
  \end{align}
We denote the optimal solutions to Problem $\textbf{UP}_i^{\omega}$  under a given price $q$ as $\big(x_i^{\omega\ast}(q),p_i^{\omega,m*}(q),\bm{p}_i^{\omega,r,u*}(q),$ $ \bm{p}_i^{\omega,ch*}(q), \bm{p}_i^{\omega,dis*}(q),  \bm{e}_i^{\omega*}(q)\big)$. 

Similarly, by introducing the auxiliary  variable $p_i^{\omega,m}:= \max_{t} \{ p_i^{\omega,g}[t]\}$ and  eliminating the variable $p_i^{\omega,g}$  via Equation \text{(2)}, we can rewrite Problem $\textbf{UPP}_i^{\omega}$ into a quadratic programming problem as follows. Such a reformulation is equivalent to the original problem, and is easier to analyze due to the elimination of the term    $\max_{t} \{ 	 p_i^{\omega,g}[t]\}$.
  \begin{align}
    \ \min \ 	&qx_i^\omega+\pi_b\sum_{t\in \mathcal{T}}(-p_i^{\omega,dis}[t]+p_i^{\omega,ch}[t])+\pi_p p_i^{\omega,m}+\notag\\&(\pi_s-\pi_b) \sum_{t\in \mathcal{T}}{ p_i^{\omega,r,u}[t]}+\varepsilon\sum_{t\in \mathcal{T}} ((p_i^{\omega,ch}[t])^2+(p_i^{\omega,dis}[t])^2)\notag\\
    \text{s.t.} \ \ & \text{(3)}, \text{(4)},\text{(5)},\text{(7)}, \eqref{eq:37}, \eqref{eq:38},\eqref{eq:39},\notag\\
    \text{var:}\ \  &x_i^\omega ,p_i^{\omega,m},\bm{p}_i^{\omega,r,u},\bm{p}_i^{\omega,ch},\bm{p}_i^{\omega,dis},  \bm{e}_i^\omega.\notag
  \end{align}
In the later analysis, we will let the penalty coefficient $\varepsilon$ approach zero. Thus, Problem $\textbf{UPP}_i^{\omega}$ will approach the linear programming problem $\textbf{UP}_i^{\omega}$ that is a linear programming problem. We denote the optimal solutions to Problem $\textbf{UPP}_i^\omega$ under a given price $q$ and parameter $\varepsilon$ as $\big(x_i^{\omega\star}(q,\varepsilon),$ $p_i^{\omega,m\star}(q,\varepsilon),$ $\bm{p}_i^{\omega,r,u\star}(q,\varepsilon),$ $\bm{p}_i^{\omega,ch\star}(q,\varepsilon),$ $\bm{p}_i^{\omega,dis\star}(q,\varepsilon),$ $\bm{e}_i^{\omega\star}(q,\varepsilon)\big)$.

For the later analysis of the problem of Stage 2, we will focus on these two equivalent problems.
\vspace{8mm}

\section*{{Appendix B: Conclusions of parametric linear programming}}
For later proof and analysis, we will use \textbf{Lemma 1} for parametric linear programming \cite{LQPara}, as introduced below.

Consider the following set of  linear programming problems  $L_2(\phi)$  parameterized by $\phi$, where $\boldsymbol{b}$ and $\boldsymbol{\beta}$ are constant vectors. 
  \begin{align*}
    L_2(\phi): \ \ \ z(\phi):= \min  &\ \boldsymbol{c}^T\boldsymbol{x}\\
    \text{s.t.}\ &A\boldsymbol{x}=\boldsymbol{b}+\phi \boldsymbol{\beta},\\
    &\boldsymbol{x}\geq 0.
  \end{align*}
Note that the right-hand-side of the constraint changes with $\phi$.  Based on the results of parametric linear programming \cite{LQPara}, we have the following lemma:

\vspace{2mm}
\noindent \textbf{Lemma 1}: The optimal value of the objective
function $z(\phi)$ in Problem $L_2(\phi)$ is a continuous, piecewise linear, and  convex
function of the parameter $\phi$. Furthermore, the number of  transition points of the function $z(\phi)$ is finite.
\vspace{2mm}

Lemma 1 lays the foundation for the later analysis of the linear programming  problem $\textbf{UP}_i^\omega$  in Stage 2 as proposed in Appendix A.

Note that when $\phi=0$, $L_2(\phi)$ reduces to the following linear programming problem, denoted as $L_1$:
  \begin{align*}
    L_1: \ \ \ \min  &\ \boldsymbol{c}^T\boldsymbol{x}\\
    \text{s.t.}\ &\boldsymbol{A}\boldsymbol{x}=\boldsymbol{b},\\
    &\boldsymbol{x}\geq \boldsymbol{0}.
  \end{align*}

The corresponding dual problem of Problem $L_1$ is $DL_1$:
  \begin{align*}
    DL_1: \ \ \ \max  &\ \boldsymbol{b}^T\boldsymbol{y}\\
    \text{s.t.}\ &A^T\boldsymbol{y}+\boldsymbol{s}=\boldsymbol{c},\\
    &\boldsymbol{s}\geq 0.
  \end{align*}

Later in Appendix D, we will present the algorithm for computing  the transition points of $L_2(\phi)$  based on the problem $L_1$ and its dual problem $DL_1$ according to\cite{LQPara}.
\vspace{6mm}

\section*{Appendix C: Proof of  Proposition 1}
We divide the proof Proposition 1 into 3 stages: (a) we first show that $x_i^{\omega*}(q)$ is a non-increasing function of the  price $q$; (b) we then prove that there exists an upper bound of $x_i^{\omega*}(q)$; (c) finally, we  show that $x_i^{\omega*}(q)$ is a stepwise function of the  price $q$.

\subsection*{A. The optimal capacity $x_i^{\omega*}(q)$ is non-increasing in price $q$.}
Recall that the optimal capacity  $x_i^{\omega*}(q)$ is the solution to Problem ${\textbf{UP}_i^{\omega}}$ under a given price $q$.  We now show that when the price $q$ increases from $q_1$ to $q_2>q_1$, any  increased  capacity $x_i^{\omega*}(q_1)+\Delta x$ (with  any $\Delta x>0$) is a worse solution for $q_2$ compared with the original solution $x_i^{\omega*}(q_1)$.

First, we analyze user $i$'s  optimization problem  ${\textbf{UP}_i^{\omega}}$ as shown in Appendix A by tentatively fixing the variable $x_i^\omega$.  This leads to a new optimization problem ${\textbf{UP}_i^{\omega,L}}{(x_i^\omega)}$ as follows: 
  \begin{align}
    \ \ f_i(x_i^\omega) := \min ~	&\pi_b\sum_{t\in \mathcal{T}}(-p_i^{\omega,dis}[t]+p_i^{\omega,ch}[t])\notag\\&+\pi_p p_i^{\omega,m}+(\pi_s-\pi_b) \sum_{t\in \mathcal{T}}{ p_i^{\omega,r,u}[t]}\tag{30}\label{eq:y01}\\
    \text{s.t.} \ \ & \text{(3)},\text{(4)},\text{(5)},\text{(7)},\eqref{eq:37}, \eqref{eq:38}, \eqref{eq:39},\notag\\
    \text{var:}\ \  &p_i^{\omega,m};\bm{p}_i^{\omega,r,u},\bm{p}_i^{\omega,ch},\bm{p}_i^{\omega,dis};  \bm{e}_i^\omega,\notag
  \end{align}
where $f_i(x_i^\omega)$ denotes the optimal value of problem  ${\textbf{UP}_i^{\omega,L}}{(x_i^\omega)}$ under the parameter $x_i^\omega$, i.e.,  user $i$'s  minimal electricity bill by utilizing the storage capacity $x_i^{\omega}$.

Second, for any  price $q_1$,  recall that the capacity $x_i^{\omega*}(q_1)$ is the optimal solution to problem $\textbf{UP}_i^{\omega}$ given the price $q_1$. For any $\Delta x>0$, let us consider Problem ${\textbf{UP}_i^{\omega,L}}{(x_i^{\omega*}(q_1)+\Delta x)}$, whose objective function is $f_i{(x_i^{\omega*}(q_1)+\Delta x)}$. Since $x_i^{\omega*}(q_1)$ is the optimal solution to Problem $\textbf{UP}_i^{\omega}$ given the price $q_1$, $f_i(x_i^{\omega*}(q_1))+q_1 x_i^{\omega*}(q_1)$  must be the optimal objective value of Problem $\textbf{UP}_i^{\omega}$ under the  price $q_1$. Therefore, for any other $x_i^{\omega*}(q_1)+\Delta x$, the overall cost can only be higher. Thus, we have
  \begin{align}
    &f_i(x_i^{\omega*}(q_1))+q_1 x_i^{\omega*}(q_1) \notag \\ 	&~~~~~~~~~~~~~\leq f_i(x_i^{\omega*}(q_1)+\Delta x)+q_1 (x_i^{\omega*}(q_1)+\Delta x). \tag{31} \label{eq:40}
  \end{align}
Note that $(q_2-q_1)x_i^{\omega*}(q_1) \leq (q_2-q_1)(x_i^{\omega*}(q_1)+\Delta x) $ due to $q_2>q_1$ and $\Delta x>0$. Adding this inequality to both sides of  \eqref{eq:40}, we have
  \begin{align}
    &f_i(x_i^{\omega*}(q_1))+q_2 x_i^{\omega*}(q_1) \notag \\ 	&~~~~~~~~~~~~~\leq f_i(x_i^{\omega*}(q_1)+\Delta x)+q_2 (x_i^{\omega*}(q_1)+\Delta x). \tag{32}\label{eq:41}
  \end{align}
This means that at the higher price  $q_2$, for any  $ \Delta x>0$, the capacity $x_i^{\omega*}(q_1)+\Delta x$ increases user $i$'s cost and is a  worse solution  than $x_i^{\omega*}(q_1)$.

Therefore, the optimal capacity will not increase with the  price $q$.

\subsection*{B. There exists an upper bound on the capacity $x_i^{\omega}$  that user $i$ will purchase given  any price.}
To prove this statement, for any price,  we first find a lower-bound for the optimal value $f_i(x_i^\omega)$ (which represents the electricity bill) in Problem ${\textbf{UP}_i^{\omega,L}}{(x_i^\omega)}$ as defined in the previous sub-section. Then, based on Lemma 1, we show that the function  $f_i(x_i^\omega)$ is non-increasing and piecewise linear. Finally, we can show that there exists a maximum capacity $x_i^{\omega_0}$ beyond which user $i$ can not benefit more under  any price $q$.

Towards this end, suppose that Problem $\textbf{UP}_i^\omega$ has been solved at a given price $q$. Based on the constraints \text{(3)} and \text{(5)}, we have
  \begin{align}
    \sum_{t\in \mathcal{T}}(\eta^cp_i^{\omega,ch}[t]-p_i^{\omega,dis}[t]/\eta^d)= 0. \tag{33}\label{eq:totalcd}
  \end{align}
Since $0\leq \eta^c\leq 1$ and  $0\leq \eta^d \leq 1$, we obtain
  \begin{align}
    \sum_{t\in \mathcal{T}}(p_i^{\omega,ch}[t]-p_i^{\omega,dis}[t])\geq 0.\tag{34}\label{eq:x01}
  \end{align}
Due to the constraints \eqref{eq:38} and  \eqref{eq:x01}, we have
  \begin{align}
    p_i^{\omega,m}&\geq \sum_{t\in \mathcal{T}}\ (P_i^{\omega,l}[t]-p_i^{\omega,r,u}[t]-p_i^{\omega,dis}[t]+p_i^{\omega,ch}[t])/T\notag\\
	&\geq \sum_{t\in \mathcal{T}}\ (P_i^{\omega,l}[t]-p_i^{\omega,r,u}[t])/T.\tag{35}\label{eq:x02}
  \end{align}
Based on  the constraints \text{(7)} and \eqref{eq:x02}, we further have 
  \begin{align}
    p_i^{\omega,m}\geq \sum_{t\in \mathcal{T}}\ (P_i^{\omega,l}[t]-P_i^{\omega,r}[t])/T\tag{36} \label{eq:x0}.
  \end{align}
Substituting $\eqref{eq:x01}$ and $\eqref{eq:x0}$ into  \eqref{eq:y01}, we have 
  \begin{align*}
    f_i(x_i^\omega)&\geq \pi_p p_i^{\omega,m}+(\pi_s-\pi_b) \sum_{t\in \mathcal{T}}{ p_i^{\omega,r,u}[t]}\\
    &\geq \frac{\pi_p\sum_{t}\ (P_i^{\omega,l}[t]-P_i^{\omega,r}[t])}{T}+(\pi_s-\pi_b) \sum_{t\in \mathcal{T}}P_i^{\omega,r}[t],
  \end{align*}
which shows that $f_i(x_i^\omega)$ is lower bounded. 

Then,  we can show that the function $f_i(x_i^\omega)$  is non-increasing because the increased $x_i^\omega$ relaxes the constraint \text{(3)} in Problem ${\textbf{UP}_i^{\omega,L}}{(x_i^\omega)}$.  We can also show that  $f_i(x_i^\omega)$ is  a piecewise linear function  with a finite number of transition points according to {Lemma 1}. 

Finally, the above properties (i.e., the existence of the  lower bound, the non-increasing piecewise linearity, and the finite number of transition points  of $f_i(x_i^\omega)$) imply that the slope of the piecewise function $f_i(x_i^\omega)$ will be zero if the virtual capacity goes beyond  a threshold $x_i^{\omega_0}$.  Otherwise,  $f_i(x_i^\omega)$ cannot be lower-bounded. Therefore, user $i$'s maximum capacity to purchase is bounded by  $x_i^{\omega_0}$ since acquiring more capacity will not reduce his electricity bill $f_i(x_i^\omega)$.

\subsection*{C. The optimal capacity $x_i^{\omega*}(q)$ is a stepwise correspondence in  price $q$.}
We next show that the optimal capacity $x_i^{\omega*}(q)$ is a stepwise correspondence in  price $q$. To this end, we analyze the minimal electricity bill $f_i(x_i^\omega)$ under the capacity $x_i^\omega$ compared with the virtual storage cost $q x_i^\omega$.

Specifically, since the  objective of Problem $\textbf{UP}_i^\omega$ is equivalent to 
  \begin{align*}
	\min_{{x_i^\omega}}\min_{\substack{{p_i^{\omega,m},\bm{p}_i^{\omega,r,u}},\\{\bm{p}_i^{\omega,ch},\bm{p}_i^{\omega,dis},  \bm{e}_i^\omega}}}&\pi_b\sum_{t\in \mathcal{T}}(-p_i^{\omega,dis}[t]+p_i^{\omega,ch}[t])\notag\\&+\pi_p p_i^{\omega,m}+(\pi_s-\pi_b) \sum_{t\in \mathcal{T}}{ p_i^{\omega,r,u}[t]}+qx_i^\omega,
  \end{align*}
\vspace{-1mm}
we can further rewrite the objective  as follows:
  \begin{align}
	\min_{{x_i^\omega}}  f_i(x_i^\omega)+q x_i^\omega. \tag{37} \label{eq:eqo}
  \end{align}   

We have that  $f_i(x_i^\omega)$ is a continuous, non-increasing,  piecewise linear convex function with $x_i^\omega$ according to Lemma 1. Furthermore, we have shown that there exists a maximum capacity $x_i^{\omega_0}$ beyond which the electricity bill   $f_i(x_i^\omega)$ cannot be further reduced. Therefore, the function $f_i(x_i^\omega)$ can be expressed as follows:
\vspace{1mm}
  \begin{align}
    f_i(x_i)=\left \{
    \begin{aligned}
    &-q_i^{\omega_{K_i^\omega}} x_i+s_i^{\omega_{K_i^\omega}},\ x_i \in[x_i^{\omega_{K_i^\omega}},x_i^{\omega_{K_i^\omega-1}}),\\
    &-q_i^{\omega_{K_i^\omega-1}} x_i+s_i^{\omega_{K_i^\omega-1}},\ x_i \in[x_i^{\omega_{K_i^\omega-1}},x_i^{\omega_{K_i^\omega-2}}),\\
    &\ \ \ \ ...\\
    & -q_i^{\omega_0} x_i+s_i^0,\ x_i \in[x_i^{\omega_0},\infty),
\end{aligned}
\right.\tag{38}\label{eq:f}
  \end{align}
where $q_i^{\omega_{K_i^\omega}}> q_i^{\omega_{K_i^\omega-1}}>\cdots > q_i^{\omega_0}=0$ and  $x_i^{\omega_0}>x_i^{\omega_1}>\cdots > x_i^{\omega_{K_i^\omega}}=0$. We denote the set of slopes  of the piece-wise function as  $\mathcal{Q}_i^\omega=\{ q_i^{\omega_1},\cdots, q_i^{\omega_{K_i^\omega-1}}, q_i^{\omega_{K_i^\omega}}\}$, and the set of threshold points of the function as $\mathcal{X}_i^\omega=\{x_i^{\omega_0}, x_i^{\omega_1},\cdots, x_i^{\omega_{K_i^\omega-1}}, x_i^{\omega_{K_i^\omega}}\}$. {Both sets of $\mathcal{Q}_i^\omega$ and $\mathcal{X}_i^\omega$ are finite based on Lemma 1.}

Finally, for the objective function \eqref{eq:eqo},  {by comparing the slopes of the electricity bill $f_i(x_i^\omega)$ and the storage cost $qx_i$ with respect to $x_i^\omega$}, we obtain the optimal solutions $ x_i^{\omega\ast}(q)$  of Problem  $\textbf{UP}_i^{\omega}$ as follows:
  \begin{align}
     x_i^{\omega\ast}(q)=\left \{
     \begin{aligned}
     &x_i^{\omega_0},q \in (0,q_i^{\omega_1}),\\
    &x_i^{\omega_1},q \in (q_i^{\omega_1},q_i^{\omega_2}),\\
    &...\\
    &x_i^{\omega_{K_i^\omega}},q \in (q_i^{\omega_{K_i^\omega}},\infty),\\
    \end{aligned}
    \right.\tag{39}\label{eq:xc}
  \end{align}
and for any $ q=q_i^{\omega_k}\in \mathcal{Q}_i^\omega$,  $x_i^\ast(q)$  can be any value in the set $ [x_i^{\omega_{k-1}},x_i^{\omega_k}]$. 

Based on Section C.A, C.B and C.C in this appendix, we have Proposition 1 proved. $\qed$

\vspace{8mm}

\section*{Appendix D: The algorithm to compute the sets  $\mathcal{Q}_i^\omega$ and $\mathcal{X}_i^\omega$}

We design Algorithm 3 to compute the set of threshold prices $\mathcal{Q}_i^\omega$ and the set of  optimal capacities $\mathcal{X}_i^\omega$ stated in Proposition 1.

We first present Subroutine 2 that is used in Algorithm 3. First, recall that we have defined the linear programming problem $L_1$, its dual problem $DL_1$, and the parametric linear programming problem  $L_2(\theta)$ with the optimal value $z(\theta)$ in Appendix B.  For the parametric linear programming problem $L_2(\phi)$,  we present Subroutine 2  (proposed  in  Section 4.4 of \cite{LQPara}) to compute the transition points and slopes of the  optimal value $z(\theta)$. It is proved (shown in Theorem 34 of \cite{LQPara}) that Subroutine 2 terminates after a finite number of iterations. If $K$ is the number of iterations upon termination, then $\phi(1)$, $\phi(2)$,...,$\phi(K)$ are the successive transition-points of $z(\phi)$ (from the starting parameter $\phi(0)$ extending to the nonnegative part of the real line) with the slope  $z'(\phi(k))$ on the interval $(\phi(k),\phi(k+1)) (1\leq k<K)$.  The main idea of this subroutine is that the domain of $z(\phi)$ can be partitioned in a finite set of linear subintervals such that the dual optimal set is constant on each linear subinterval. The details about this subroutine can be found in the paper \cite{LQPara}.

Then, based on Subroutine 2, we design Algorithm 3 for any user $i$ to compute the set of slopes and threshold points of the optimal value $f_i(x_i^\omega)$ of Problem ${\textbf{UP}_i^{\omega,L}}{(x_i^\omega)}$, where the threshold points form the set $\mathcal{X}_i^\omega$ and the absolute value of slopes form the set $\mathcal{Q}_i^\omega$. {Note that in \eqref{eq:f}, the slopes of the optimal value $f_i(x_i^\omega)$ are the opposite values  of the threshold prices, which is why there is a negative sign in Step 4.}\footnote{Note that in our problem, we will also include the starting parameter $x_i^\omega=0$ as the threshold price.}

	\begin{algorithm}  
	\caption*{Algorithm 3: User $i$ computes the sets  $\mathcal{Q}_i^\omega$ and $\mathcal{X}_i^\omega$}  
	\begin{algorithmic}[1]  
		\STATE Input: the parametric linear programming problem ${\textbf{UP}_i^{\omega,L}}{(x_i^\omega)}$;
		\STATE  Transform Problem ${\textbf{UP}_i^{\omega,L}}{(x_i^\omega)}$ into the form of Problem $L_2(\phi)$ and construct the corresponding primal problem $L_1$ and dual problem $DL_1$ without the perturbation parameter (i.e., $x_i^\omega=0$);
	    \STATE Implement Subroutine 2 $Td({\textbf{UP}_i^{\omega,L}}{(x_i^\omega)})=(\mathcal{Z},{\Phi})$ to compute the transition points and slopes;
		\STATE $\mathcal{Q}_i^\omega=-\mathcal{Z}$, $\mathcal{X}_i^\omega={\Phi}$;
		\STATE {\textbf{output}}: $(\mathcal{Q}_i^\omega,\mathcal{X}_i^\omega)$
	\end{algorithmic}  
   \end{algorithm} 

	\begin{algorithm}  
	\caption*{Subroutine 2 $Td(L_2(\phi))$ for computing the transition points and slopes}  
	\begin{algorithmic}[1]  
		\STATE \textbf{initialization}: $k=0$; $ready$:=false;  $x(0):=x^*$ which is an optimal solution  of the primal  problem $L_1$; $(y(0),s(0)):=(y^*,s^*)$ which is an optimal solution of the dual  problem $DL_1$; a perturbation vector $\Delta b$.
		\STATE \textbf{Original slope}: Solve $z'(\phi(k))=\max_{y,s}\{\Delta b^T y: A^Ty+s=c,s\geq0, s^Tx(0)=0\}$;
		\REPEAT 
		\STATE	Solve  $\max_{\phi,x}\{\phi: Ax=b+\phi \Delta b,x\geq0, x^Ts(k)=0\}$. If it is unbounded,  $ready$:=true; otherwise, its optimal solution is assigned to  $(\phi(k+1),x(k+1))$; 
		\STATE Solve $z'(\phi(k+1))=\max_{y,s}\{\Delta b^T y: A^Ty+s=c,s\geq0, s^Tx(k+1)=0\}$. If it is unbounded,  $ready$:=true; otherwise, its optimal solution is assigned to  $(y(k+1),s(k+1))$; 
         \STATE k:=k+1;
		\UNTIL{	$ready$ }
		\STATE {\textbf{output}}:$(\mathcal{Z},{\Phi})$,where $\mathcal{Z}:=\{z'(\phi(i)),i=0,2,...k-1\}$; ${\Phi}:\{\phi(i),i=0,2,...k-1\}$.
	\end{algorithmic}  
   \end{algorithm}

\vspace{8mm}

\section*{Appendix E: Proof of  Proposition 2}

To prove Proposition 2, we first prove the uniqueness of the optimal charge and discharge decisions  $\bm{p}_i^{\omega,ch\star}$ and $\bm{p}_i^{\omega,dis\star}$. Then, based on the uniqueness of  $\bm{p}_i^{\omega,ch\star}$ and $\bm{p}_i^{\omega,dis\star}$, we further show the uniqueness of other optimal decision variables, i.e.,  $x_i^{\omega\star}$ , $\bm{e}_i^{\omega\star}$, $p_i^{\omega,m\star}$ and $\bm{p}_i^{\omega,r,u\star}$.

For simplicity of notations, we collect all decision variables of Problem $\textbf{UPP}_i^{\omega}$ to a  vector   ${\textbf{x}_i^\omega}$, i.e.,
$\textbf{x}_i^\omega$ $=(x_i^\omega,p_i^{\omega,m},\bm{p}_i^{\omega,r,u},$ $\bm{p}_i^{\omega,ch},$ $\bm{p}_i^{\omega,dis}, $ $ \bm{e}_i^\omega)$.
We denote the optimal solution to Problem $\textbf{UPP}_i^\omega$ by  $\textbf{x}_i^{\omega\star}(q,\varepsilon)$, and the optimal objective value of Problem $\textbf{UPP}_i^\omega$ by $G(\textbf{x}_i^{\omega\star}(q,\varepsilon))$. For simplicity, we use $\textbf{x}_i^{\omega\star}$  to denote  $\textbf{x}_i^{\omega\star}(q,\varepsilon)$ for the later discussion whenever $q$ and $\varepsilon$ are clearly fixed.

\subsection*{A. Uniqueness of the optimal solutions $\bm{p}_i^{\omega,ch\star}$ and $\bm{p}_i^{\omega,dis\star}$} 

We first show by contradiction that the optimal decisions of  $\bm{p}_i^{\omega,ch\star}$ and $\bm{p}_i^{\omega,dis\star}$ must be unique. Suppose that there are two optimal solutions  $\textbf{x}_i^{\omega\star}$ and  $\textbf{x}_i^{\omega'}$,  such that there exits {a time slot  $t$}  making ${p}_i^{\omega,ch\star}[t]\neq {p}_i^{\omega,ch'}[t]$ or ${p}_i^{\omega,dis\star}[t]\neq {p}_i^{\omega,dis'}[t]$.

Since the constraints of Problem  $\textbf{UPP}_i^{\omega}$ are convex, we can construct another feasible solution   $\textbf{x}_i^{\omega\dagger}=(\textbf{x}_i^{\omega\star} +\textbf{x}_i^{\omega'})/{2}$ to Problem  $\textbf{UPP}_i^{\omega}$.			
 We must have $G(\textbf{x}_i^{\omega'})=G(\textbf{x}_i^{\omega\star})$, i.e.,
  \begin{align*}
    &qx_i^{\omega\star}+\pi_b\sum_{t\in \mathcal{T}}(-p_i^{\omega,dis\star}[t]+p_i^{\omega,ch\star}[t])+\pi_p p_i^{\omega,m\star}\\&+\hspace{-1mm}(\pi_s-\pi_b) \hspace{-1mm}\sum_{t\in \mathcal{T}}{ p_i^{\omega,r,u\star}[t]}+\varepsilon\hspace{-1mm}\sum_{t\in \mathcal{T}} \big(({p}_i^{\omega,ch\star}[t])^2+({p}_i^{\omega,dis\star}[t])^2\big)\\=&qx_i^{\omega'}+\pi_b\sum_{t\in \mathcal{T}}(-p_i^{\omega,dis'}[t]+p_i^{\omega,ch'}[t])+\pi_p p_i^{\omega,m'}\\&+\hspace{-1mm}(\pi_s-\pi_b) \hspace{-1mm}\sum_{t\in \mathcal{T}}{ p_i^{\omega,r,u'}[t]}+\varepsilon\hspace{-1mm}\sum_{t\in \mathcal{T}}\big( ({p}_i^{\omega,ch'}[t])^2+({p}_i^{\omega,dis'}[t])^2\big).
  \end{align*} 

Then we have
  \begin{align*}
	G(\textbf{x}_i^{\omega\dagger})-G(\textbf{x}_i^{\omega\star})&=-\varepsilon\sum_{t\in \mathcal{T}} \left(\frac{1}{2}({p}_i^{\omega,ch\star}[t]^\star-{p}_i^{\omega,ch\dagger}[t])\right)^2\\&-\varepsilon\sum_{t\in \mathcal{T}} \left(\frac{1}{2}({p}_i^{\omega,dis\star}[t]^\star-{p}_i^{\omega,dis\dagger}[t])\right)^2 <0,
  \end{align*}
which contradicts the assumption that $\textbf{x}_i^{\omega\star}$ is optimal. Therefore, we have ${p}_i^{\omega,ch\star}[t]= {p}_i^{\omega,ch'}[t]$ and ${p}_i^{\omega,dis\star}[t]={p}_i^{\omega,dis'}[t]$, for any $t \in\mathcal{T}$.

\subsection*{B. Uniqueness of the  optimal solutions $x_i^{\omega\star}$ and $\bm{e}_i^{\omega\star}$} We next show that the optimal capacity $x_i^{\omega\star}$ can be uniquely determined given the optimal  charge and discharge profiles $\bm{p}_i^{\omega,ch\star}$ and $\bm{p}_i^{\omega,dis\star}$.

Based on the constraints \text{(3)} and \text{(4)}, we have
  \begin{align}
	&e_i^{\omega\star}[t]=e_i^{\omega\star}[0]+\psi[t],\ \forall t \in \mathcal{T},\tag{40}\label{eq:ee}\\ 
	&0 \leq e_i^{\omega\star}[t]\leq {x_i^{\omega\star}},\ \forall t \in \mathcal{T}',\tag{41}\label{eq:newx}
  \end{align}
where we have let  
  \begin{align}
    \psi[t]=\sum_{\tau=1}^t(\eta_i^c p_i^{\omega,ch\star}[\tau]-p_i^{\omega,dis\star}[\tau]/\eta_i^d),~\forall t \in \mathcal{T}.\tag{42}\label{eq:fi}
  \end{align}

For  constraint \eqref{eq:newx}, $ e_i^{\omega\star}[t]=0$ must be satisfied in some time slots and $e_i^{\omega\star}[t]= {x_i^{\omega\star}}$ must be satisfied in some other time slots. Otherwise, we can always reduce  $x_i^{\omega\star}$  to reduce the cost. Thus, we obtain that
  \begin{align}
    x_i^{\omega\star}&= \max_{t\in \mathcal{T}} \left\{e_i^{\omega\star}[t]\right\}-\min_{t\in \mathcal{T}}  \left\{ e_i^{\omega\star}[t]\right\}\notag\\&=\max_{t\in \mathcal{T}} \left\{e_i^{\omega\star}[0]+\psi[t]\right\}-\min_{t\in \mathcal{T}} \left\{ e_i^{\omega\star}[0]+\psi[t]\right\}\notag \\&=\max_{t\in \mathcal{T}} \left\{\psi[t]\right\}-\min_{t\in \mathcal{T}}  \left\{ \psi[t]\right\}, \tag{43}\label{eq:43}
  \end{align}
with
  \begin{align}
    &\max_{t\in \mathcal{T}} \left\{ e_i^{\omega\star}[0]+\psi[t]\right\}=x_i^{\omega\star}\tag{44}, \\&\min_{t\in \mathcal{T}} \left\{ e_i^{\omega\star}[0]+\psi[t]\right\}=0. \tag{45}
  \end{align}
Therefore,
  \begin{align}
    e_i^{\omega\star}[0]=-\min_{t\in \mathcal{T}} \left\{ \psi_t\right\}.\tag{46}\label{eq:44}
  \end{align}

Since $\bm{p}_i^{\omega,ch\star}$ and $\bm{p}_i^{\omega,dis\star}$ are unique, $\psi[t]$ is also uniquely determined according to \eqref{eq:fi}. Then,  the optimal capacity $x_i^{\omega\star}$  is unique according to \eqref{eq:43}. The storage level $\bm{e}_i^{\omega\star}$ is also uniquely determined according to \eqref{eq:44} and \eqref{eq:ee}.

\subsection*{C. Uniqueness of optimal solutions $p_i^{\omega,m\star}$ and $\bm{p}_i^{\omega,r,u\star}$}
We next show that the optimal solutions $p_i^{\omega,m\star}$ and $\bm{p}_i^{\omega,r,u\star}$ can be uniquely determined given other unique decisions  $x_i^{\omega\star}$,$\bm{e}_i^{\omega\star}$, $\bm{p}_i^{\omega,ch\star}$ and $\bm{p}_i^{\omega,dis\star}$.

For  $p_i^{\omega,m\star}$ and $\bm{p}_i^{\omega,r,u\star}$, we have the constraints \eqref{eq:38} and \text{(7)} as follows for all $t\in \mathcal{T}$:
  \begin{align*}
    &0\leq P_i^{\omega,l}[t]-p_i^{\omega,r,u\star}[t]-p_i^{\omega,dis\star}[t]+p_i^{\omega,ch\star}[t] \leq p_i^{\omega,m\star},\\
    & 0\leq p_i^{\omega,r,u\star}[t] \leq P_i^{\omega,r}[t].
  \end{align*}

First, given other unique decisions  $x_i^{\omega\star}$,$\bm{e}_i^{\omega\star}$, $\bm{p}_i^{\omega,ch\star}$ and $\bm{p}_i^{\omega,dis\star}$, we can determine the optimal solution $\bm{p}_i^{\omega,r,u\star}$ as follows. In the objective function of Problem $\textbf{UPP}_i^\omega$, the item $(\pi_s-\pi_b) \sum_t p_i^{\omega,r,u\star}[t]$ is always negative due to $\pi_s<\pi_b$. Hence,  $p_i^{\omega,r,u\star}[t]$ should be as large as possible in order to reduce the item $(\pi_s-\pi_b) \sum_t p_i^{\omega,r,u\star}[t]$ in the objective and reduce the shaved peak $p_i^{\omega,m\star}$ in the constraint \eqref{eq:38}. Then, we can show that the optimal solution $\bm{p}_i^{\omega,r,u\star}$  satisfies the following properties:
\begin{itemize}
	\item If $P_i^{\omega,l}[t]-P_i^{\omega,r}[t]-p_i^{\omega,dis\star}[t]+p_i^{\omega,ch\star}[t]\geq 0$, then $p_i^{\omega,r,u\star}[t]=P_i^{\omega,r}[t]$. Otherwise, if $p_i^{\omega,r,u\star}[t]<P_i^{\omega,r}[t]$, we can always increase $p_i^{\omega,r,u\star}[t]$ to reduce user's cost.
	\item If $P_i^{\omega,l}[t]-P_i^{\omega,r}[t]-p_i^{\omega,dis\star}[t]+p_i^{\omega,ch\star}[t]< 0$, then 
	$p_i^{\omega,r,u\star}[t]=P_i^{\omega,l}[t]-p_i^{\omega,dis\star}[t]+p_i^{\omega,ch\star}[t]$. Otherwise, if	$p_i^{\omega,r,u\star}[t]<P_i^{\omega,l}[t]-p_i^{\omega,dis\star}[t]+p_i^{\omega,ch\star}[t]$, we can always increase $p_i^{\omega,r,u\star}[t]$ to reduce user's cost. Note that according to  the constraint \eqref{eq:37}, $p_i^{\omega,r,u\star}[t]\leq P_i^{\omega,l}[t]-p_i^{\omega,dis\star}[t]+p_i^{\omega,ch\star}[t]$.
\end{itemize}
Since $\bm{p}_i^{\omega,ch\star}$ and $\bm{p}_i^{\omega,dis\star}$ are unique and $\bm{p}_i^{\omega,r,u\star}$  is uniquely determined by $\bm{p}_i^{\omega,dis\star}$ and $\bm{p}_i^{\omega,ch\star}$ as above,   it implies that $\bm{p}_i^{\omega,r,u\star}$ is unique. 

Then, we can  show that ${p}_i^{\omega,m\star}$ is uniquely determined by $\bm{p}_i^{\omega,r,u\star}$, $\bm{p}_i^{\omega,dis\star}$ and $\bm{p}_i^{\omega,ch\star}$ according to the constraint \eqref{eq:38} as follows:
  \begin{align}
    {p}_i^{\omega,m\star}=\min_{t\in \mathcal{T}}\{ P_i^{\omega,l}[t]-p_i^{\omega,r,u\star}[t]-p_i^{\omega,dis\star}[t]+p_i^{\omega,ch\star}[t]\}, \tag{47}
  \end{align}
which implies that  the optimal solution ${p}_i^{\omega,m\star}$ is also unique.
\vspace{2mm}

Combining the results in Sections D.A, D.B, and D.C of this appendix, we have proved that the optimal solution $\textbf{x}_i^{\omega\star}(q,\varepsilon)$ is unique. $\qed$

\vspace{3mm}

\section*{{Appendix F: The Maximum Theorem}}

{For later proof and analysis, we first introduce the Maximum Theorem   in \cite{maximum} as follows.}

Consider the following set of optimization problem ${MO}$:
  \begin{align*}
    {MO}: ~~&\sup_{\boldsymbol{y}}  f(\boldsymbol{z},\boldsymbol{y})\\
    \text{s.t.} \ &\boldsymbol{y} \in D(\boldsymbol{z}).
  \end{align*}
Here, $\boldsymbol{z}$ is a parameter chosen from a set  $Z\subset \mathbb{R}^L$, and $\boldsymbol{y}$ is the variable chosen from a  set $Y\subset \mathbb{R}^K$.  Suppose $f: Z \times Y \rightarrow \mathbb{R}$ is a function and $D: Z \rightarrow Y$ is a
non-empty correspondence that describes the feasibility constraints. For each $\boldsymbol{z}$, the problem $MO$ finds the optimal $\boldsymbol{y}$ in $D(\boldsymbol{z})$ to maximize $f(\boldsymbol{z},\boldsymbol{y})$.

For given $\boldsymbol{z}$, we denote  $u(\boldsymbol{z})$ as the optimal value of the objective, i.e., 
$$u(\boldsymbol{\boldsymbol{z}}):=\sup_{y\in D(\boldsymbol{z})}f(\boldsymbol{z},\boldsymbol{y}),$$
and we denote $S(\boldsymbol{z})$ as the solution set, i.e.,
$$S(\boldsymbol{z}):=\{\boldsymbol{y}\in D(\boldsymbol{z}):f(\boldsymbol{z},\boldsymbol{y})=u(\boldsymbol{z})\}.$$
Note that $S(\cdot)$ is a correspondence due to the possibility of  multiple solutions.
\vspace{2mm}

\noindent \textbf{The  Maximum Theorem }: If the function $f$ is continuous and $D$ is a compact-valued and continuous correspondence, then we have 
\begin{enumerate}
	\item The objective function $u:  Z \times Y \rightarrow \mathbb{R}$  is continuous;
	\item The solution correspondence $S : Z \rightarrow Y$ is nonempty, compact valued, and upper hemi-continuous (u.h.c).
\end{enumerate}
\vspace{3mm}	

The Maximum Theorem lays the foundation for the later analysis when we let the penalty coefficient $\varepsilon$ approach zero in Problem $\textbf{UPP}_i^{\omega}$.

\vspace{8mm}
\section*{Appendix G: Proofs of  Theorem 1}
For the convenience of the later proof, we first let 
  \begin{align}
     &v_i^\omega=\pi_b\sum_{t\in \mathcal{T}}(-p_i^{\omega,dis}[t]+p_i^{\omega,ch}[t])\notag\\&~~~~~~~~~~~~~~~~~~+\pi_p p_i^{\omega,m}+(\pi_s-\pi_b) \sum_{t\in \mathcal{T}}{ p_i^{\omega,r,u}[t]}.\tag{48}\label{eq:v}
  \end{align} 
 We can write  Problem $\textbf{UPP}_i^{\omega}$ as 
  \begin{align}
    \min  ~&q x_i^\omega+v_i^\omega+\varepsilon\sum_{t\in \mathcal{T}} ((p_i^{\omega,ch}[t])^2+(p_i^{\omega,dis}[t])^2),\tag{49} \label{eq:v2}\\
    \text{s.t.} \ \ & \text{(3)}, \text{(4)},\text{(5)},\text{(7)}, \eqref{eq:37}, \eqref{eq:38},\eqref{eq:39} ~\text{and}~ \eqref{eq:v}.\notag\\
    \text{var:}\ \  &x_i^\omega ,p_i^{\omega,m},\bm{p}_i^{\omega,r,u},\bm{p}_i^{\omega,ch},\bm{p}_i^{\omega,dis} ~\text{and}~   v_i^\omega.\notag
  \end{align}   
We can also rewrite  Problem $\textbf{UP}_i^{\omega}$ as
  \begin{align}
    \min  ~&q x_i^\omega+v_i^\omega, \tag{50} \label{eq:v3}\\
    \text{s.t.} \ \ & \text{(3)}, \text{(4)},\text{(5)},\text{(7)}, \eqref{eq:37}, \eqref{eq:38},\eqref{eq:39} ~\text{and}~ \eqref{eq:v}.\notag\\
    \text{var:}\ \  &x_i^\omega ,p_i^{\omega,m},\bm{p}_i^{\omega,r,u},\bm{p}_i^{\omega,ch},\bm{p}_i^{\omega,dis} ~\text{and}~ v_i^\omega.\notag
  \end{align} 
 
For Problem $\textbf{UPP}_i^{\omega}$ of optimizing the objective \eqref{eq:v2}, given a certain price $q$ and $\varepsilon$,  we denote the corresponding  optimal value of $v_i^\omega$ as $v_i^{\omega\star}(q,\varepsilon)$. Similarly, for Problem $\textbf{UP}_i^{\omega}$ of optimizing the objective \eqref{eq:v3}, given a certain price $q$,  we denote the corresponding optimal value of $v_i^\omega$ as $v_i^{\omega*}(q)$.

To prove Theorem 1, we will first characterize the limit of the optimal solutions $x_i^{\omega\star}(q,\varepsilon)$ as $\varepsilon$ approaches zero  based on the Maximum Theorem \cite{maximum}. 
Then, we consider the limit of $v_i^{\omega\star}(q,\varepsilon)$ as $\varepsilon$ approaches zero.  Finally, we show the existence of the right limits of  $\bm{p}_i^{\omega,dis\star}(q,\varepsilon)$ and  $\bm{p}_i^{\omega,ch\star}(q,\varepsilon)$ as $\varepsilon$ approaches zero.

\subsection*{A. The limit of $x_i^{\omega\star}(q,\varepsilon)$ as $\varepsilon$ approaches zero } To characterize the limit of $x_i^{\omega\star}(q,\varepsilon)$ as $\varepsilon$ approaches zero, we will first show that the optimal capacity $x_i^{\omega\star}(q,\varepsilon)$ of Problem $\textbf{UPP}_i^\omega$ is single-valued over  $\varepsilon\in [0,\bar{\varepsilon}]$ given  any $ q \notin \mathcal{Q}_i^\omega$ and  the fixed upper bound $\bar{\varepsilon}>0$. Then, based on {the Maximum Theorem}, we show that for any $ q \notin \mathcal{Q}_i^\omega$, as $\varepsilon$ approaches zero,  $x_i^{\omega\star}(q,\varepsilon)$ approaches the limit $x_i^{\omega*}(q)$, which is the optimal capacity of Problem $\textbf{UP}_i^\omega$ (i.e.,when $\varepsilon=0$).

First, given a certain price $q$ and an upper bound $\bar{\varepsilon}>0$ , for any $\varepsilon \in (0,\bar{\varepsilon}]$, the optimal capacity $x_i^{\omega\star} (q,\varepsilon)$ of Problem $\textbf{UPP}_i^\omega$ is unique according to Proposition 2.   For any price $ q \notin \mathcal{Q}_i^\omega$, when $\varepsilon=0$, the optimal capacity $x_i^{\omega*}(q)$ of Problem $\textbf{UP}_i^\omega$ is also unique according to  Proposition 1. Note that Problem $\textbf{UP}_i^\omega$ is a special case of Problem $\textbf{UPP}_i^\omega$ at $\varepsilon=0$. Hence, we can simply use $x_i^{\omega\star} (q,\varepsilon)$ at $\varepsilon=0$  to represent the solution  $x_i^{\omega*}(q)$.  Thus, for any $ q \notin \mathcal{Q}_i^\omega$, the optimal capacity $x_i^{\omega\star}(q,\varepsilon)$ is single-valued for any $\varepsilon\in [0,\bar{\varepsilon}]$.  

Second, in Problem $\textbf{UPP}_i^\omega$, we can think of the coefficient $\varepsilon$ as the parameter $z$ in Problem ${MO}$ as defined in the Maximum Theorem in Appendix $F$. 
In order to use {the Maximum Theorem} \cite{maximum}, we can verify that  Problem $\textbf{UPP}_i^\omega$ satisfies the following conditions: (i) the objective function is continuous, and (ii) the feasibility constraint, which is a correspondence  of the  parameter $\varepsilon$, is compact-valued and continuous. Therefore, according to  {the Maximum Theorem} \cite{maximum}, the optimal solution $x_i^\omega(q,\varepsilon)$ is upper hemi-continuous in $[0,\bar{\varepsilon}]$. Since the optimal capacity $x_i^\omega(q, \varepsilon)$ is single-valued in $[0,\bar{\varepsilon}]$ for any $ q \notin \mathcal{Q}_i^\omega$, the upper hemi-continuity implies that  $x_i^\omega(q, \varepsilon)$ is continuous in $[0,\bar{\varepsilon}]$. Further, this continuity implies that for any $ q \notin \mathcal{Q}_i^\omega$, as $\varepsilon$ approaches zero,  $x_i^{\omega\star}(q,\varepsilon)$ approaches the limit $x_i^{\omega*}(q)$.  Note that this implies that the set $\mathcal{X}_i^\omega$ includes all the limiting optimal capacity over each threshold price interval $(0,q_i^{\omega_1}),(q_i^{\omega_1},q_i^{\omega_2})...$ that are defined in \eqref{eq:xc}.

\subsection*{B. The limit of $v_i^{\omega\star}(q,\varepsilon)$ as $\varepsilon$ approaches zero }
Similar to the proof of  the limit of $x_i^{\omega\star}(q,\varepsilon)$, to show the limit of $v_i^{\omega\star}(q,\varepsilon)$ as $\varepsilon$ approaches zero, we will first show that the optimal $v^{\star}(q,\varepsilon)$ of Problem $\textbf{UPP}_i^\omega$ is single-valued over  $[0,\bar{\varepsilon}]$ for any $ q \notin \mathcal{Q}_i^\omega$. Then based on {the Maximum Theorem}, we show that for any $ q \notin \mathcal{Q}_i^\omega$, as $\varepsilon$ approaches zero,  $v_i^{\omega\star}(q,\varepsilon)$ approaches the limit $v_i^{\omega*}(q)$. Finally, we also show that the limit  $v_i^{\omega*}(q)$ is stepwise over the threshold price set $\mathcal{Q}_i^\omega$.

First,  for any $\varepsilon \in (0,\bar{\varepsilon}]$ with fixed $\bar{\varepsilon}>0$, and given a certain price $q$, the optimal value  $v_i^{\omega\star} (q,\varepsilon)$ of Problem $\textbf{UPP}_i^\omega$ is unique according to Proposition 2.  Further, when $\varepsilon=0$, for any price $ q \notin \mathcal{Q}_i^\omega$, we already have that the optimal capacity $x_i^{\omega*}(q)$ of Problem $\textbf{UP}_i^\omega$ is unique. Since there must be a unique optimal value of the objective $\eqref{eq:v3}$, we obtain that the optimal value of  $v_i^{\omega*} (q)$ of Problem $\textbf{UP}_i^\omega$ is also unique  for any price $ q \notin \mathcal{Q}_i^\omega$.  Note that Problem $\textbf{UP}_i^\omega$ is a special case of Problem $\textbf{UPP}_i^\omega$ at $\varepsilon=0$. Hence, we can simply use $v_i^{\omega\star} (q,\varepsilon)$ at $\varepsilon=0$  to represent the solution  $v_i^{\omega*}(q)$.  Combining the above discussion, we conclude that, for any $ q \notin \mathcal{Q}_i^\omega$, the optimal value $v_i^{\omega\star}(q,\varepsilon)$ is single-valued for any $\varepsilon\in [0,\bar{\varepsilon}]$.  

Second, similar to the proof of  the limit of $x_i^{\omega\star}(q,\varepsilon)$, in Problem $\textbf{UPP}_i^\omega$, we regard the coefficient $\varepsilon$ as the parameter $z$ in Problem ${MO}$ as defined in the Maximum Theorem in Appendix $F$.
According to {the Maximum Theorem} \cite{maximum}, the optimal value $v_i^{\omega\star}(q,\varepsilon)$ is upper hemi-continuous in $[0,\bar{\varepsilon}]$. Since the optimal value $v_i^{\omega\star}(q, \varepsilon)$ is single-valued in $[0,\bar{\varepsilon}]$ for any $ q \notin \mathcal{Q}_i^\omega$, the upper hemi-continuity implies that $v_i^{\omega\star}(q, \varepsilon)$ is continuous in $[0,\bar{\varepsilon}]$. Further, this continuity implies that for any $q \notin \mathcal{Q}_i^\omega$, as $\varepsilon$ approaches zero,  $v_i^{\omega\star}(q,\varepsilon)$ approaches the limit $v_i^{\omega*}(q)$.

Third, we show that the optimal value $v_i^{\omega*} (q)$ is also stepwise over the threshold price set $\mathcal{Q}_i^\omega$. For Problem $\textbf{UP}_i^\omega$, we fix the optimal solution  $x_i^{\omega*}(q)$ and regard it as a parameter, which leads to a new optimization problem  $\textbf{V}_i^\omega(x_i^{\omega*}(q))$ as follows.

\textbf{Problem~}  $\textbf{V}_i^\omega(x_i^{\omega*}(q))$:
\begin{align}
 &\ \min \ v_i^\omega\notag\\
\text{s.t.} \ \ & \text{(3)},\text{(5)},\text{(7)},\eqref{eq:37}, \eqref{eq:38}, \eqref{eq:39},\notag\\
	&0 \leq e_i^\omega[t]\leq x_i^{\omega*}(q),~\forall t \in \mathcal{T'},\tag{51}\label{eq:x*}　\\
\text{var:}\ \  &p_i^{\omega,m},\bm{p}_i^{\omega,r,u},\bm{p}_i^{\omega,ch},\bm{p}_i^{\omega,dis}, \bm{e}_i^\omega.\notag
\end{align}
 Since  $x_i^{\omega*}(q)$ is stepwise over the threshold price set $\mathcal{Q}_i^\omega$, the optimal value $v_i^{\omega*} (q)$  also shows the  stepwise property over $\mathcal{Q}_i^\omega$.

\subsection*{C. The limit of $\bm{p}_i^{\omega,ch\star}(q,\varepsilon)$ and  $\bm{p}_i^{\omega,dis\star}(q,\varepsilon)$ as $\varepsilon$ approaches zero}
 To prove this statement, we follow the steps below. 
 
First, we rewrite the  optimal solutions $\bm{p}_i^{\omega,ch\star}(q,\varepsilon)$ and $\bm{p}_i^{\omega,dis\star}(q,\varepsilon)$  as continuous functions of  $\left(x_i^{\omega\star}(q,\varepsilon),v_i^{\omega\star}(q,\varepsilon)\right)$. Second,
we  show the existence of the right limit of $\left(\bm{p}_i^{\omega,ch\star}(q,\varepsilon),\bm{p}_i^{\omega,ch\star}(q,\varepsilon)\right)$ as $\varepsilon$ approaches zero  based on the limit of $\left(x_i^{\omega\star}(q,\varepsilon),v_i^{\omega\star}(q,\varepsilon)\right)$. Finally, we show that the limit of $\left(\bm{p}_i^{\omega,ch\star}(q,\varepsilon), \bm{p}_i^{\omega,ch\star}(q,\varepsilon)\right)$ is stepwise over price $q$.

\subsubsection*{1)  Rewriting the optimal charge and discharge decision $\bm{p}_i^{\omega,ch\star}(q,\varepsilon)$ and  $\bm{p}_i^{\omega,dis\star}(q,\varepsilon)$ as continuous functions of $\left(x_i^{\omega\star}(q,\varepsilon),v_i^{\omega\star}(q,\varepsilon)\right)$}  Given certain price $q \notin \mathcal{Q}_i^\omega$ and $\varepsilon>0$, by fixing  the optimal $\left(x_i^{\omega\star}(q,\varepsilon),v_i^{\omega\star}(q,\varepsilon)\right)$, we can write  Problem $\textbf{UPP}_i^\omega$ into Problem $\textbf{CD}_i^\omega\left(x_i^{\omega\star}(q,\varepsilon),v_i^{\omega\star}(q,\varepsilon)\right)$ as follows.

\textbf{Problem}  $\textbf{CD}_i^\omega\left(x_i^{\omega\star}(q,\varepsilon),v_i^{\omega\star}(q,\varepsilon)\right)$:
\begin{align}
\ \min \ 	&\sum_{t\in \mathcal{T}} ((p_i^{\omega,ch}[t])^2+(p_i^{\omega,dis}[t])^2)\notag\\
\text{s.t.} \ \ & \text{(3)},\text{(5)},\text{(7)}, \eqref{eq:37}, \eqref{eq:38},\eqref{eq:39},\notag\\
	&0 \leq e_i^\omega[t]\leq x_i^{\omega\star}(q,\varepsilon),~\forall t \in \mathcal{T'},\tag{52}\label{eq:x}　\\
	&\pi_b\sum_{t\in \mathcal{T}}(-p_i^{\omega,dis}[t]+p_i^{\omega,ch}[t])+\pi_p p_i^{\omega,m}\notag\\&~~~~~~~~~~~+(\pi_s-\pi_b) \sum_{t\in \mathcal{T}}{ p_i^{\omega,r,u}[t]}=v_i^{\omega\star}(q,\varepsilon)\tag{53}\\
\text{var:}\ \  &p_i^{\omega,m},\bm{p}_i^{\omega,r,u},\bm{p}_i^{\omega,ch},\bm{p}_i^{\omega,dis},  \bm{e}_i^\omega.\notag
\end{align}

According to {the Maximum Theorem }\cite{maximum}, the optimal solution $\big(\bm{p}_i^{\omega,ch\star}, \bm{p}_i^{\omega,ch\star}\big)$ is upper hemi-continuous  with respect to $\left(x_i^{\omega\star}(q,\varepsilon),v_i^{\omega\star}(q,\varepsilon)\right)$. Since the optimal solution  $\big(\bm{p}_i^{\omega,ch\star}, \bm{p}_i^{\omega,ch\star}\big)$ is unique in Problem $\textbf{CD}_i^\omega$ (due  to the strictly convex objective function), we conclude that the optimal solution  $\big(\bm{p}_i^{\omega,ch\star}, \bm{p}_i^{\omega,ch\star}\big)$ is continuous with respect to $\big(x_i^{\omega\star}(q,\varepsilon),v_i^{\omega\star}(q,\varepsilon)\big)$.

\subsubsection*{2) The existence of the right limit of the optimal charge and discharge decision $\left(\bm{p}_i^{\omega,ch\star}(q,\varepsilon),\bm{p}_i^{\omega,dis\star}(q,\varepsilon)\right)$ as $\varepsilon$ approaches zero} We have shown in Sections G.A and G.B of this proof that the optimal solution  $\big(x_i^{\omega\star}(q,\varepsilon), v_i^{\omega\star}(q,\varepsilon)\big)$ approaches a limit  as $\varepsilon$ approaches  zero for any  $q \notin \mathcal{Q}_i^\omega$. Since  the optimal solution $(\bm{p}_i^{\omega,ch\star},\bm{p}_i^{\omega,dis\star})$ is a continuous  function of   $\big(x_i^{\omega\star}(q,\varepsilon), v_i^{\omega\star}(q,\varepsilon)\big)$, we conclude that $(\bm{p}_i^{\omega,ch\star},\bm{p}_i^{\omega,dis\star})$ also approaches a limit as $\varepsilon$ approaches $0^+$ for any $q \notin \mathcal{Q}_i^\omega$. 

Further, this limit can be determined from the solution of   Problem $\textbf{CD}_i^\omega$ where we replace $(x_i^{\omega\star}(q,\varepsilon),v_i^{\omega\star}(q,\varepsilon))$ by their limits  $(x_i^{\omega*}(q),v_i^{\omega*}(q))$ as $\varepsilon\rightarrow 0^+$, i.e., from Problem $\textbf{CD}_i^\omega(x_i^{\omega*},v_i^{\omega*})$.  This also implies that 
 	\begin{align*}
    \hspace{-1mm} &\lim_{\varepsilon \rightarrow 0^+}\big(\bm{p}_i^{\omega,ch\star}(q,\varepsilon),\bm{p}_i^{\omega,dis\star}(q,\varepsilon)\big)\hspace{-1mm}\in\hspace{-1mm} \big(\bm{P}_i^{\omega,ch*}(q), \bm{P}_i^{\omega,dis*}(q)\big),
    \end{align*}
where that $(\bm{P}_i^{\omega,ch*}(q),\bm{P}_i^{\omega,dis*}(q))$ is the optimal charge and discharge set under $\varepsilon=0$ as defined in  Section \uppercase\expandafter{\romannumeral4}.A.2) of the main text.

\subsubsection*{3) The stepwise property of the right limit $\lim_{\varepsilon \rightarrow 0^+}\big(\bm{p}_i^{\omega,ch\star}(q,\varepsilon),\bm{p}_i^{\omega,dis\star}(q,\varepsilon)\big)$} Since the limit of the optimal solution $\left(x_i^{\omega\star}(q,\varepsilon),v_i^{\omega,m\star}(q,\varepsilon)\right)$ is stepwise over price $q$ as $\varepsilon$ approaches zero and $(\bm{p}_i^{\omega,ch\star},\bm{p}_i^{\omega,dis\star})$ is  continuous in  $\big(x_i^{\omega\star}(q,\varepsilon), v_i^{\omega\star}(q,\varepsilon)\big)$, the limit of $\left(\bm{p}_i^{\omega,ch\star}(q,\varepsilon),\bm{p}_i^{\omega,dis\star}(q,\varepsilon)\right)$  is also stepwise over price $q$.

Combining the results in Sections G.A, G.B and G.C of this appendix, we have proved Theorem 1. $\qed$

\vspace{7mm}

\section*{Appendix H:  The algorithm to compute the limit of $\bm{p}_i^{\omega,ch\star}(q,\varepsilon)$ and $\bm{p}_i^{\omega,dis\star}(q,\varepsilon)$}

The analysis in Appendix G also suggests  Algorithm 4 that can be used by any user $i$ to compute the $\lim_{\varepsilon \rightarrow 0^+}(\bm{p}_i^{\omega,ch\star}(q,\varepsilon),\bm{p}_i^{\omega,dis\star}(q,\varepsilon))$.  The idea is as follows. Since  the optimal solution $(\bm{p}_i^{\omega,ch\star},\bm{p}_i^{\omega,dis\star})$ is a continuous  function of   $(x_i^{\omega\star}(q,\varepsilon), v_i^{\omega,m\star}(q,\varepsilon))$, we will first compute the limit of $(x_i^{\omega\star}(q,\varepsilon), v_i^{\omega,m\star}(q,\varepsilon))$. Then, we further   compute the limit of $\big(\bm{p}_i^{\omega,ch\star}(q,\varepsilon),\bm{p}_i^{\omega,dis\star}(q,\varepsilon)\big)$.

\begin{algorithm}  
	\caption*{Algorithm 4: User $i$ computes $\lim_{\varepsilon \rightarrow 0^+}\bm{p}_i^{\omega,dis\star}(q,\varepsilon)$ and $\lim_{\varepsilon \rightarrow 0^+}\bm{p}_i^{\omega,ch\star}[t](q,\varepsilon)$ of scenario $\omega$ }  
	\label{alg:4}  
	\begin{algorithmic}[1]  
		\STATE compute the  threshold prices set $\mathcal{Q}_i^\omega$ and the  optimal capacity set $\mathcal{X}_i^\omega$ by Algorithm 3;
		\FOR {\textbf{each} $x_i^{\omega*}\in\mathcal{X}_i^\omega$}
		\STATE  solve Problem ${\textbf{V}_i^{\omega}}{(x_i^{\omega*})}$ (presented in Section G.B) and achieve the corresponding optimal   $v_i^{\omega*}$;
		\STATE  compute the charge and discharge decision limit  $(\bm{p}_i^{\omega,ch\star},\bm{p}_i^{\omega,dis\star})$ by solving Problem $\textbf{CD}_i^\omega(x_i^{\omega*},v_i^{\omega*})$ (presented in G.C);
		\ENDFOR
		\STATE {\textbf{output}}: $(\bm{p}_i^{\omega,dis\star},\bm{p}_i^{\omega,ch\star})$.
    \end{algorithmic}  
\end{algorithm}

\vspace{7mm}

\section*{Appendix I: Proof of  Proposition 3}

To prove Proposition 3, the high level intuition is that both  the revenue and the cost of the aggregator are continuous functions of $\left(x_i^{\omega\star}(q,\varepsilon),\bm{p}_i^{ch,\omega\star}(q,\varepsilon),\bm{p}_i^{dis,\omega\star}(q,\varepsilon)\right)$. Since we have derived  $\lim_{\varepsilon \rightarrow 0^+} \left(x_i^{\omega\star}(q,\varepsilon),\bm{p}_i^{ch,\omega\star}(q,\varepsilon),\bm{p}_i^{dis,\omega\star}(q,\varepsilon)\right)$, we can also obtain the limit of the revenue, the cost as well as the profit of the aggregator as $\varepsilon$ approaches zero.

\subsubsection*{1) The limit of revenue} Since we have proved in Theorem 1 that for any $q \notin \mathcal{Q}_i^\omega$,  $\lim_{\varepsilon \rightarrow 0^+}x_i^{\omega\star}(q,\varepsilon)=x_i^{\omega*}(q)$, we have $$\lim_{\varepsilon \rightarrow 0^+}\sum_{\omega}\rho^{\omega}q \sum_i x_i^{\omega\star}(q,\varepsilon)=\sum_{\omega}\rho^{\omega}q \sum_ix_i^{\omega*}(q).$$
\subsubsection*{2) The limit of cost} We first prove that  the aggregator's cost is a continuous function in $\varepsilon \in (0,\bar{\varepsilon}]$ and its right limit  exists at  $\varepsilon=0$. Then, we show that the cost  approaches  a stepwise function as $\varepsilon$ goes to zero.

First, for any price $ q$, we have shown in Appendix G that the optimal solution  $\left(\bm{p}_i^{ch,\omega\star}(q,\varepsilon),\bm{p}_i^{dis,\omega\star}(q,\varepsilon)\right)$  to Problem $\textbf{UPP}_i^\omega$ is a continuous function in $\varepsilon \in (0,\bar{\varepsilon}]$ and its right limit also exists at  $\varepsilon=0$. Furthermore, for any user $i$, the aggregate charge and discharge requirement $(\bm{p}_a^{\omega,ch}(q,\varepsilon),\bm{p}_a^{\omega,dis}(q,\varepsilon))$, which is determined by \text{(11)} and \text{(11)}, is a continuous function in  $\left(\bm{p}_i^{ch,\omega\star}(q,\varepsilon),\bm{p}_i^{dis,\omega\star}(q,\varepsilon)\right)$. Therefore,  the aggregate charge and discharge demand $(\bm{p}_a^{\omega,ch}(q,\varepsilon),\bm{p}_a^{\omega,dis}(q,\varepsilon))$ is a continuous function  with respect to  $\varepsilon \in (0,\bar{\varepsilon}]$ and its right limit  exists at  $\varepsilon=0$.    For the aggregator's optimization Problem $\mathbf{CO}$ in Proposition 3(b),  the optimal cost  ${C}_a$ is a continuous function of $(\bm{p}_a^{\omega,ch}(q,\varepsilon),\bm{p}_a^{\omega,dis}(q,\varepsilon))$  according to {the  Maximum Theorem}\cite{maximum}. Then, we have that the optimal cost  ${C}_a$ is a continuous function  in $\varepsilon \in (0,\bar{\varepsilon}]$ and its right limit exists at  $\varepsilon=0$. We can achieve the limiting cost $C_a(q)$ by computing the optimal objective value of Problem $\mathbf{CO}$ given user $i$'s limiting charge and discharge decision $\left(\lim_{\varepsilon \rightarrow 0^+}\bm{p}_i^{\omega,ch\star}(q,\varepsilon), \lim_{\varepsilon \rightarrow 0^+}\bm{p}_i^{\omega,dis\star}(q,\varepsilon)\right)$, for all $i,\omega$.

Then, in Stage 2,  the limits  $\lim_{\varepsilon \rightarrow 0^+}\bm{p}_i^{\omega,ch\star}(q,\varepsilon)$ and $\lim_{\varepsilon \rightarrow 0^+}\bm{p}_i^{\omega,ch\star}(q,\varepsilon)$  are both stepwise  over the threshold-price interval  $(0,q_i^{\omega_1}),(q_i^{\omega_1},q_i^{\omega_2})...$ as defined in \eqref{eq:xc}. 
Thus, in Stage 1, the limits of $\bm{p}_a^{\omega,ch}(q,\varepsilon)$ and $\bm{p}_a^{\omega,dis}(q,\varepsilon)$ are also stepwise  over the price set $\bigcup_i \mathcal{Q}_i^\omega$ as $\varepsilon \rightarrow 0^+$.   According to {the  Maximum Theorem}\cite{maximum}, the aggregator's cost $C_a$ is continuous in  $(\bm{p}_a^{\omega,ch}(q,\varepsilon),  \bm{p}_a^{\omega,dis}(q,\varepsilon))$, which implies that as $\varepsilon\rightarrow 0^+$, the total cost also approaches a function that is stepwise over $\mathcal{Q}_a=\bigcup_i \bigcup_\omega \mathcal{Q}_i^\omega$, which is denoted as  $C_a(q)$:

$$  C_a(q)= \left \{
\begin{aligned}
&C^0,q \in (0,q_a^1),\\
&C^1,q \in (q_a^1,q_a^2),\\
&...\\
&C^{K_a},q \in (q_i^{K_a},\infty),\\
\end{aligned}
\right.$$
where $C^0,C^1,..C^{K-1},C^{K_a}=0$ are constant.

\subsubsection*{3) The limit of profit}

We characterize the limit of the profit based on the limits of the revenue and cost as $\varepsilon$ approaches zero.

Since
	\begin{align*}
     &\lim_{\varepsilon \rightarrow 0^+}R_a^v(q,\varepsilon)=R_a^v(q),\\
     & \lim_{\varepsilon \rightarrow
	 0^+}{C}_a(q,\varepsilon)=  C_a(q),
    \end{align*}
 
we have that
 	\begin{align*}
   \lim_{\varepsilon \rightarrow 0^+} R_a^{pf}(q,\varepsilon)=\lim_{\varepsilon \rightarrow 0^+}(R_a^v(q,\varepsilon)-{C}_a(q,\varepsilon))= R_a^v(q)- C_a(q).
    \end{align*}
 \qed

\vspace{7mm}
\section*{Appendix J: Computing  the lowest-nonnegative-profit price}

In this appendix, we develop a solution method to compute the lowest-nonnegative-profit price (LNP price) from the aggregator's perspective. The lowest-nonnegative-profit price is defined as the minimal price that keeps the aggregator’s profit nonnegative. In Section J.A, we will present the formulation of Aggregator's Lowest-nonnegative-profit Price (ALP) problem. Then, in Section J.B, we will present the solution method, the main idea of which is similar to the solution method of computing the optimal-profit price in Algorithm 1.  Specifically, we first characterize the LNP price when $\varepsilon$ goes to zero using the piecewise linear property of the aggregator's profit (that has been proved in Proposition 3).  Then, due to the multi-optima problem as we explained in Section IV.A, the aggregator cannot choose $\varepsilon = 0$. Therefore, we develop an iterative algorithm in Algorithm 2 to compute a near-LNP  price by adjusting $\varepsilon$, such that the near-optimal solution can approximate the LNP price in the limiting case within any given accuracy.

\subsection*{A. The formulation of Problem $\textbf{ALP}$}
As we briefly introduced in Section \uppercase\expandafter{\romannumeral3}.B of the main body of our paper, the LNP price is the minimal price that  keeps the aggregator's profit nonnegative. We can compute the LNP price by solving Problem $\textbf{ALP}$ as follows.

\textbf{Stage 1: Aggregator's Lowest-nonnegative-profit Price Problem ($\textbf{ALP}$)}\footnote{Note that the LNP price $q^l$ must be strictly positive. This is because, when $q=0$, the aggregator's profit  must be strictly negative, i.e., $R_a^{pf}(q,\varepsilon)<0$, which contradicts the constraint of Problem \textbf{ALP}. To see this, note that  the aggregator receives no revenue  at $q=0$, but she still needs to bear the cost of the physical storage to satisfy users' demand. Another reason for excluding  $q=0$ from the feasible set of Problem \textbf{ALP} is due to the multi-solution issue related to Proposition 2.}
	\begin{align*}
	\min_{q>0} &~~ q ~~~\text{s.t.} \  R_a^{pf}(q,\varepsilon)\geq 0.
	\end{align*}

To solve the LNP price in Problem $\textbf{ALP}$, we will utilize the {piecewise linear} structure of  $R_a^{pf}(q,\varepsilon)$ when $\varepsilon$ approaches zero.  We present the detailed solution method next. 
	
\subsection*{B. The solution method for Problem $\textbf{ALP}$}

In Proposition 3, we have characterized the limiting value of the aggregator's profit (when $\varepsilon$ approaches zero) as a piecewise linear function $R_a^{pf}(q)$ in price $q$ (see Figure 6(b)). Based on this piecewise linear structure, in Subsection 1), we can efficiently compute the LNP price $q^{l}$ that makes the limiting profit $R_a^{pf}(q)$ nonnegative. However, this LNP price $q^l$ in the limiting case cannot be used directly because the aggregator cannot choose $\varepsilon = 0$. In Subsection 2), we solve a near-LNP price $\hat{q}^l$ under a small $\varepsilon > 0$ to approximate $q^l$ within any given accuracy.

\subsubsection*{1) The LNP price $q^{l}$ making the  limiting profit $R_a^{pf}(q)$ nonnegative} Our solution method consists of the following steps. First, considering the fact that the limiting profit ${R}_a^{pf}(q)$ is not continuous at each threshold price, we will construct a correspondence $\tilde{R}_a^{pf}(q)$ that is well defined at each threshold price. Second, based on the correspondence $\tilde{R}_a^{pf}(q)$, we formulate a new problem $\mathbf{NALP}$ for solving the LNP price $q^{l}$ that makes $\tilde{R}_a^{pf}(q) \geq 0$.  Third, we analyze the structure of  the LNP price $q^l$ and characterize the necessary and sufficient conditions for computing the LNP price $q^l$. Last but not least, we present a holistic solution method in Algorithm 2 to compute the LNP price $q^l$ based on the necessary and sufficient conditions. Next, we provide the details of these steps.
 
First, we construct a correspondence $\tilde{R}_a^{pf}(q)$ that is well defined at each threshold price.  Recall in Proposition 3 that the piecewise linear function ${R}_a^{pf}(q)$ is discontinuous and not well-defined at the threshold prices.  For any threshold price $q_a \in \mathcal{Q}_a$, we denote the left-handed limit of ${R}_a^{pf}(q_a)$ as $\tilde{R}_a^{pf}(q_a^{-})$ and the right-handed limit as $\tilde{R}_a^{pf}(q_a^{+})$. Then, we construct a  correspondence  $\tilde{R}_a^{pf}(q)$ as follows. For any threshold price $q_a$,  the correspondence $\tilde{R}_a^{pf}(q_a)$ will take both values of the left-handed limit and right-handed limit, i.e.,  $\tilde{R}_a^{pf}(q_a^{-})$ and $\tilde{R}_a^{pf}(q_a^{+})$.\footnote{For the threshold price $q_a=0$, we only choose the right-handed limit, such that $q>0$ is satisfied.} For any price between two consecutive threshold prices, i.e., $q \in (q_a^k,q_a^{k+1})$,\footnote{Recall that the threshold prices are sorted in an increasing order by $0<q_a^1<q_a^2<\cdots<q_a^{K_a}$ as shown in Figure 6(b).}  the  correspondence  $\tilde{R}_a^{pf}(q)$ is equivalent to ${R}_a^{pf}(q)$ which is single-valued, continuous, and linearly increasing. One example of the correspondence is shown in Figure 	\ref{fig:zeroill1}, where $\tilde{R}_a^{pf}(q)$ can take two values  at each threshold price.
   
Second,  we formulate a new problem  $\mathbf{NALP}$ for solving the LNP price as follows.
   \begin{align*}
  \text{Problem}~  \mathbf{{NALP}}:~ \min_{q> 0} &~~ q ~~~\text{s.t.} \  \tilde{R}_a^{pf}(q)\geq 0.
   \end{align*}
Note that $\tilde{R}_a^{pf}(q)$ can take two values  at  any threshold price $q_a$. We define that $\tilde{R}_a^{pf}(q_a)\geq 0$ is satisfied if either $\tilde{R}_a^{pf}(q_a^{-}) \geq 0$ or $\tilde{R}_a^{pf}(q_a^{+}) \geq 0$, i.e., if  at least one limit is nonnegative. We  show one example in Figure 	\ref{fig:zeroill3}, where the LNP price $q^{l}$ is one of the threshold prices, with $\tilde{R}_a^{pf}(q^{l+})>0$ and $\tilde{R}_a^{pf}(q^{l-})<0$.

Third, we characterize the necessary and sufficient conditions to obtain the LNP price $q^{l}$ as follows. Since  $q^{l}$ is the minimal price making $\tilde{R}_a^{pf}(q)\geq 0$,   it is necessary  and sufficient to show that i) $\tilde{R}_a^{pf}(q)<0$ for any price $q<q^{l}$, and ii) $\tilde{R}_a^{pf}(q^{l})\geq 0$. Recall  that the  correspondence  $\tilde{R}_a^{pf}(q)$ is continuous and linearly increasing over any  interval $(q_a^k,q_a^{k+1})$ between threshold prices. Thus, we can focus on the limiting profits at the threshold prices, and consider a total of three cases for the computation of  the price $q^l$.  In Case 1, the  LNP price $q^{l}$ strictly lies between two adjacent threshold prices (as shown in Figure \ref{fig:zeroill1}). In Case 2 and Case 3, the  LNP price $q^{l}$ is exactly located at one of the threshold prices. The difference between the latter two cases is that,  in Case 2,  $\tilde{R}_a^{pf}(q^l)\geq 0$ is satisfied at the left-handed limit with equality (as shown in Figure \ref{fig:zeroill2}), while in Case 3, $\tilde{R}_a^{pf}(q^l)\geq 0$ is satisfied at the right-handed limit but not the left-handed limit (as shown in Figure \ref{fig:zeroill3}). These three cases include all the possibilities for the LNP price $q^{l}$.   Specifically, we summarize the  necessary and sufficient conditions  for each case as follows.
\begin{itemize}
	\item \textbf{Case 1}: The LNP price $q^{l}$ exists between two adjacent threshold prices, i.e., $q^{l}\in (q_a^k,q_a^{k+1})$, as shown in Figure \ref{fig:zeroill1}.\\
	   \textbf{Necessary and sufficient conditions for Case 1}: i) Both $\tilde{R}_a^{pf}(q_a^-)<0$ and  $\tilde{R}_a^{pf}(q_a^+)<0$ for any  threshold price ${q_a}<q_a^{k+1}$; ii) for the  threshold price $q_a^{k+1}$, the left-handed limit  $\tilde{R}_a^{pf}(q_a^{({k+1})-})>0$; iii) $\tilde{R}_a^{pf}(q^l)=0$.
	   \setcounter{figure}{8}
		   \begin{figure}[h]
		   \centering
		   \includegraphics[width=2.3in]{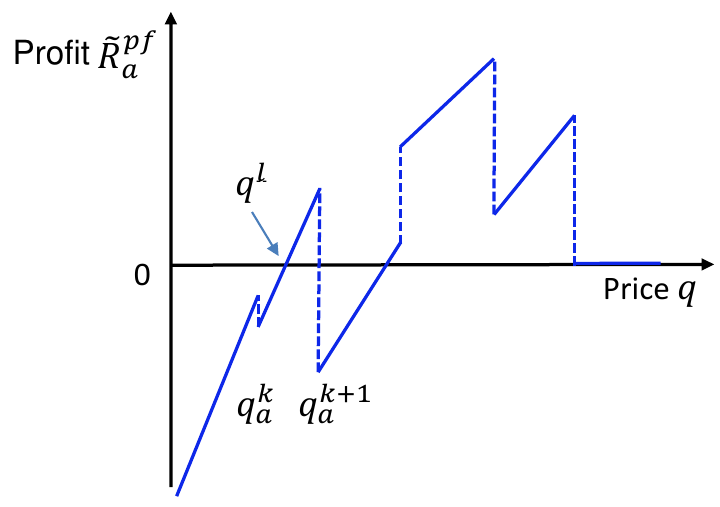}
		   \caption{\small The LNP price in Case 1}
		    \label{fig:zeroill1} 
        	\end{figure}
	\item \textbf{Case 2}: The LNP price $q^{l}$ is exactly one of the threshold prices and the left-handed limit $\tilde{R}_a^{pf}(q^{l-})=0$, as shown in Figure \ref{fig:zeroill2}.\\
		\textbf{Necessary and sufficient conditions  for Case 2}:   i) Both $\tilde{R}_a^{pf}(q_a^-)<0$ and  $\tilde{R}_a^{pf}(q_a^+)<0$ for any  threshold price $q_a<q^{l}$; ii) the left-handed limit  $\tilde{R}_a^{pf}(q^{l-})=0$.
			\begin{figure}[h]
			\centering
			\includegraphics[width=2.3in]{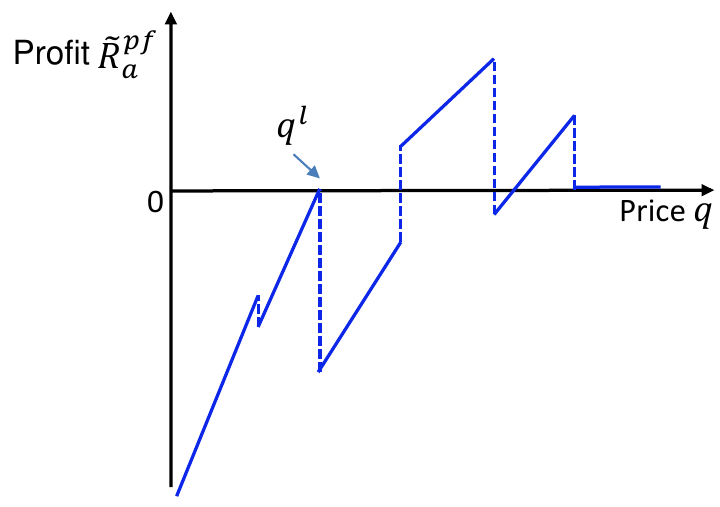}
			\caption{\small The LNP price in Case 2}
			\label{fig:zeroill2}
		    \end{figure}
	\item \textbf{Case 3}: The LNP price $q^{l}$ is exactly one of the threshold prices, the right-handed limit satisfies  $\tilde{R}_a^{pf}(q^{l+})\geq 0$, and the left-handed limit satisfies $\tilde{R}_a^{pf}(q^{l-})< 0$, as shown in Figure \ref{fig:zeroill3}.\\
	    \textbf{Necessary and sufficient conditions  for Case 3}:  i) Both $\tilde{R}_a^{pf}(q_a^-)<0$ and  $\tilde{R}_a^{pf}(q_a^+)<0$ for any  threshold price $q_a<q^{l}$; ii) the left-handed limit $\tilde{R}_a^{pf}(q^{l-})<0$ and the right handed-limit $\tilde{R}_a^{pf}(q^{l+})\geq 0$.
			\begin{figure}[h]
		    \centering
		    \includegraphics[width=2.3in]{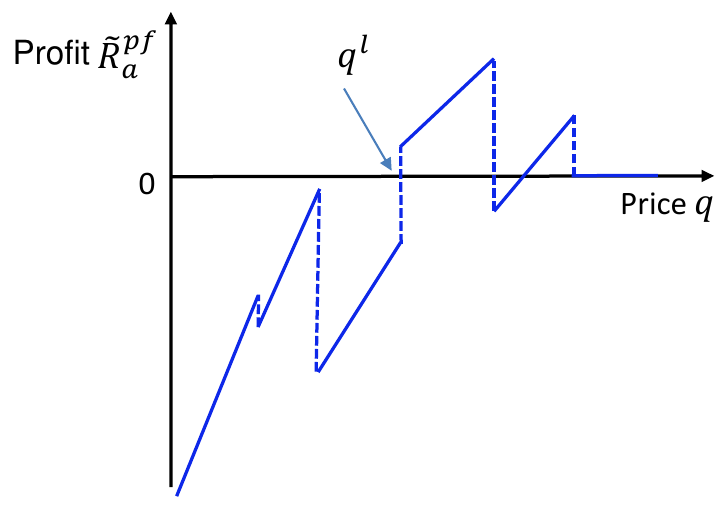}
		    \caption{\small The LNP price in Case 3}
	     	\label{fig:zeroill3}
	       \end{figure}
\end{itemize}

Finally, we present a search algorithm to compute  the LNP price $q^{l}$ in Algorithm 2 based on the three cases. In Lines \ref{al2:user1}-\ref{al2:user2} of Algorithm 2, users compute and report the threshold prices and capacities to the aggregator. Within the for-loop in Lines 8-25, we search the threshold price set $\mathcal{Q}_a$ for the LNP price  in an increasing order. Specifically, we compute the left-handed and right-handed limits  $\tilde{R}_a^{pf}(q_a^{k-})$ and $\tilde{R}_a^{pf}(q_a^{k+})$ at each threshold price in Line \ref{al2:pf1} and Line \ref{al2:pf2}. Then, we will examine the corresponding profits to determine the LNP price $q^{l}$ as  follows.
\begin{itemize}
	\item  If the condition in Line \ref{al:case1} is satisfied in the $k$th iteration in the for-loop of Lines 8-25, the necessary and sufficient conditions (i) and (ii) of Case 1 will be satisfied, which implies that  the LNP price $q^{l}$ lies in the threshold price interval $(q_a^{k-1},q_a^k)$. To further satisfy the necessary and sufficient condition (iii) of Case 1,  we  execute Subroutine 3.1 in Line 14  and compute the LNP price $q^{l}$ that makes $\tilde{R}_a^{pf}(q^l)=0$  in Line \ref{sb31:near}  of Subroutine 3.1, where we  utilize the linear structure over the price interval as shown in Figure \ref{fig:zeroill1}. 
	\item  If the condition in Line \ref{al:case2}  is satisfied in the $k$th iteration in the for-loop  of Lines 8-25, the necessary and sufficient conditions (i) and (ii) of Case 2  will be  both satisfied for the threshold price $q_a^k$. Thus,  we attain the LNP price $q^{l}$ at the threshold price $q_a^k$ as in Case 2.
	\item  If the condition in Line \ref{al:case3}  is satisfied in the $k$th iteration in the for-loop of Lines 8-25,  the necessary and sufficient conditions (i) and (ii) of Case 3 will be both satisfied for the threshold price $q_a^k$. Thus, we attain the LNP price $q^{l}$ at  the threshold price  $q_a^k$ as in Case 3.
\end{itemize}

\subsubsection*{2) The near-LNP price $\hat{q}^{l}$  under a positive $\varepsilon$}
Since the aggregator cannot choose $\varepsilon=0$ due to the multi-optima problem as we explained in Section \uppercase\expandafter{\romannumeral4}.A, we cannot directly use the LNP price $q^l$ in the limiting case where $\varepsilon$ approaches zero.  Thus, we propose an iterative procedure in Subroutines 3.1-3.3 of Algorithm 2 (for Cases 1-3, respectively) to compute a near-LNP price $\hat{q}^{l}$ by adjusting $\varepsilon$, such that i) we can still utilize the piecewise linear solution structure in the limiting case; ii) we can compute a near-LNP price to approximate the  LNP price in the limiting case within a given arbitrary accuracy. Furthermore, after computing the  near-LNP price $\hat{q}^{l}$  and $\varepsilon$, we can compute the  investment decision  accordingly by Subroutine 1.
\begin{itemize}
	\item For Case 1, in Subroutine 3.1, we compute a  small $\varepsilon$  in an iterative fashion (by decreasing $\varepsilon$ iteratively), such that ${R_a^{pf}({q}^{l},\varepsilon)}$ is close to zero within an acceptable  accuracy $err_4$ (in Lines \ref{sb31:eps1}-\ref{sb31:eps2} of Subroutine 3.1). In this way, we use the LNP price $q^{l}$ in the limiting case as the near-LNP price $\hat{q}^{l}$. 
	\item For Case 2, in Subroutine 3.2, the LNP price $q^{l}$ is attained at one threshold price $q_a^k$ and  $\tilde{R}_a^{pf}(q_a^{k-})=0$. Note that we cannot directly choose the threshold price due to the discontinuity of $\tilde{R}_a^{pf}(q)$. Instead, in Subroutine 3.2, since profit $\tilde{R}_a^{pf}(q)$ is linearly increasing  over $(q_a^{k-1},q_a^k)$, we let the aggregator choose a near-LNP price $\hat{q}^{l}$ slightly lower than $q_a^{k}$,  such that the profit ${R}_a^{pf}(\hat{q}^{l})$ approximates  the  profit $\tilde{R}_a^{pf}(q_a^{k-})$ within a given accuracy $err_3$ (in Lines \ref{sb32:near1}-\ref{sb32:near2} of Subroutine 3.2).\footnote{Recall that ${R}_a^{pf}(q)$ is equivalent to $\tilde{R}_a^{pf}(q)$ at non-threshold prices.} Then, the aggregator can compute a small $\varepsilon$ in an iterative fashion so that the profit ${R}_a^{pf}(\hat{q}^{l},\varepsilon)$ approximates ${R}_a^{pf}(\hat{q}^{l})$ within the accuracy $err_4$ (in Lines \ref{sb32:eps1}-\ref{sb32:eps2} of Subroutine 3.2). 
	\item For Case 3, in Subroutine 3.3, the LNP price $q^{l}$ is attained at one threshold price $q_a^k$ and  $\tilde{R}_a^{pf}(q_a^{k+})\geq 0$. Similar to Case 2, we cannot directly choose the threshold price due to the discontinuity of $\tilde{R}_a^{pf}(q)$. In Subroutine 3.3, since profit $\tilde{R}_a^{pf}(q)$ is linearly increasing  over $(q_a^{k},q_a^{k+1})$, we let the aggregator choose a near-LNP price $\hat{q}^{l}$  slightly higher than $q_a^{k}$, such that the profit ${R}_a^{pf}(\hat{q}^{l})$ approximates  the  profit $\tilde{R}_a^{pf}(q_a^{k+})$ within a given accuracy $err_3$ (in Lines \ref{sb33:near1}-\ref{sb33:near2}  of Subroutine 3.3). Then, the aggregator can compute a small $\varepsilon$ in an iterative fashion so that the profit ${R}_a^{pf}(\hat{q}^{l},\varepsilon)$ approximates ${R}_a^{pf}(\hat{q}^{l})$  within the accuracy $err_4$ (in Lines \ref{sb33:eps1}-\ref{sb33:eps2} of Subroutine 3.3).  
\end{itemize}

\begin{algorithm}  
	\caption*{Algorithm 2: Search of the near-LNP price $\hat{q}^l$}
	\label{alg:2}  
	\begin{algorithmic}[1]  
		\STATE {\textbf{initialization}: set  iteration index $k=0$;}
		\FOR  {each user $i\in \mathcal{I}$ \textbf{in parallel}} \label{al2:user1}
		\STATE Compute the  set   $\mathcal{Q}_i^\omega$ and  $\mathcal{X}_i^\omega$ by Algorithm 3 of Appendix D;
		\STATE Compute the limiting optimal charge/discharge decision corresponding to each element in $\mathcal{X}_i^\omega$ by Algorithm 4 of Appendix H, for all $\omega\in {\Omega}$;
		\STATE Report all the computation results to the aggregator;
		\ENDFOR  \label{al2:user2}
		\STATE The aggregator sorts the threshold price set $\mathcal{Q}_a=\bigcup_{i,\omega}\mathcal{Q}_i^\omega$ by an increasing order: $0=q_a^0<q_a^1<\cdots<q_a^{K_a}$. 
		\FOR {$k=0$ to $K_a$}
		\IF {$q_a^k>0$}
		\STATE The aggregator computes $\tilde{R}_a^{pf}(q_a^{k-})=R_a^{v}(q_a^{k-})-C_a(q_a^{k-})$, where  $R_a^{v}(q_a^{k-})$ is the  left-handed limit of the revenue at $q_a^k$ computed by  \text{(20)} and $C_a(q_a^{k-})$ is the  left-handed limit of the cost at $q_a^k$ computed by solving Problem \textbf{CO} as in Proposition 3(b);  \label{al2:pf1}
		\ENDIF
		\STATE The aggregator computes  $\tilde{R}_a^{pf}(q_a^{k+})=R_a^{v}(q_a^{k+})-C_a(q_a^{k+})$, where  $R_a^{v}(q_a^{k+})$ is the  right-handed limit of the revenue at $q_a^k$ computed by \text{(20)} and $C_a(q_a^{k+})$ is the  right-handed limit of the cost at $q$ computed by solving Problem \textbf{CO} as in Proposition 3(b);  \label{al2:pf2}
		\IF {$\tilde{R}_a^{pf}(q_a^{k-})> 0$}  \label{al:case1}
		\STATE The aggregator executes  Subroutine 3.1 and computes
		$\big(\hat{q}^{l}, \varepsilon^{j},  P(\hat{q}^{l},\varepsilon^{j}),  X(\hat{q}^{l},\varepsilon^{j})\big)=QZ^{int}(q_a^k);$
		\STATE \textbf{break};
		\ENDIF
			\IF {$\tilde{R}_a^{pf}(q_a^{k-})= 0$}   \label{al:case2}
		\STATE The aggregator executes  Subroutine 3.2 and computes
	$\big(\hat{q}^{l}, \varepsilon^{j},   P(\hat{q}^{l},\varepsilon^{j}),  X(\hat{q}^{l},\varepsilon^{j})\big)=QZ^{-}(q_a^k);$
		\STATE \textbf{break};
		\ENDIF
	 
		\IF {$\tilde{R}_a^{pf}(q_a^{k+})\geq 0$}    \label{al:case3}
		\STATE The aggregator executes  Subroutine 3.3 and computes
	$\big(\hat{q}^{l}, \varepsilon^{j},  P(\hat{q}^{l},\varepsilon^{j}),X(\hat{q}^{l},\varepsilon^{j})\big)=QZ^{+}(q_a^k);$
		\STATE \textbf{break};
		\ENDIF
		\ENDFOR
		\STATE {\textbf{output}}: \big($\hat{q}^{l}, \varepsilon^{j},  P(\hat{q}^{l},\varepsilon^{j} ), \ X(\hat{q}^{l},\varepsilon^{j})$\big); 
	\end{algorithmic}  
\end{algorithm} 

\begin{algorithm}  
	\caption*{Subroutine 3.1:  $QZ^{int}(q_a^k)$ for Case 1} 
	\label{alg:s3}  
	\begin{algorithmic}[1]  
		\STATE \textbf{input}: price $q_a^k$;
    	\STATE {\textbf{initialization}: set  iteration index $j=0$, $\varepsilon^0>0$,   absolute error $err_3>0$ and $err_4>0$};
		\STATE The aggregator computes the near-LNP price $\hat{q}^{l}$ (which is also the LNP price ${q}^{l}$ in the limiting case)  between the adjacent threshold prices   $q_a^{k-1}$ and $q_a^k$ as follows:
		\begin{align*}
		&\hat{q}^l=\frac{q_a^{k-1} \tilde{R}_a^{pf}(q_a^{k-})-q_a^k\tilde{R}_a^{pf}(q_a^{(k-1)+})}{\tilde{R}_a^{pf}(q_a^{k-})-\tilde{R}_a^{pf}(q_a^{(k-1)+})};
		\end{align*} \label{sb31:near}
	
		\REPEAT   \label{sb31:eps1}
		\STATE $j=j+1$;
		\STATE $\varepsilon^{j}=\varepsilon^{j-1}/10$;
		\STATE The aggregator executes Subroutine 1 and computes $$\left(\hspace{-0.3mm}R_a^{pf}(\hat{q}^l,\varepsilon^{j}),X(\hat{q}^l,\varepsilon^{j}),P(\hat{q}^l,\varepsilon^{j})\hspace{-0.3mm}\right )={CU}(\hat{q}^{l},\varepsilon^{j});$$		
		\UNTIL{	$\mid {R_a^{pf}(\hat{q}^{l},\varepsilon^{j})\mid} \leq err_4;$ } \label{sb31:eps2}
		\STATE {\textbf{output}}: \big($\hat{q}^{l},  \varepsilon^{j}, P(\hat{q}^{l},\varepsilon^{j}),  X(\hat{q}^{l},\varepsilon^{j})$\big); 
	\end{algorithmic}  
\end{algorithm}

\begin{algorithm}  
	\caption*{Subroutine 3.2:  $QZ^-(q_a^k)$ for Case 2}
	\label{alg:s4}  
	\begin{algorithmic}[1]  
		\STATE \textbf{input}: price $q_a^k$;
	\STATE {\textbf{initialization}: set  iteration index $j=0$, $\varepsilon^0>0$,   absolute error $err_3>$ and $err_4>$};
		\STATE The aggregator calculates the slope  $slp(q_a^{k})$  of $R_a^{pf}(q)$ over  the threshold price interval  $(q_a^{k-1},q_a^{k})$:
		$$slp(q_a^{k})=\frac{ \tilde{R}_a^{pf}(q_a^{k-})-\tilde{R}_a^{pf}(q_a^{(k-1)+})}{q_a^{k}-q_a^{k-1}};$$\par \label{sb32:near1}
		\STATE The aggregator  computes the near-LNP  price $\hat{q}^{l}$ (smaller than $q_a^{k}$) such that  $R_a^{pf}(\hat{q}^{l})$  is close enough to zero (within the absolute error $err_3$): \label{sb32:near2}
	\vspace{-1mm}
		\begin{align*}
		&\hat{q}^l= q_a^{k}-\frac{err_3}{slp(q_a^{k})};
		\end{align*}
		\REPEAT   \label{sb32:eps1}
		\STATE $j=j+1$;
		\STATE $\varepsilon^{j}=\varepsilon^{j-1}/10$;
		\STATE The aggregator executes Subroutine 1 and computes $$\left(\hspace{-0.3mm}R_a^{pf}(\hat{q}^l,\varepsilon^{j}),X(\hat{q}^l,\varepsilon^{j}),P(\hat{q}^l,\varepsilon^{j})\hspace{-0.3mm}\right )={CU}(\hat{q}^{l},\varepsilon^{j});$$	
		\UNTIL{	$\mid R_a^{pf}(\hat{q}^{l},\varepsilon^j)-R_a^{pf}(\hat{q}^{l})\mid \leq err_4;$ } \label{sb32:eps2}
		\STATE {\textbf{output}}: \big($\hat{q}^{l}, \varepsilon^{j},  P(\hat{q}^{l},\varepsilon^{j} ), X(\hat{q}^{l},\varepsilon^{j})$\big); 
	\end{algorithmic}  
\end{algorithm}

\begin{algorithm}  
	\caption*{Subroutine 3.3:  $QZ^+(q_a^k)$ for Case 3}
	\label{alg:s5}  
	\begin{algorithmic}[1]  
		\STATE \textbf{input}: price $q_a^k$;
	\STATE {\textbf{initialization}: set  iteration index $j=0$, $\varepsilon^0>0$,   absolute error $err_3>0$ and $err_4>0$};
		
		\STATE The aggregator calculates the slope  $slp(q_a^{k+1})$  of $R_a^{pf}(q)$ over  the threshold price interval $(q_a^{k},q_a^{k+1})$:
	
		$$slp(q_a^{k+1})=\frac{ \tilde{R}_a^{pf}(q_a^{(k+1)-})-\tilde{R}_a^{pf}(q_a^{k+})}{q_a^{k+1}-q_a^{k}};$$ \label{sb33:near1} \par  
		\STATE The aggregator  computes  the near-LNP  price $\hat{q}^{l}$  (larger than $q_a^{k}$) such that  $R_a^{pf}(\hat{q}^{l} )$ is close enough to $\tilde{R}_a^{pf}(q_a^{k+})$ (within the absolute error $err_3$): \label{sb33:near2}
		\vspace{-1mm}
		\begin{align*}
		&\hat{q}^l= q_a^{k}+\frac{err_3}{slp(q_a^{k+1})};
		\end{align*}
		\REPEAT   \label{sb33:eps1}
		\STATE $j:=j+1$;
		\STATE $\varepsilon^{j}=\varepsilon^{j-1}/10$;
		\STATE The aggregator executes Subroutine 1 and computes $$\left(\hspace{-0.3mm}R_a^{pf}(\hat{q}^l,\varepsilon^{j}),X(\hat{q}^l,\varepsilon^{j}),P(\hat{q}^l,\varepsilon^{j})\hspace{-0.3mm}\right )={CU}(\hat{q}^{l},\varepsilon^{j});$$	
		\UNTIL{	$\mid R_a^{pf}(\hat{q}^{l},\varepsilon^j)-R_a^{pf}(\hat{q}^{l})\mid \leq err_4;$} \label{sb33:eps2}
		\STATE {\textbf{output}}: \big($\hat{q}^{l}, \varepsilon^{j} ,  P(\hat{q}^{l},\varepsilon^{j} ), X(\hat{q}^{l},\varepsilon^{j})$\big); 
	\end{algorithmic}  
\end{algorithm} 

\vspace{7mm}

\section*{Appendix K: Simulation}

\subsection*{A. Benchmark}

For the benchmark system for comparison, we consider a system (without virtual storage) where each user invests in a physical storage product (e.g., Tesla Powerwall) by himself.
He will optimize the (fixed) capacity of the physical storage, and can only use his own storage during the lifespan of the storage. Apart from the capacity cost, each user also bears the power rating cost and the operational cost by himself. User $i$ solves an optimization problem $\textbf{BM}_i$ as follows  to minimize his  cost and determines the optimal storage size over the investment phase. 

  \begin{align}
    \min  \ &\sum_{\omega}\rho^\omega\big[
    C_i^e(\bm{p}_i^{\omega,r,u},\bm{p}_i^{\omega,ch},\bm{p}_i^{\omega,dis})-R_i^r(\bm{p}_i^{\omega,r,u})\notag\\
    &+{c^s\sum_{t\in \mathcal{T}}({p_i^{\omega,ch}[t]+p_i^{\omega,dis}[t]})\big]+\kappa c_u^x x_i+\kappa c_u^p p_i}\notag\\
    \hspace{-5mm } \text{s.t.} \  &\text{(3)},\text{(5)},\text{(6)},\text{(7)}, ~\forall \omega \in \Omega \notag;\\ 
    & \gamma^{\text{min}} x_i\leq e_i^\omega[t]\leq {\gamma^{\text{max} }x_i},~\forall t \in \mathcal{T'},\forall \omega \in \Omega \tag{54} \label{eq:34};\\
    &{  0 \leq p_i^{\omega,ch}[t],  p_i^{\omega,dis}[t] \leq p_i},\forall t \in \mathcal{T},~\forall \omega \in \Omega\tag{55}\label{eq:35};\\
    \text{var:}\ \  &x_i,p_i,\{\bm{p}_i^{\omega,r,u},\bm{p}_i^{\omega,ch},\bm{p}_i^{\omega,dis},  \bm{e}_i^\omega, \forall \omega \in \Omega\}\notag,
  \end{align}
where $c_u^x$ is the unit capacity cost  and $c_u^p$ is the unit power rating cost. For Tesla Powerwall, customers need to pay about {\$7000} (including installation fee) for the storage capacity of {14KWh} and an extra inverter cost\cite{Teslap}. \footnote{At present, many companies only provide several  capacity choices for consumers. For example, Tesla Powerwall only offers fixed capacity of 13.5KWh to consumers. In benchmark $\textbf{BM}_i$, we let users flexibly determine the capacity.} 
Compared with user's Problem $\textbf{UP}_i^\omega$ in our virtualization model, in Problem $\textbf{BM}_i$ each user's decision of storage capacity $x_i$ and power rating $p_i$ are fixed for all scenarios as shown in \eqref{eq:34} and \eqref{eq:35}. Apart from the capacity cost $c_u^x x_i$, each user $i$ also bears the operational cost incurred by the charge  decision $\bm{p}_i^{\omega,ch}$ and discharge decision  $\bm{p}_i^{\omega,dis}$ as well as  the inverter/converter cost for the power rating $p_i$.

\subsection*{B. Simulation parameter}

We consider the lithium-ion battery as the energy storage technology. We use realistic load data  from PG\&E Corporation in 
2012 \cite{loaddata} to simulate users' load profiles, and we use wind speed and solar radiation data from Hong Kong Observatory\cite{hkob} to calculate users' renewable generations. To illustrate our storage virtualization model, we simulate a system with three users of different types, and choose 7 typical scenarios to approximate the original scenario set.

{We use the electricity price data from  \cite{dech}, which is a monthly demand charge tariff. However, note that in users' models, we consider a  daily billing period and thus adopt the daily demand charge tariff. Under a monthly demand charge tariff, the peak power consumption is charged only once over 28-31 days (depending on which month is concerned). {However, in a daily demand charge tariff, the peak power consumption is charged on each day, and thus 28-31 times in a month. Therefore, we need to scale down the monthly peak price to the daily peak price, so that users pay the same or similar bill under the daily and monthly demand charge tariff in each month. {We note that it is nontrivial regarding how to scale down and obtain an accurate daily peak price for users since different users  have different daily and monthly peak load patterns.}  Instead, for an illustrative purpose, we calculate the peak price of the  daily demand charge tariff  in our simulations as follows.} Specifically,}
 \begin{figure}[t]
 	\centering
 	\includegraphics[width=3.3in]{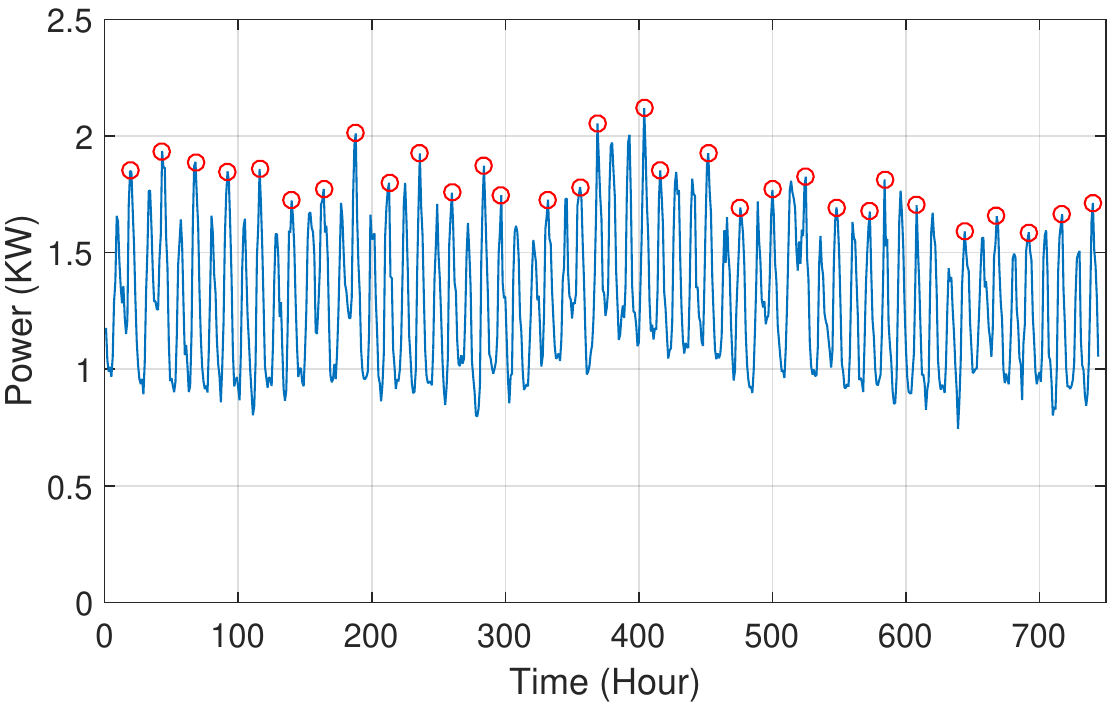}
 	\vspace{-2mm}
 	\caption{\small Type-3 user's load in a month.}
 	\label{fig:loadmonth}
 	\vspace{-5mm}
 \end{figure}
{ we denote the peak price for the {monthly demand charge tariff} and {daily demand charge tariff} by $\pi_p^M$ and $\pi_p^D$, respectively. We denote the peak power in one month of $M$ days (e.g., 30 days) by $p^m$ and the peak power in each day $d$ by $p^d$, where $d=1,2,3,...,M$. 	
 In order to scale $\pi_p^M$ to $\pi_p^D$, we argue that users maintain the same total electricity bill under the daily and monthly demand charge, i.e.,
 $$\sum_{d=1}^{M}p^d \pi_p^D=p^m \pi_p^M.$$
 Thus, we obtain the scaled-down peak price of the daily demand charge tariff as   $$  \pi_p^D=\pi_p^M \frac{p^m}{\sum_{d=1}^{M}p^d}.$$
 In practice, the daily peak power $p^d$ of users in one month can be similar across days.\footnote{In Figure \ref{fig:loadmonth}, we show the hourly load of Type-3 user in the month of 2012 January (with total 744 hours), where the peak power of each day is highlighted in red circle. We can see that the daily peak power is similar, and is around 1.75KW.}  Therefore, for simplicity and for the illustrative purpose,   we consider the case where each day's peak power is the same, so we let  $p^d=p^m$ for all $d$. By choosing $M=30$,  we then obtain the scaling factor $\frac{p^m}{\sum_{d=1}^{M}p^d}=1/30$. Thus, we obtain the peak price as $\pi_p^D=0.4$\$/KW for the \emph{daily} demand charge tariff  based on the realistic peak price $\pi_p^M=10.8$\$/KW of the \emph{monthly} demand charge tariff.
Furthermore, although we choose a daily operational horizon in our model, our framework can be extended to different timescales of the operational horizon and billing cycles, e.g., one month.
}

Therefore, we scale down and obtain the daily peak price $\pi_p$ at
$0.4$\$/kW.  The energy charge price $\pi_b$ is $0.03\$ $/kWh \cite{dech}. We set the renewable energy selling price $\pi_s$ at $0.01\$ $/kWh. For the storage cost, we set the capacity cost $c^X$=160\$/KWh, power rating cost $c^P$=55\$/KW, \footnote{The lithium-ion battery production cost is estimated from 145\$/KWh to 227\$/KWh \cite{batteryc}\cite{batterycc}\cite{batterycost3} in 2017. } and the operational cost $c^s=0.001$\$/\text{kWh}\cite{hao2}. For other technical parameters of the storage, we set efficiency rates $\eta^c =\eta^d=\eta_a^c =\eta_a^d=0.95$, and the effective capacity rate  $\gamma^{\text{min}}=0.9, \gamma^{\text{max}}=1$\cite{overview1}. For the additional resources, we set $c_a^{ch}=0$ and $c_a^{dis}=0.1$\$/KWh \cite{disg}. We choose a sufficiently small penalty coefficient $\varepsilon=3 \times 10^{-7} \$ /(\text{kWh})^2$.

{In addition, our optimization framework is applicable to any parameter configuration of the system.  Interested users can also use other parameter values of the system to test the performance of our framework.}
\vspace{7mm}

\section*{Appendix L: The daily capital recovery factor  $\kappa$ }
The aggregator's investment cost is
  \begin{align*}
    C_a^{cap}(X,P)=\kappa c^{X} X+\kappa c^{P} P.
  \end{align*}

Note  that in practice $c^{X}$ and  $c^{P}$ are given as unit costs over an investment phase of several years. Here, we scale them into the operational horizon of one day by the daily capital recovery factor $\kappa$. {For the factor $\kappa$, we first calculate the present value of an annuity (a series of equal annual cash flows) with the  annual interest rate $r$, and then we divide the annuity equally to each day.} This leads to the formulation of the factor $\kappa$  as follows\cite{NearSitSizeSto},
  \begin{align}
    \kappa=\frac{r(1+r)^y}{(1+r)^y-1}\cdot \frac{1}{Y_d},\tag{56}\label{eq:factor}
  \end{align}
where  $y$ is the number of years over the total time horizon, and $Y_d$ is the number of days (e.g., 365) in one year.

\section*{{Appendix M: Simulation of the case without wind energy}}

We conduct the simulations where one user has no renewable generation and the other two users have solar energy. It can be another common case in practice especially for some areas that are unsuitable to install wind turbines. Recall that we consider three types of users with seven scenarios in the simulation part of Section \uppercase\expandafter{\romannumeral5}: Type-1 user has the wind energy while Type-2 and Type-3 users have solar energy. In this part, we conduct the simulation for the case where Type-1 user has no wind energy while Type-2 and Type-3 users have solar energy. For the parameter setup, we keep the parameters of the load and renewable generation the same as the setup in  Section \uppercase\expandafter{\romannumeral5} except that Type-1 user has no wind energy. In Figure \ref{fig:loadd}, we show the seven scenarios of load and renewable generation for Type-1 user (without wind), Type-2 user and  Type-3 user, respectively.

 We can show that at the  optimal-profit price $q^\star$, the invested physical capacity is reduced by 35\% compared with the sold virtual capacity. In Figures \ref{fig:cost_reduction_without_wind2} and \ref{fig:cost_reduction_without_wind1}, we also show users' cost  reduction in our model compared with the benchmarks (where users install their own physical storage at the production cost $c^p$ and the market price $c^r$, respectively) under the price $q^l$ and $q^\star$. Type-2 and Type-3 users' costs can be reduced by up to 20.4\%, while Type-1 user's cost can only be reduced by up to 3.1\%. This shows that even though there is no wind energy considered for Type-1 user, our model can still work well for significantly reducing the invested physical storage capacity and reducing the cost of Type-2 and Type-3  users with solar energy. Note that in Section \uppercase\expandafter{\romannumeral5} where Type-1 user has the wind generation, Type-1 user's cost can be reduced by up to 34.7\% (instead of 3.1\% when he has no wind generation). This shows that users without renewable energy may benefit less compared to users with renewable energy in our framework.


\begin{figure}[ht]
	\centering
	\subfigure[]{
		\label{fig:load1e} 
		\raisebox{-4mm}{\includegraphics[width=2.1in]{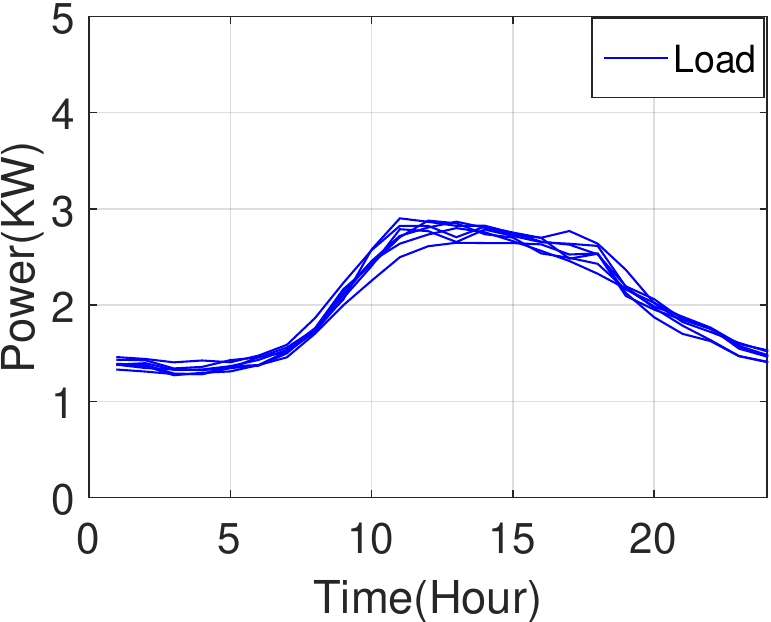}}}
	\hspace{-1ex}
	\subfigure[]{
		\label{fig:load2e} 
		\raisebox{-4mm}{\includegraphics[width=2.1in]{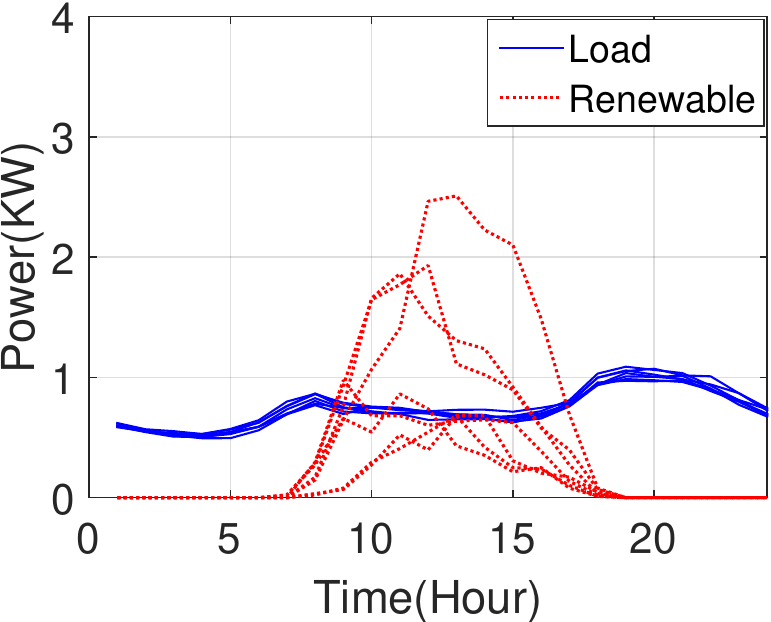}}}
	\hspace{-1ex}
	\subfigure[]{
		\label{fig:load3e} 
		\raisebox{-4mm}{\includegraphics[width=2.1in]{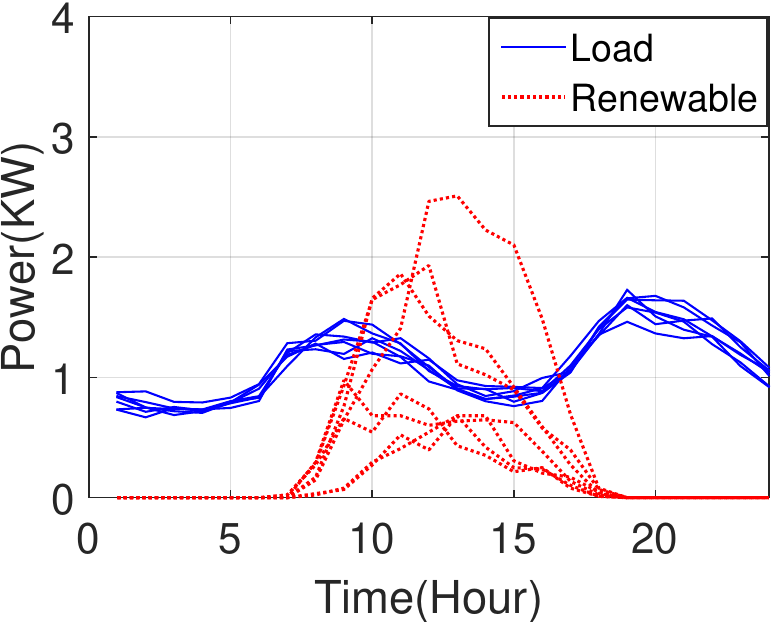}}}
	\vspace{-2mm}
	\caption{\small 7 typical load and renewable generation scenarios for (a) Type-1 user (without wind); (b) Type-2 user; (c) Type-3 user. }
	\label{fig:loadd}
	\vspace{-3mm}
\end{figure}

\begin{figure}[!hpt]
	\centering
	\subfigure[]{
		\label{fig:cost_reduction_without_wind2} 
		\raisebox{-8mm}{\includegraphics[width=2.1in]{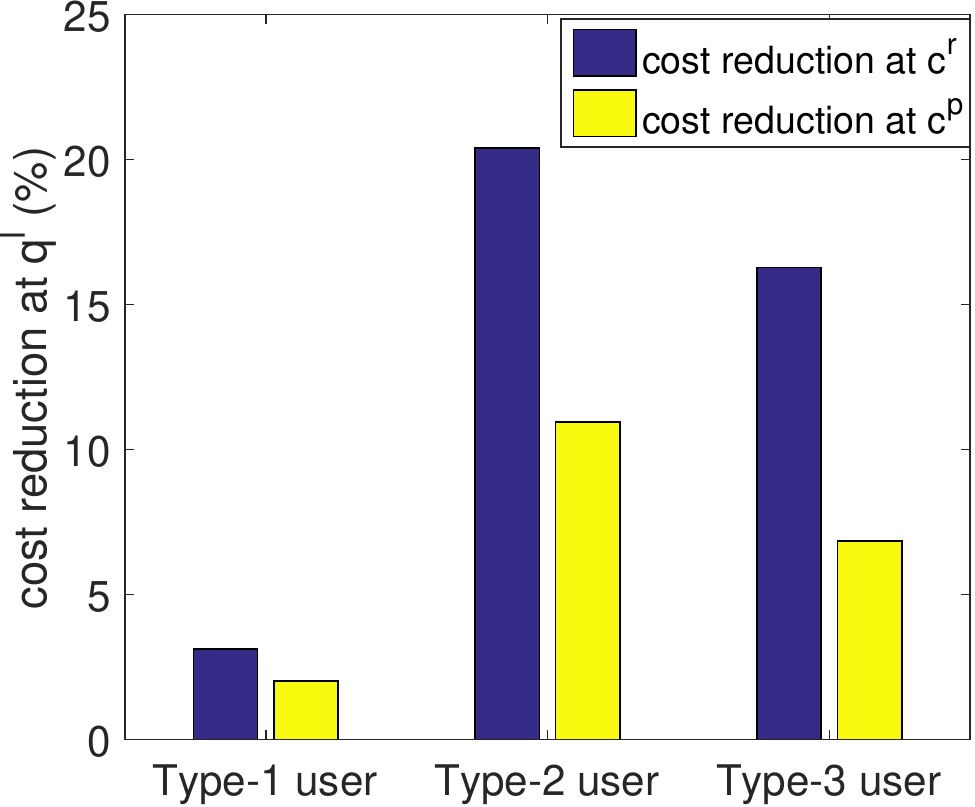}}}
	\hspace{-1ex}
	\subfigure[]{
		\label{fig:cost_reduction_without_wind1} 
		\raisebox{-8mm}{\includegraphics[width=2.1in]{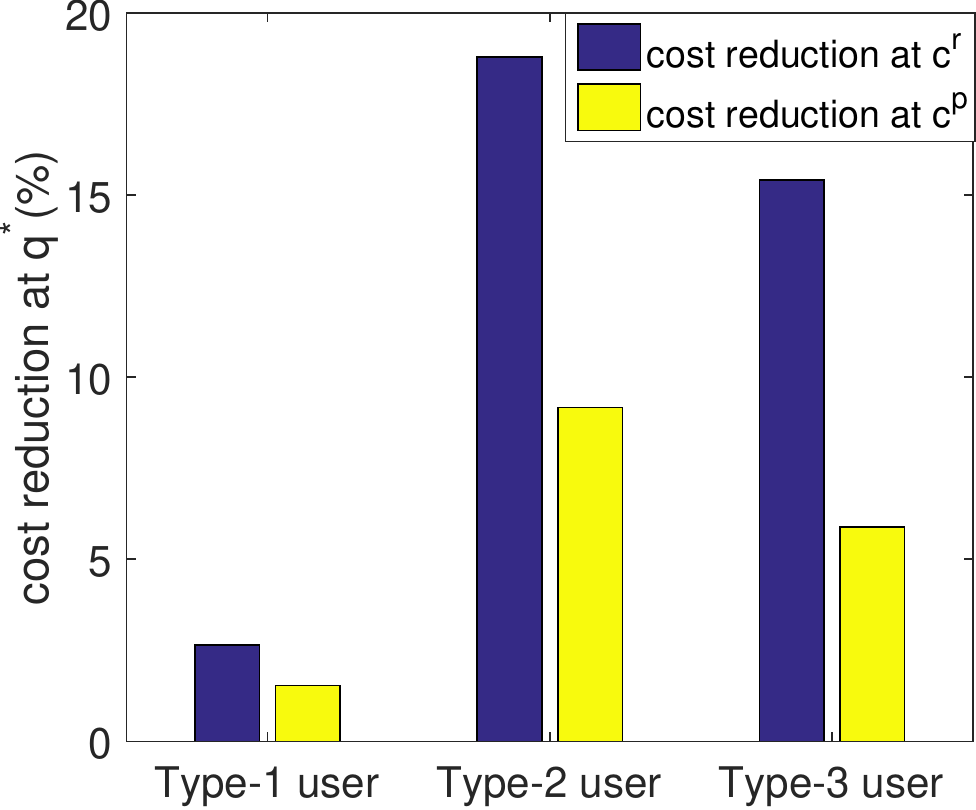}}}
	\vspace{-2mm}
	\caption{\small (a) Cost reduction at $q^l$; (b) Cost reduction at $q^\star$.}
	\label{fig:propp}
	\vspace{-3mm}
\end{figure} 

\section*{{Appendix N: Discussion of the uncertainty of the load and renewable generation}}

 Next we discuss the impact of uncertainties of renewable generations and loads on the virtual storage operation. In our model, users make purchase decisions on the virtual capacity as well as the charge and discharge decisions at the beginning of each day, based on the prediction of their loads and renewable generations for the next day. There have been extensive studies on the prediction of wind power generation \cite{windz},  solar power generation \cite{solarforecast}, and power consumption \cite{homeforecast}. Since the focus of our work is on the design of the virtual storage sharing framework, we have initially chosen to assume that users can perfectly predict their renewable generations and loads. Here we further provide the discussions about the impact of uncertainties in the day-ahead prediction on the users' purchase of virtual capacities and their virtual storage operation.

To study the impact of uncertainties in loads and renewable generations, we incorporate prediction errors in the simulations for users' load and renewable profiles. Recall that we consider three types of users with seven scenarios in the simulation part of Section \uppercase\expandafter{\romannumeral5}. For the impact of uncertainties in loads and renewable profiles, we focus on one type of users and consider one scenario of the load and renewable profiles as the benchmark of the day-ahead prediction.  Specifically, we choose Type-2 user of Scenario 1, whose (day-ahead predicted) load and renewable profiles  are shown  as the solid lines in Figure \ref{fig:random1powere}. We assume that the prediction error $err^l$ of the load profile is characterized by the uniform distribution  $U (-\beta P_i^{\omega,l}[t],\beta P_i^{\omega,l}[t])$ over intervals $[-\beta P_i^{\omega,l}[t],\beta P_i^{\omega,l}[t]]$, where the coefficient $\beta$ denotes the maximum deviation range of the error $err^l$.  Thus, in the real-time operation, the  load  profile is characterized by  $P_i^{\omega,l}[t]+err^l$ when considering the prediction error. Similarly, for the renewable generation, we assume that the prediction error $err^r$ is characterized by the uniform distribution  $U (-\beta P_i^{\omega,r}[t],\beta P_i^{\omega,r}[t])$. Then, in the real-time operation, the renewable generation is characterized by  $P_i^{\omega,r}[t]+ err^r$ when considering the prediction error. For the real-time load and renewable profiles, we randomly generate 50 realizations of the random load  $P_i^{\omega,l}[t]+err^l$ and renewable generation $P_i^{\omega,r}[t]+ err^r$ with $\beta=10\%$, respectively, the maximum deviation ranges (away from day-ahead prediction) of which are depicted by the dashed lines in Figure \ref{fig:random1powere}.

We first solve the users'  optimization problem  $\textbf{UPP}_i^\omega$ with the day-ahead prediction load and renewable generation, and obtain the optimal decisions of the charge/discharge  and the purchased virtual capacities as a benchmark.  Then,  given 50 realizations of the real-time load  and renewable profiles, we also solve the users'  optimization problem  $\textbf{UPP}_i^\omega$, and obtain the corresponding optimal decisions of the  charge/discharge and the purchased capacity.   We regard the optimal decisions under the day-ahead predicted profiles as the benchmark. For  the optimal decisions under 50 realizations  of the real-time load  and renewable profiles, we show their maximum deviation from the benchmark in Figure \ref{fig:prope}. 

Specifically, for users' optimal  decisions, we assume that the aggregator chooses the optimal-profit price $p^\star$ computed under the day-ahead prediction. In Figure \ref{fig:random2e}, we show the optimal charge and discharge decision at the price $p^\star$  under the day-ahead prediction as the black curve (where the positive values represent charge and the negative values represent discharge), and we show the maximum deviations of the optimal decisions under the prediction error away from the decisions of the day-ahead prediction as the blue error bars. In Figure \ref{fig:random3e}, we show the optimal purchased virtual capacities from the price zero to price $p^\star$ under the day-ahead prediction as the  black curve, and we show the maximum deviations under the prediction error away from the day-ahead prediction as the blue error bars.

In Figure \ref{fig:random2e},  we can see that, when considering the prediction error, the whole-day  optimal charge and discharge decisions will not deviate more than 0.3KW compared with the optimal decisions under the day-ahead prediction. As shown in Figure \ref{fig:random3e}, the optimal purchased capacity considering the prediction error will not deviate more than $9\%$ compared with the decisions under the  day-ahead prediction. Furthermore, the cost of Type-2 user under the prediction error of $\beta=10\%$ is within the range $[47.5,~53.1]$ cents, with a maximum deviation of 6\% from the minimized cost 49.87 cents under the day-ahead prediction.

\begin{figure}[ht]
	\centering
	\includegraphics[width=2.4in]{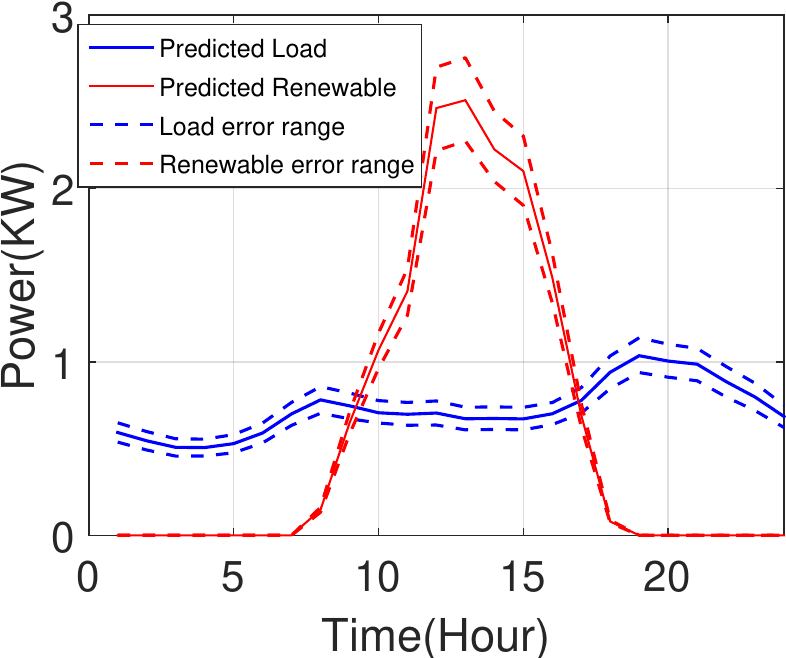}
	\vspace{-2mm}
	\caption{\small Type-2 user's load and renewable profiles.}
	\label{fig:random1powere}
	
\end{figure}

\begin{figure}[ht]
	\centering
	\subfigure[]{
		\label{fig:random2e} 
		\raisebox{-8mm}{\includegraphics[width=2.4in]{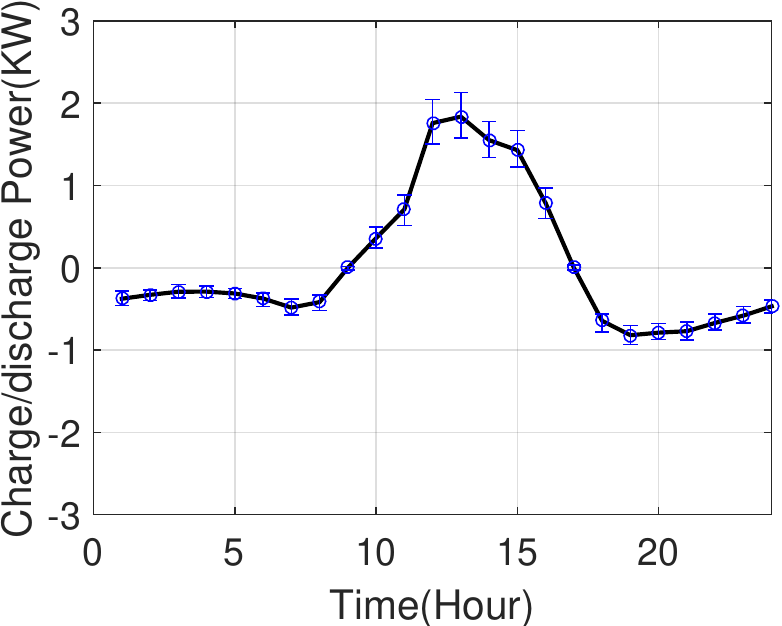}}}
	\hspace{-1ex}
	\subfigure[]{
		\label{fig:random3e} 
		\raisebox{-8mm}{\includegraphics[width=2.5in]{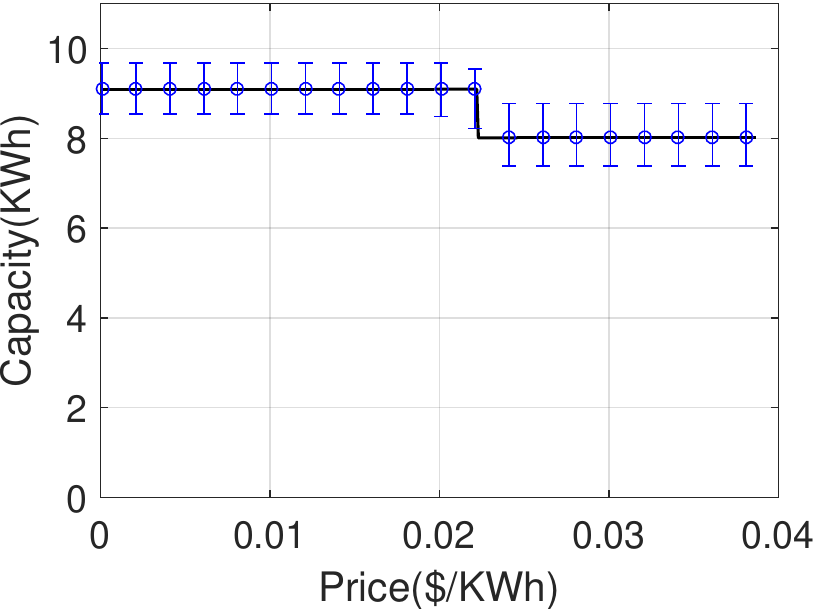}}}
	\vspace{-2mm}
	\caption{\small (a) Type-2 user's optimal charge and discharge decisions; (b) Type-2 user's optimal purchased virtual capacity. }
	\label{fig:prope}
	\vspace{-3mm}
\end{figure} 

\section*{{Appendix O: Diversity of users' load profiles in a localized region}}

 {We refer to the residential load data from   \cite{diverseload} and demonstrate that users' load profiles can be significantly diverse in a localized region. }

{Specifically,   we acquired residential load data from   \cite{diverseload}, which is provided by the Pecan Street Smart Grid Demonstration Program \cite{pecan}.} Over 1000 households in the Mueller community of Austin, Texas, U.S.,  participated in the program and shared their electricity consumption data with the project. We take the  hourly load profiles of four different households on  January 1, 2017 as an illustrative example to show diverse electricity consumption in  Figure \ref{fig:loaddiverse}. We see that Household 1's load profile is flat across the whole day, but Household 2 has a peak load at late night (around 22:00). In contrast, the peak load of Household 3 appears in  the evening (around 17:00), and the peak load of Household 4 occurs around noon (around 11:00). 

Thus, we see that the electricity consumption of different households in one community can be considerably different. Such different load profiles  lead to diverse charge and discharge decisions of users, which can potentially reduce the required capacity  of the physical storage and save the investment cost as demonstrated in the simulations in Section \uppercase\expandafter{\romannumeral 5} of the main text of our paper. 

\begin{figure}[ht]
	\centering
	\includegraphics[width=2.6in]{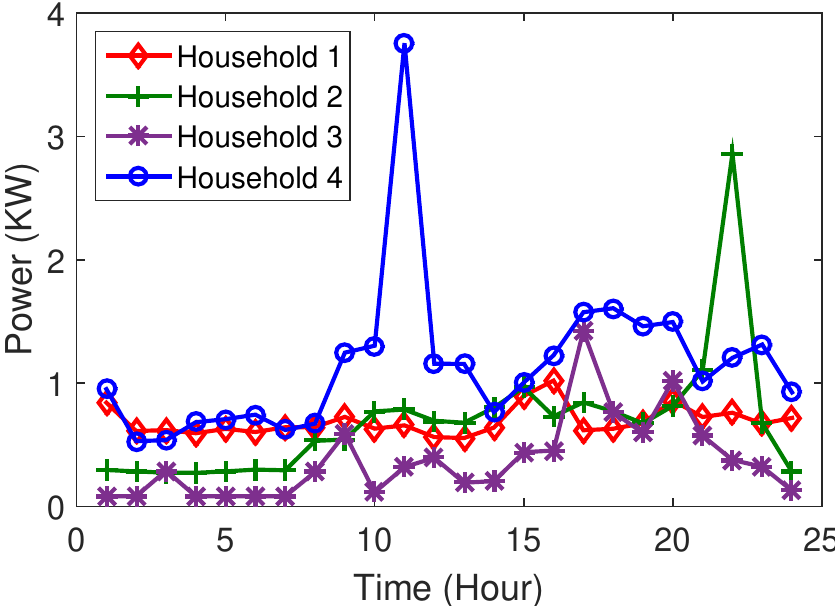}
	\vspace{-2mm}
	\caption{\small Load profiles of four households the Mueller community of Austin,Texas U.S. \cite{pecan}.}
	\label{fig:loaddiverse}
\end{figure}

\section*{{Appendix P: Peak reduction in the system}}

 	\begin{figure}[ht]
	\centering
	\subfigure[]{
		\label{fig:subfig:load11} 
		\raisebox{-1mm}{\includegraphics[width=2.3in]{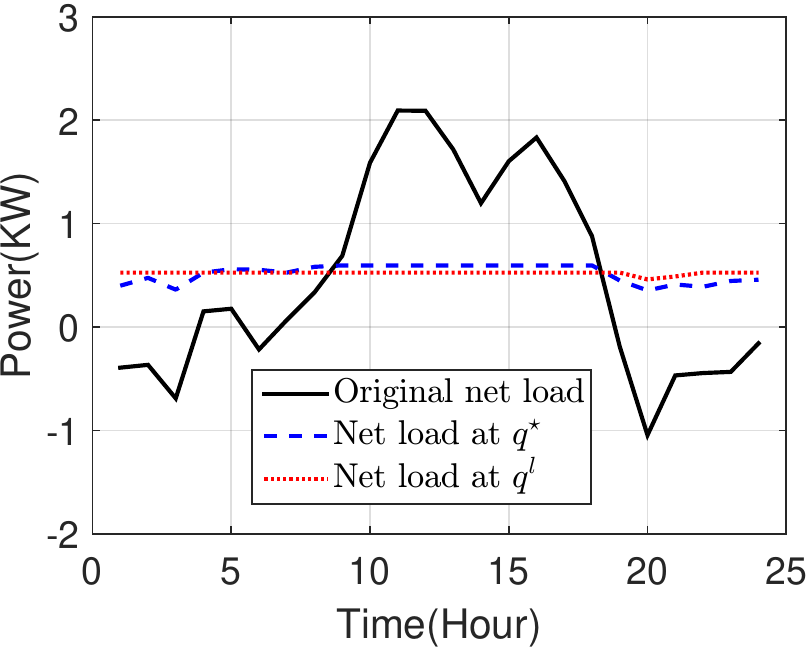}}}
	\hspace{-3mm}
	\subfigure[]{
		\label{fig:subfig:load22} 
		\raisebox{-1mm}{\includegraphics[width=2.3in]{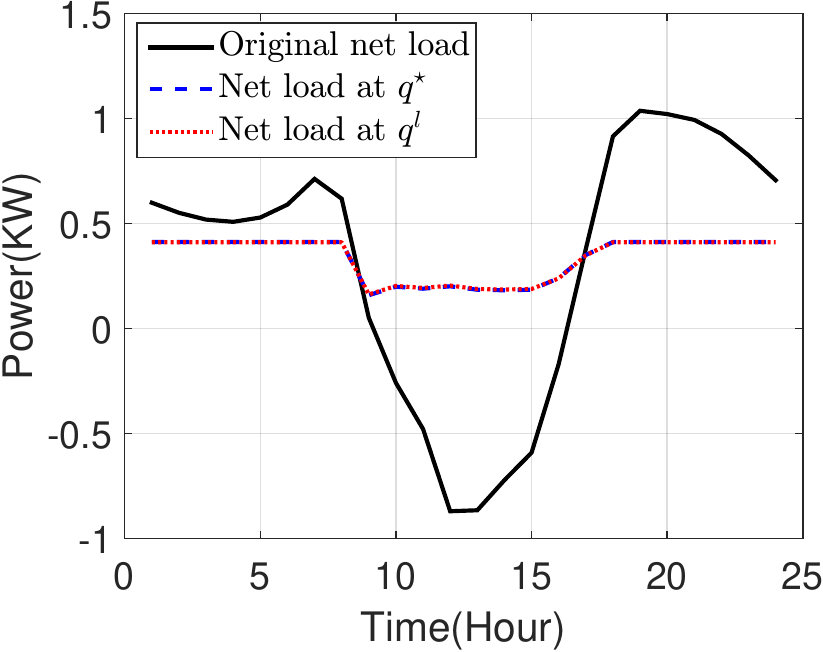}}}
	\hspace{-3mm}
	\subfigure[]{
		\label{fig:subfig:load33} 
		\raisebox{-1mm}{\includegraphics[width=2.3in]{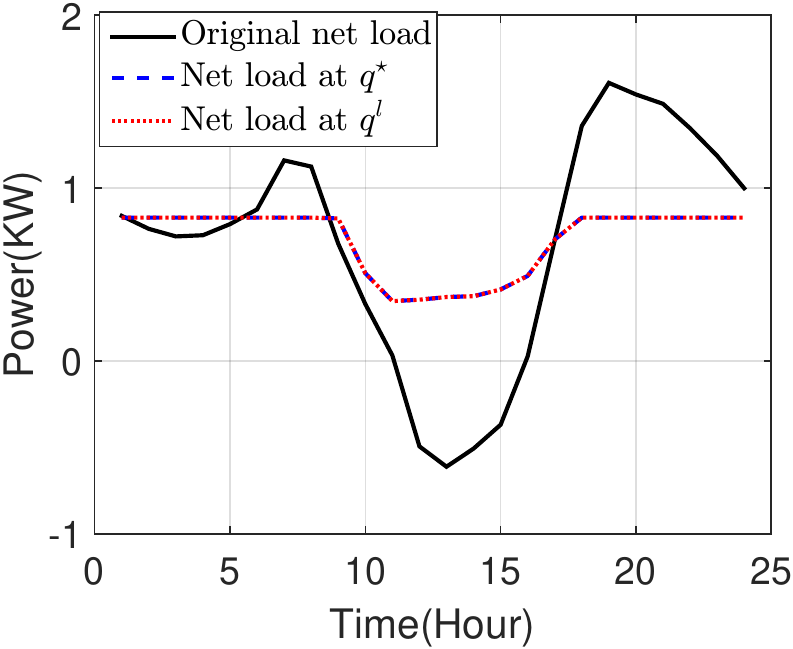}}}
	\subfigure[]{
		\label{fig:sub:peak1} 
		\raisebox{-1mm}{\includegraphics[width=2.3in]{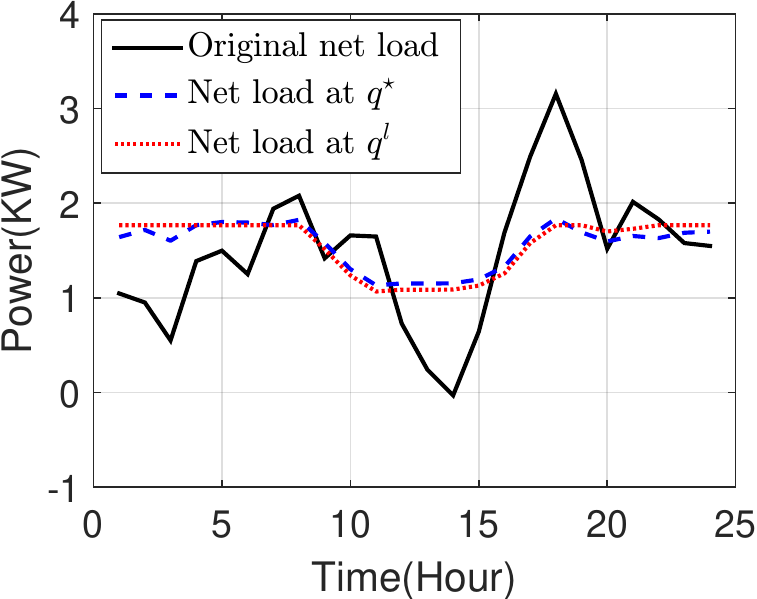}}}
	\vspace{-3mm}
	\caption{\small Net load of (a) Type-1 user; (b) Type-2 user; (c) Type-3 user;  (d) the system. }
	\label{fig:loadxx}
	\vspace{-6mm}
\end{figure} 
Even though we do not explicitly consider minimizing the system peak load as the objective in our model, we numerically demonstrate that  the virtual charge and discharge of users not only helps users cut their electricity bill but also leads to a peak load reduction in the system

For the purpose of illustration, we depict the results of the peak-load reduction of individual users and the peak-load reduction of the system in Figure \ref{fig:loadxx}, considering a three-user system. In  Figure \ref{fig:subfig:load11}, \ref{fig:subfig:load22} and \ref{fig:subfig:load33}, we show three types of  users' original net daily load (i.e., load minus the renewable generation) compared with the net load after utilizing the virtual storage (where the blue dashed curve corresponds to the optimal-profit (OP) price $q^\star$ and the red dotted curve  corresponds to the lowest-nonnegative-profit (LNP) price $q^l$). We  see that each user's peak net-load is shaved significantly by more than 40\%.\footnote{Note that for Type-2 and Type-3 users, their decisions of virtual storage are the same at OP price $q^\star$ and at LNP price $q^l$, and thus the net load curves overlap with each other.} In Figure \ref{fig:sub:peak1}, we show the aggregated original system net load (i.e., aggregated load minus the aggregated renewable generation of all three users) compared with the aggregated system net load after utilizing the virtual storage (where the blue dashed curve  corresponds to the OP price $q^\star$ and the red dotted curve corresponds to the LNP price $q^l$). We see that the aggregated peak load of the system in one day is reduced by 41.7\% at $q^\star$ and  44.0\% at   $q^l$, compared with the system peak load without virtual storage. Thus, the results shown in Figure \ref{fig:loadxx} demonstrate  that our model not only reduces users' peak load,  but also reduces the system peak, and therefore benefits the whole system.

\bibliographystyle{IEEEtran}
\bibliography{storage,bilevel}
	\begin{IEEEbiography}[{\includegraphics[width=1in,height=1.25in,clip,keepaspectratio]{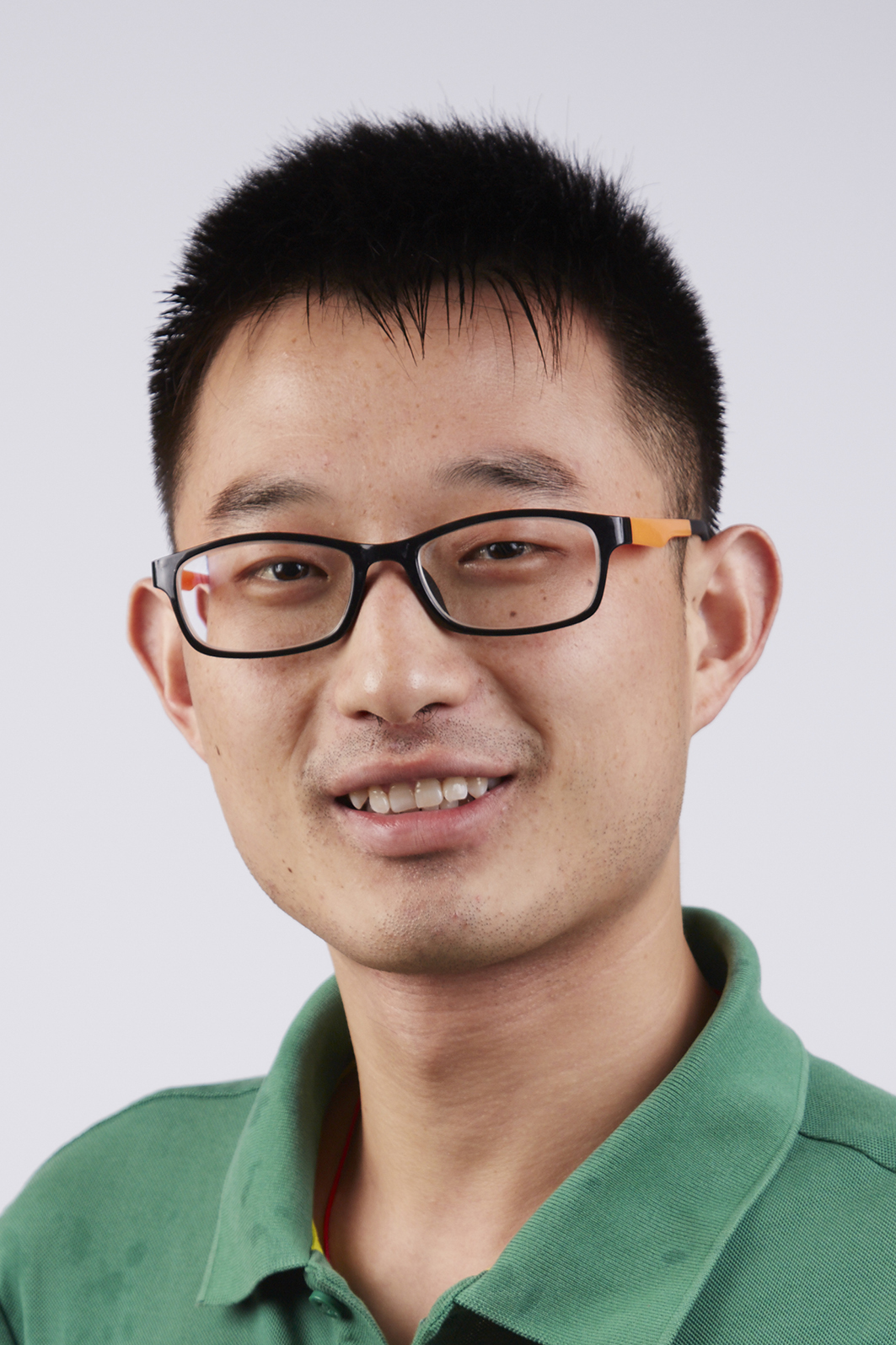}}]{Dongwei Zhao}
	Dongwei Zhao (S'16) received the B.E. degree from the College of Electrical Engineering, Zhejiang University, Hangzhou, China, in 2015. He is currently pursuing the Ph.D.
	degree with the Department of Information Engineering, The Chinese University of Hong Kong.
	His research areas include optimization and game theory of power and energy systems.
\end{IEEEbiography}

\begin{IEEEbiography}[{\includegraphics[width=1in,height=1.25in,clip,keepaspectratio]{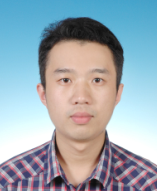}}]{Hao Wang}
	Hao Wang (M'16) is a Postdoctoral Scholar at Stanford University. He received his Ph.D. from The Chinese University of Hong Kong and has been a Washington Research Foundation Innovation Fellow at University of Washington, Seattle. His main research interests are in the optimization, machine learning, and data analytics of power and energy systems. More information at https://web.stanford.edu/~haowang6/.
\end{IEEEbiography}

\begin{IEEEbiography}[{\includegraphics[width=1in,height=1.25in,clip,keepaspectratio]{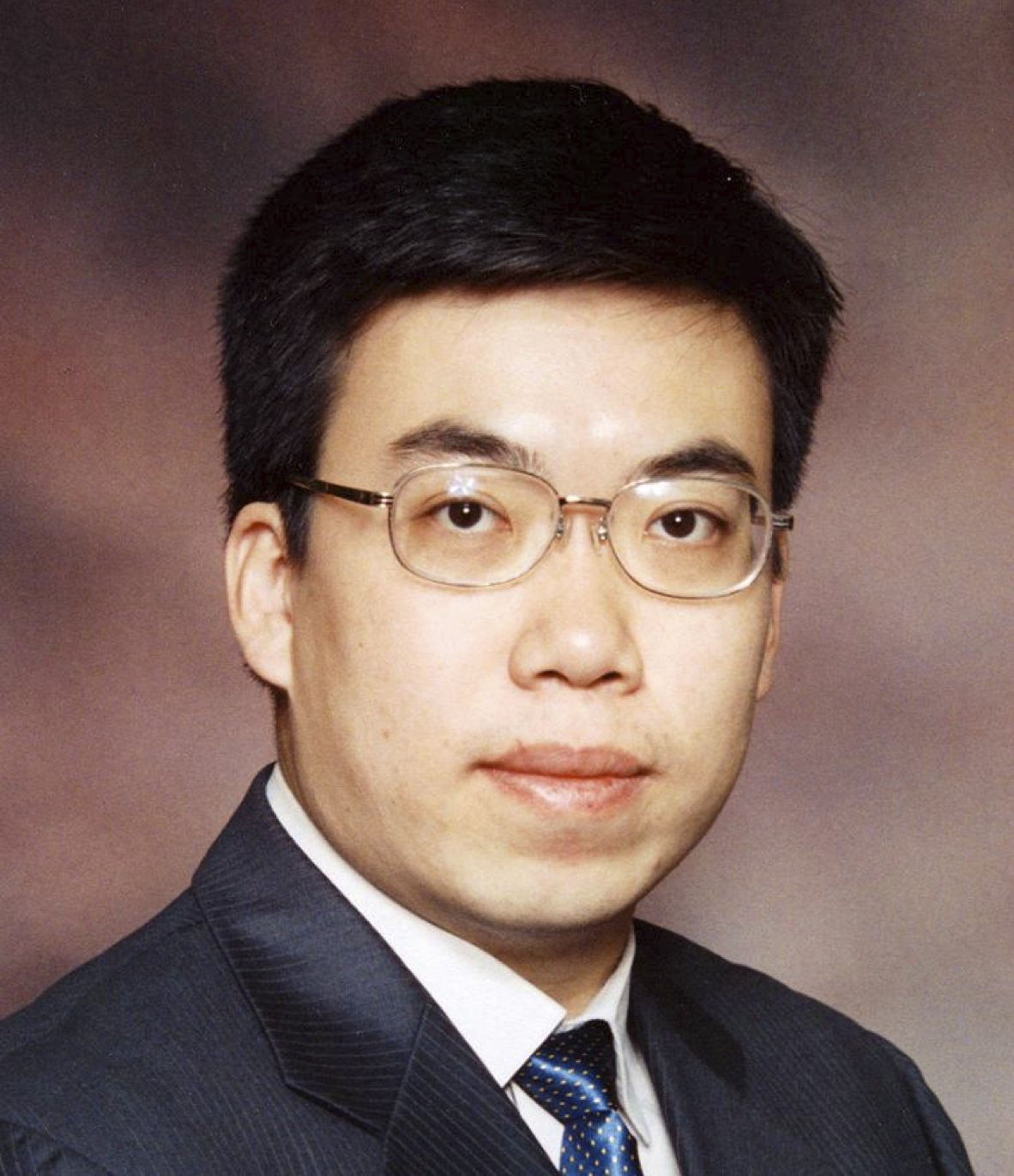}}]{Jianwei Huang}
	Jianwei Huang (F'16) is a Presidential Chair Professor and the Associate Dean of the School of Science and Engineering, The Chinese University of Hong Kong, Shenzhen. He is also a Professor in the Department of Information Engineering, The Chinese University of Hong Kong. He is the co-author of 9 Best Paper Awards, including IEEE Marconi Prize Paper Award in Wireless Communications 2011. He has co-authored six books, including the textbook on "Wireless Network Pricing”. He has served as the Chair of IEEE ComSoc Cognitive Network Technical Committee and Multimedia Communications Technical Committee. He has been an IEEE Fellow, an IEEE ComSoc Distinguished Lecturer, and a Clarivate Analytics Highly Cited Researcher. More  information at http://jianwei.ie.cuhk.edu.hk/.
\end{IEEEbiography}

\begin{IEEEbiography}[{\includegraphics[width=1in,height=1.25in,clip,keepaspectratio]{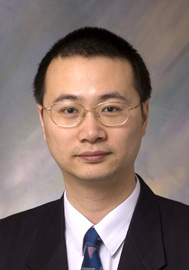}}]{Xiaojun
		Lin} (S'02 M'05 SM'12 F'17) received his B.S. from Zhongshan
	University, Guangzhou, China, in 1994, and his M.S. and Ph.D.
	degrees from Purdue University, West Lafayette, IN, in 2000 and
	2005, respectively. He is currently a Professor of
	Electrical and Computer Engineering at Purdue University.
	
	Dr. Lin's research interests are in the analysis, control and
	optimization of large and complex networked systems, including
	both communication networks and power grid. He received the IEEE
	INFOCOM 2008 best paper and 2005 best paper of the year award
	from Journal of Communications and Networks.  He received the
	NSF CAREER award in 2007. He was the Workshop co-chair for IEEE
	GLOBECOM 2007, the Panel co-chair for WICON 2008, the TPC
	co-chair for ACM MobiHoc 2009, the Mini-Conference co-chair for
	IEEE INFOCOM 2012, and General co-chair for ACM e-Energy 2019.
	He is currently serving as an Area Editor for (Elsevier)
	Computer Networks Journal, and has served as an Associate Editor
	for IEEE/ACM Transactions on Networking and a Guest Editor for
	(Elsevier) Ad Hoc Networks journal.  
\end{IEEEbiography}

\end{document}